\documentclass[aps,prc,showpacs,preprintnumbers,
               nofootinbib,float,superscriptaddress,longbibliography]{revtex4-1}

\usepackage{epsfig}
\usepackage{color}
\usepackage[dvipsnames]{xcolor}
\usepackage{float,amsmath,amssymb,diagbox}
\usepackage{graphicx}
\usepackage{url}
\usepackage{bbold}
\usepackage{ulem}
\usepackage[utf8]{inputenc}
\usepackage{hyperref}
\usepackage{lineno}

\definecolor{applegreen}{rgb}{0.55, 0.71, 0.0}

\definecolor{dgreen}{cmyk}{1.,0.,1.,0.2}
\definecolor{orange}{cmyk}{0.,0.353,1.,0.}

\newcommand{\emphasize}[1]{{#1}}

\newcommand{ \be }{\begin{eqnarray}}
\newcommand{ \ee }{\end{eqnarray}}

\newcommand{ \dgv }{(\Delta\gamma/v_2)}
\newcommand{ \dgamma }{\Delta\gamma}
\newcommand{ \sss }{\scriptscriptstyle}
\newcommand{\ru}{\rm Ru+Ru}
\newcommand{\zr}{\rm Zr+Zr}
\newcommand{\Ru}{\rm Ru+Ru}
\newcommand{\Zr}{\rm Zr+Zr}

\begin{document}

\title{Search for the Chiral Magnetic Effect with Isobar Collisions at $\sqrt{s_{_{\rm NN}}} = 200$ GeV by the STAR  Collaboration at RHIC}

\affiliation{Abilene Christian University, Abilene, Texas   79699}
\affiliation{AGH University of Science and Technology, FPACS, Cracow 30-059, Poland}
\affiliation{Alikhanov Institute for Theoretical and Experimental Physics NRC "Kurchatov Institute", Moscow 117218, Russia}
\affiliation{Argonne National Laboratory, Argonne, Illinois 60439}
\affiliation{American University of Cairo, New Cairo 11835, New Cairo, Egypt}
\affiliation{Brookhaven National Laboratory, Upton, New York 11973}
\affiliation{University of California, Berkeley, California 94720}
\affiliation{University of California, Davis, California 95616}
\affiliation{University of California, Los Angeles, California 90095}
\affiliation{University of California, Riverside, California 92521}
\affiliation{Central China Normal University, Wuhan, Hubei 430079 }
\affiliation{University of Illinois at Chicago, Chicago, Illinois 60607}
\affiliation{Creighton University, Omaha, Nebraska 68178}
\affiliation{Czech Technical University in Prague, FNSPE, Prague 115 19, Czech Republic}
\affiliation{Technische Universit\"at Darmstadt, Darmstadt 64289, Germany}
\affiliation{ELTE E\"otv\"os Lor\'and University, Budapest, Hungary H-1117}
\affiliation{Frankfurt Institute for Advanced Studies FIAS, Frankfurt 60438, Germany}
\affiliation{Fudan University, Shanghai, 200433 }
\affiliation{University of Heidelberg, Heidelberg 69120, Germany }
\affiliation{University of Houston, Houston, Texas 77204}
\affiliation{Huzhou University, Huzhou, Zhejiang  313000}
\affiliation{Indian Institute of Science Education and Research (IISER), Berhampur 760010 , India}
\affiliation{Indian Institute of Science Education and Research (IISER) Tirupati, Tirupati 517507, India}
\affiliation{Indian Institute Technology, Patna, Bihar 801106, India}
\affiliation{Indiana University, Bloomington, Indiana 47408}
\affiliation{Institute of Modern Physics, Chinese Academy of Sciences, Lanzhou, Gansu 730000 }
\affiliation{University of Jammu, Jammu 180001, India}
\affiliation{Joint Institute for Nuclear Research, Dubna 141 980, Russia}
\affiliation{Kent State University, Kent, Ohio 44242}
\affiliation{University of Kentucky, Lexington, Kentucky 40506-0055}
\affiliation{Lawrence Berkeley National Laboratory, Berkeley, California 94720}
\affiliation{Lehigh University, Bethlehem, Pennsylvania 18015}
\affiliation{Max-Planck-Institut f\"ur Physik, Munich 80805, Germany}
\affiliation{Michigan State University, East Lansing, Michigan 48824}
\affiliation{National Research Nuclear University MEPhI, Moscow 115409, Russia}
\affiliation{National Institute of Science Education and Research, HBNI, Jatni 752050, India}
\affiliation{National Cheng Kung University, Tainan 70101 }
\affiliation{Nuclear Physics Institute of the CAS, Rez 250 68, Czech Republic}
\affiliation{Ohio State University, Columbus, Ohio 43210}
\affiliation{Institute of Nuclear Physics PAN, Cracow 31-342, Poland}
\affiliation{Panjab University, Chandigarh 160014, India}
\affiliation{Pennsylvania State University, University Park, Pennsylvania 16802}
\affiliation{NRC "Kurchatov Institute", Institute of High Energy Physics, Protvino 142281, Russia}
\affiliation{Purdue University, West Lafayette, Indiana 47907}
\affiliation{Rice University, Houston, Texas 77251}
\affiliation{Rutgers University, Piscataway, New Jersey 08854}
\affiliation{Universidade de S\~ao Paulo, S\~ao Paulo, Brazil 05314-970}
\affiliation{University of Science and Technology of China, Hefei, Anhui 230026}
\affiliation{Shandong University, Qingdao, Shandong 266237}
\affiliation{Shanghai Institute of Applied Physics, Chinese Academy of Sciences, Shanghai 201800}
\affiliation{Southern Connecticut State University, New Haven, Connecticut 06515}
\affiliation{State University of New York, Stony Brook, New York 11794}
\affiliation{Instituto de Alta Investigaci\'on, Universidad de Tarapac\'a, Arica 1000000, Chile}
\affiliation{Temple University, Philadelphia, Pennsylvania 19122}
\affiliation{Texas A\&M University, College Station, Texas 77843}
\affiliation{University of Texas, Austin, Texas 78712}
\affiliation{Tsinghua University, Beijing 100084}
\affiliation{University of Tsukuba, Tsukuba, Ibaraki 305-8571, Japan}
\affiliation{Valparaiso University, Valparaiso, Indiana 46383}
\affiliation{Variable Energy Cyclotron Centre, Kolkata 700064, India}
\affiliation{Warsaw University of Technology, Warsaw 00-661, Poland}
\affiliation{Wayne State University, Detroit, Michigan 48201}
\affiliation{Yale University, New Haven, Connecticut 06520}

\author{M.~S.~Abdallah}\affiliation{American University of Cairo, New Cairo 11835, New Cairo, Egypt}
\author{B.~E.~Aboona}\affiliation{Texas A\&M University, College Station, Texas 77843}
\author{J.~Adam}\affiliation{Brookhaven National Laboratory, Upton, New York 11973}
\author{L.~Adamczyk}\affiliation{AGH University of Science and Technology, FPACS, Cracow 30-059, Poland}
\author{J.~R.~Adams}\affiliation{Ohio State University, Columbus, Ohio 43210}
\author{J.~K.~Adkins}\affiliation{University of Kentucky, Lexington, Kentucky 40506-0055}
\author{G.~Agakishiev}\affiliation{Joint Institute for Nuclear Research, Dubna 141 980, Russia}
\author{I.~Aggarwal}\affiliation{Panjab University, Chandigarh 160014, India}
\author{M.~M.~Aggarwal}\affiliation{Panjab University, Chandigarh 160014, India}
\author{Z.~Ahammed}\affiliation{Variable Energy Cyclotron Centre, Kolkata 700064, India}
% ...

\collaboration{STAR Collaboration}\noaffiliation
\date{\today}

\begin{abstract}

\vspace{10pt}
The chiral magnetic effect (CME) is predicted to occur as a consequence of a local violation of $\cal P$ and $\cal CP$ symmetries
of the strong interaction amidst a strong electro-magnetic field generated in relativistic heavy-ion collisions.
Experimental manifestation of the CME involves a separation of positively and negatively charged hadrons along the direction of the magnetic field.
Previous measurements of the CME-sensitive charge-separation observables remain inconclusive because of large background contributions. In order to better control the influence of signal and backgrounds, the STAR Collaboration
 performed a blind analysis of a large data sample of approximately 3.8 billion isobar collisions of {$^{96}_{44}$Ru+$^{96}_{44}$Ru and $^{96}_{40}$Zr+$^{96}_{40}$Zr} 
at $\sqrt{s_{_{\rm NN}}}=200$~GeV. 
Prior to the blind analysis, the CME signatures are predefined as a significant excess of the CME-sensitive observables
in Ru+Ru collisions over those in Zr+Zr collisions, owing to a larger magnetic field in the former. 
A precision down to 0.4\% is achieved, as anticipated, in the relative magnitudes of the pertinent observables between the two isobar systems. 
Observed differences in the multiplicity and flow harmonics at the matching centrality indicate that the magnitude of the CME background is different between the two species. 
No CME signature that satisfies the predefined criteria has been observed in isobar collisions in this blind analysis.
\end{abstract}

\maketitle

\tableofcontents

\clearpage

\section{Introduction}

In heavy-ion collisions, an exciting possibility is that regions may be briefly formed in which parity ($\cal P$) and charge-parity ($\cal CP$) symmetries are locally violated by the strong interaction~\cite{Kharzeev:1998kz,Kharzeev:1999cz,Morley:1983wr}. This would lead to an imbalance between the numbers of right- and left-handed (anti-)quarks. It is demonstrated that if a sufficiently strong (electro-)magnetic field exists in such a region (as it may be in off-center heavy-ion collisions, generated chiefly by the protons in the two nuclei ~\cite{Skokov:2009qp,Bzdak:2011yy,Deng:2012pc,Bloczynski:2012en,Tuchin:2013ie,Bloczynski:2013mca,McLerran:2013hla,Sun:2019hao}) the net effect would be a separation of charges along the direction of the magnetic field~\cite{Fukushima:2008xe,Kharzeev:2007jp,Kharzeev:2004ey}. This separation of charges is called the Chiral Magnetic Effect (CME). If an observation of the CME could be clearly established in heavy-ion collisions, it would imply the existence of these {$\cal CP$}-violating regions, the restoration of the approximate chiral symmetry in the Quark Gluon Plasma (QGP) medium, and the action of an ultra-strong magnetic field on the collision region (see Refs.~\cite{Kharzeev:2020jxw,Kharzeev:2015znc} for reviews). A precision experimental test of the CME has been an important scientific goal of Brookhaven National Laboratory's Relativistic Heavy-Ion Collider (RHIC) program over the past decade. CME is also being explored in condensed matter systems~\cite{Li:2014bha,PhysRevB.99.075150}.

Over the years, extensive efforts have been invested to measure the CME-sensitive charge separation perpendicular to the reaction plane (RP, defined by the collision impact parameter and the beam direction) in heavy-ion collisions~\cite{Abelev:2009ac,Abelev:2009ad,Abelev:2012pa,Adamczyk:2013kcb,Adamczyk:2013hsi,Adamczyk:2014mzf,Khachatryan:2016got,Sirunyan:2017quh,Acharya:2017fau,STAR:2019xzd,ALICE:2020siw} (also see reviews in Refs.~\cite{Kharzeev:2013ffa,Kharzeev:2015znc,Huang:2015oca,Zhao:2018ixy,Zhao:2018skm,Zhao:2019hta,Li:2020dwr,Kharzeev:2020jxw}). In order to quantify the CME-induced charge transport and other modes of collective motion of the QGP, the azimuthal distribution of final-state particles is often Fourier-decomposed as
\begin{equation}
    \frac{dN_{\alpha}}{d\phi^*} \approx \frac{N_\alpha}{2\pi} \left[1 + 2v_{1,\alpha}\cos(\phi^*) + 2a_{1,\alpha}\sin(\phi^*) + 2v_{2,\alpha}\cos(2\phi^*) + \cdots \right]\,,
\label{equ:Fourier_expansion}
\end{equation}
where $\phi^* = \phi - {\rm \Psi_{RP}}$, with $\phi$ and $\Psi_{\rm RP}$ being the azimuthal angle of a particle and of the RP, respectively.
The subscript $\alpha$ ($+$ or $-$) denotes the charge sign of a particle. The coefficients $v_1$ and $v_2$ are called ``directed flow" and ``elliptic flow", respectively. 
The $v_n$ are functions of  transverse momentum ($p_T$) and pseudorapidity ($\eta$). The coefficient $a_1$ (with $a_{1,-} = -a_{1,+}$) characterizes the electric charge separation with respect to the RP which is correlated with the direction of magnetic field~\cite{Bzdak:2011yy,Deng:2012pc,Deng:2012pc,Bloczynski:2012en}.  The most widely used observable in the CME search is the ``$\gamma$ correlator," originally proposed in Ref.~\cite{Voloshin:2004vk}, 
\begin{equation}
    \gamma_{\alpha\beta}=\left<\cos(\phi_{\alpha}+\phi_{\beta}-2\Psi_{\textsc{rp}})\right>\,,\label{eq:gamma}
\end{equation}
where $\phi_{\alpha}$ and $\phi_{\beta}$ are the azimuthal angles of particles of interest (POIs). Here the averaging $\left<\cdots\right>$ is performed over the pairs of particles and over events. In order to eliminate charge-independent correlation backgrounds mainly from global momentum conservation~\cite{Bzdak:2010fd,Pratt:2010zn}, the difference between the opposite-sign (OS) and same-sign (SS) $\gamma$ correlators is considered,
\begin{equation}
    \Delta\gamma=\gamma_{\textsc{os}}-\gamma_{\textsc{ss}}\,.\label{eq:dgamma}
\end{equation}
The $\Delta\gamma$ is sensitive to the preferential emission of positively and negatively charged particles to the opposite sides of the RP. The first measurements of non-zero $\Delta\gamma$ from the STAR (Solenoidal Tracker at RHIC) Collaboration in Au+Au and Cu+Cu collisions at $\sqrt{s_{_{\rm NN}}}=200$ GeV are reported in Refs.~\cite{Abelev:2009ac,Abelev:2009ad}. In those publications, connections to expectations from CME-driven signals ($\Delta\gamma=2a_1^2$) and flow-induced background due to resonance decays 
are identified as possible sources that contribute to $\Delta\gamma$. Subsequent measurements from RHIC~\cite{Adamczyk:2013hsi,Adamczyk:2014mzf} and the LHC~\cite{Abelev:2012pa} at different energies 
have confirmed the observation of non-zero $\Delta\gamma$. Despite the theoretical progress, the quantification of the magnitudes of CME signals in heavy-ion collisions remains a challenge~\cite{Kharzeev:2001ev,Muller:2010jd,Liu:2011ys,Mace:2016shq,Mace:2016svc,Lappi:2017skr,Liao:2014ava,Yin:2015fca,Jiang:2016wve,Shi:2017cpu}. On the other hand, it is understood from phenomenological studies that measurements of $\Delta\gamma$ are dominated by backgrounds that are unrelated to the CME~\cite{Wang:2009kd,Bzdak:2009fc,Schlichting:2010qia,Bzdak:2010fd}. The dominant backgrounds arise from intra-cluster  correlations coupled with azimuthal anisotropy~\cite{Voloshin:2004vk,Wang:2009kd,Bzdak:2009fc,Schlichting:2010qia,Wang:2016iov,Kovner:2017gab,Schenke:2019ruo,Zhao:2019kyk}; namely,
\begin{equation}
    \Delta\gamma_{\rm bkgd} = \frac{4N_{\rm 2p}}{N^2}\left<\cos(\phi_{\alpha}+\phi_{\beta}-2\phi_{\rm 2p})\right>v_{2,{\rm 2p}}\,,
    \label{eq:bkgd}
\end{equation}
where $\phi_{\rm 2p}$ is the azimuthal angle of a correlated 2-particle cluster, $v_{2,{\rm 2p}}$ is the elliptic flow of such clusters, $N_{\rm 2p}$ is the number of those clusters, and $N$ is the multiplicity of the POIs~\cite{Voloshin:2004vk,Wang:2009kd,Zhao:2019hta,Zhao:2019kyk}. 
An example for this is the correlations among the decay daughters of resonance particles carrying elliptic flow.

Collisions of small systems are often considered to provide a data-driven baseline for a background scenario~\cite{Khachatryan:2016got}. In such collisions, the direction of the magnetic field is uncorrelated with azimuthal anisotropies, resulting in nearly vanishing CME-driven signals, while different sources of backgrounds for $\Delta\gamma$ remain~\cite{Khachatryan:2016got,Belmont:2016oqp,Kharzeev:2017uym}. Measurements performed at LHC energies by the CMS Collaboration show similar $\Delta\gamma$ signals for overlapping multiplicities in $p$+Pb and Pb+Pb collisions~\cite{Khachatryan:2016got}. Similar studies are carried out by STAR, with results that show similar (or even larger) values of $\Delta\gamma$ scaled by elliptic anisotropy in $p$+Au and $d$+Au collisions as compared to Au+Au collisions ~\cite{STAR:2019xzd}. Such measurements appear to challenge the interpretation of magnetic-field-driven sources of charge separation. However, RP-independent background from three-particle correlations can be significant in those small-system collisions and peripheral heavy-ion collisions; the same may not be true for more-central collisions~\cite{Abelev:2009ac,Abelev:2009ad,Kovner:2017gab,Zhao:2019kyk}. Extrapolation of small-system results as quantitative background baselines for different nucleus-nucleus systems, across the entire range of centrality, is not straightforward.

Over the past years, efforts have been dedicated towards developing data-driven methods and observables to isolate possible CME-driven signals from background contributions~\cite{Voloshin:2010ut,Schukraft:2012ah,Adamczyk:2013kcb,Chatterjee:2014sea,Skokov:2016yrj,Sirunyan:2017quh,Acharya:2017fau,Magdy:2017yje,Zhao:2017nfq,Xu:2017qfs,Voloshin:2018qsm,Du:2008zzb,Finch:2018ner,Tang:2019pbl}, and to applying those methods to existing data. The event-shape engineering (ESE) analyses by the CMS and ALICE Collaborations at the LHC~\cite{Sirunyan:2017quh,Acharya:2017fau} have reported a CME-induced charge separation that is consistent with zero with an upper limit (on the fraction of the $\Delta\gamma$ measurement that is due to CME) of the order of 7\% and 26\% at 95\% confidence level (CL), respectively. Measurements of the pair invariant mass dependence of the
$\Delta\gamma$ from STAR~\cite{Adam:2020zsu} have determined an upper limit of 15\% at the 95\% CL. A recent measurement by the STAR Collaboration using the spectator plane and participant plane analysis~\cite{STAR:2021pwb} has found a signal consistent with zero in peripheral collisions and a hint of finite positive signal in mid-central Au+Au collisions with a 1--3$\sigma$ significance. Possible remaining effects from non-flow correlations (two- and multi-particle correlations unrelated to a global symmetry plane) are under investigation~\cite{Feng:2021pgf}. An alternative charge-sensitive variable, $R_{\Psi_2}(\Delta S)$, has been proposed~\cite{Ajitanand:2010rc,Magdy:2017yje,Magdy:2018lwk} to aid the characterization of CME-driven charge separation. The sensitivity of the $R_{\Psi_2}(\Delta S)$ variable has been studied in different contexts and has also been compared to that for the $\Delta\gamma$ observable~\cite{Magdy:2017yje,Bozek:2017plp,Magdy:2018lwk,Sun:2018idn,Feng:2018chm,Huang:2019vfy,Shi:2019wzi,Feng:2020cgf,Magdy:2020xqs,Choudhury:2021jwd}. In a recent comprehensive
investigation of different experimental observables for CME searches, it is found that the $R_{\Psi_2}(\Delta S)$ and $\Delta\gamma$ variables provide similar
sensitivities to the CME signal and backgrounds for the two isobars~\cite{Choudhury:2021jwd}. 

In order to overcome the large backgrounds, isobar Ruthenium+Ruthenium ($^{96}_{44}$Ru+$^{96}_{44}$Ru) and Zirconium+Zirconium ($^{96}_{40}$Zr+$^{96}_{40}$Zr) collisions have been proposed~\cite{Voloshin:2010ut}.  
It is expected that the magnetic field squared would be about 15\% larger in Ru+Ru collisions due to its larger atomic number~\cite{Kharzeev:2007jp,Skokov:2009qp}, leading to a similar increase in the CME contribution in $\Delta\gamma$,  while the same mass number of these two nuclei would lead to similar flow-driven backgrounds. With 1.2 billion minimum-bias (MB) events for each collision system, a $5\sigma$ significance is expected in the CME signal difference between Ru+Ru and Zr+Zr~\cite{starbur17}. This expectation is based on the same projection scheme as in Ref.~\cite{Deng:2016knn}, assuming that the CME-related signal fraction is 20\% in $\Delta\gamma$.

Although similar, the backgrounds in Ru+Ru and Zr+Zr collisions are not expected to be identical. The difference in the nuclear deformation of the two isobars has been  estimated to yield less than 1\% difference in $\Delta\gamma$ background in peripheral to mid-central collisions. In more-central collisions from 0--20\% centrality, the difference in $\Delta\gamma$ background can be larger than 2\%~\cite{Deng:2016knn,Deng:2018dut}. Further work from sophisticated nuclear structure calculations suggests that the resulting eccentricities (hence the flow-related backgrounds) may differ by 2--3\% in mid-central collisions between the two isobars even without deformation~\cite{Xu:2017zcn,Li:2018oec}.  An approximate $4\%$ difference in flow-driven background between these two systems is found in hydrodynamic simulations which include local charge conservation~\cite{Schenke:2019ruo}. In order to account for a possible difference in $v_2$, one of the variables we will focus on in this paper is the ratio $\Delta\gamma/v_2$, assuming that background proportionality to $v_2$ is identical between the isobar systems. 
Note that although $v_2$ can be precisely measured, the elliptic anisotropy contains non-flow contributions, and the background in $\Delta\gamma$ depends also on other physical processes besides the $v_2$ (see Eq.~(\ref{eq:bkgd})). Therefore, it is crucial to minimize background contributions in order to search for the possibly small CME signal. Isobar collisions are considered to be an effective way to achieve that by studying the difference in the CME-sensitive observables between the two isobar systems.

Isobar collisions were acquired at RHIC in 2018~\cite{Marr:2019kom}. This paper reports results from a blind analysis~\cite{STAR:2019bjg} performed on the isobar data collected by the STAR Collaboration. 

\section{Isobar data and blind analysis}

\subsection{Modality of isobar running at RHIC}

The proposal for colliding isobar species is outlined in the 2017-18 RHIC beam use request by the STAR Collaboration~\cite{starbur17}. The specific request was for two 3.5-week runs in the year 2018 with collisions of isobar nuclei, $^{96}_{44}$Ru$+^{96}_{44}$Ru and $^{96}_{40}$Zr+$^{96}_{40}$Zr. 
This proposal is based on the prospect of achieving 5$\sigma$ significance in a scenario of a relative difference of the primary CME observable of 2-3\% between the two isobar species~\cite{starbur17}. 
It is estimated that with 3.5-week runs it is possible to collect more than 1.2 billion MB events for each species and achieve a statistical precision on the observable difference of about $0.5\%$. However, a special strategy is needed to minimize the systematic uncertainties. 
This required a specific plan in synergy with the RHIC Collider Accelerator Department to execute the isobar runs~\cite{Marr:2019kom}. %

Studies from previous years using Au$+$Au and U$+$U collision data ~\cite{Tribedy:2017hwn} indicate that there are several sources of systematics in the measurements of CME-sensitive observables. Two major sources are: 1) loss of detector acceptance, and 2) variation of luminosity during runs. These effect leads to run-to-run variation of the online trigger efficiency and charged-particle track reconstruction efficiency in the Time Projection Chamber (TPC)~\cite{Anderson2003659}. These two sources can lead to irreducible systematic uncertainties in CME-sensitive observables. In order to keep the systematics due to these two major sources below the aforementioned statistical precision it is necessary to minimize the differences between the run conditions for the two species. Therefore the proposed procedure is to: 1) alternate the isobar species between each store of beam in RHIC, 2) keep long stores with constant beam luminosity, 3) match luminosities between the species, and 4) adjust the luminosity in such a way that the hadronic interaction rate at STAR is close to 10 kHz. With such a strategy, it is estimated that the systematic uncertainties in the ratio of observables could be reduced to about $0.2\%$. As we discuss later, these conditions were successfully provided by the RHIC facility~\cite{Marr:2019kom} and this level of precision is indeed achieved in our measurements. 

\subsection{Detector apparatus and data quality cuts\label{sec:detector}}

STAR was the only operational detector for RHIC running in 2018. The main subsystems used for the analysis of isobar data are the TPC, the Time-of-flight detector (TOF)~\cite{Llope:2003ti}, the Event Plane Detector (EPD)~\cite{Adams:2019fpo}, the Zero-Degree Calorimeters (ZDCs)~\cite{Adler:2000bd} and the Vertex Position Detectors (VPDs)~\cite{Llope:2014nva}. 

The TPC is used to detect charged particles within the pseudorapidity range $|\eta|\!<\!1$, with full $2\pi$ azimuthal coverage and a transverse momentum lower limit of $p_T\!>\!0.2$ GeV/$c$~\cite{Anderson2003659}. The TPC is situated inside a magnet which maintained a constant solenoidal field of 0.5~T during the entire isobar runs. 
The tracking efficiency of the TPC ranges from $85\%$ to $90\%$ as determined using \textsc{geant} Monte Carlo (MC) simulations embedded into randomly sampled MB data events~\cite{Fine:2000gn}. 
We are able to exploit the advantage of having data sets for two isobars collected under similar run conditions. 
For example, in the analyses we study the ratios of measurements between the two isobars.  We do not apply efficiency corrections because the effects of inefficiency cancel out in these ratios. 

For each collision we use the TPC to reconstruct the primary vertex position ($V_{z,\textsc{tpc}}$) along the beam direction (defined as the $z$ axis) of the primary vertex as well as its radial distance from the $z$ axis ($V_r$).
For all analyses, each event is required to have a vertex position within $-35$ $<V_{z,\textsc{tpc}}<25$ cm and $V_{r}<2$ cm using a coordinate system with the origin at the TPC center. To reduce the contamination from secondary charged particles, we require tracks reconstructed in the TPC  to have a distance of closest approach (DCA) to the primary vertex of less than 3 cm. We also require each track to have at least 16 ionization points ($N_{\rm fits}$) in the TPC. 
To study the effect of track splitting and merging on different $v_n$ coefficients, we carefully study their relative pseudorapidity ($\Delta\eta$) dependence as splitting and merging will result in a peak or a dip, respectively, in this dependence~\cite{STAR:2016vqt,STAR:2017idk,STAR:2017bqr}. 
We do not see  evidence of track splitting effects; however, we observe a dip  at  low $\Delta\eta$ due to track merging  that is dominant in central events. To minimize track merging, a requirement of $\Delta\eta>0.05$ is applied. We also do this study for same-sign and opposite-sign pair correlations separately as the possible effects of track merging and splitting are expected to be different between the two cases. During the isobar run in 2018, one of the 24 sectors of the TPC was being used to commission the inner TPC (iTPC) sector and the data from this sector are not used for physics analysis. The loss of tracks due to this sector leads to an identifiable region of depletion in the $\eta$-$\phi$ acceptance map. However, the effect of this acceptance deficit in the final observables is corrected by reconstructing the harmonic flow vectors ($Q$-vectors) using re-weighting, re-centering, and shifting methods~\cite{Poskanzer:1998yz}. Such $Q$-vectors are then used for estimation of different observables and the EP in this analysis. It is important to note that this effect is consistently present over the entire period of the run and is common to both the isobar species, and therefore cancels in the ratios of physics observables between Ru+Ru and Zr+Zr. 

The MB data sample is collected with a trigger based on information from the VPDs~\cite{Llope:2014nva}. The VPDs ($4.4<|\eta|<4.9$) also provide information on primary collision vertices along the beam direction ($V_{z,\textsc{vpd}}$). For the selection of good events we require the condition of $| V_{z,\textsc{tpc}} - V_{z,\textsc{vpd}} | < 5$~cm (unless otherwise noted). Variations in luminosity are kept to a minimum during the runs, with the dominant part of the MB data set having a variation of luminosity that corresponds to a coincidence of signals from the ZDCs in the range of 9.5--11.5 kHz. The variation of luminosity affects our centrality selection and a correction for this is made. We achieve a trigger efficiency close to 100\% for events in which more than 50 tracks are reconstructed per unit pseudorapidity in the TPC (see Sec.~\ref{sec:centrality}). For events with fewer tracks, the trigger efficiency decreases and a MC Glauber model is used to estimate and correct for such inefficiencies, as discussed in Sec.~\ref{sec:centrality}.  

Our event selection techniques suffer from out-of-time pile-up that requires an offline rejection. About 0.5\% of events are identified as pile-up and removed by  excluding outliers in the correlation between the number of TPC tracks and the number of those tracks matched with a hit in the TOF detector (the TOF is a fast detector and does not suffer from out-of-time pileup). We also require at least one TPC track matched to the TOF for selecting good events. After all event selection cuts, we analyze approximately 1.8 billion MB events for Ru+Ru and 2.0 billion MB events for Zr+Zr collisions. 

Our measurement  uses the EPD detector for the first time in collider mode~\cite{Adams:2019fpo}. The EPD is used for measurements of the second- and third-harmonic event planes (EPs) at forward rapidity. The EPD consists of two segmented scintillator wheels located at $\pm 3.75$ m from the center of the TPC, along the beam direction, covering an acceptance window of approximately $2.1 <|\eta| < 5.1$ in
pseudorapidity and $2\pi$ in azimuth. Each wheel consists of 12 “supersectors” (in azimuth) that are further divided (radially) into 31 tiles  made of plastic scintillator. Each tile is connected to a silicon photomultiplier via optical fiber. Charged particles emitted in the forward and backward directions produce a signal distribution with identifiable peaks corresponding to various numbers of minimally ionizing particles in the EPD tiles. 
This information in each tile is used to reconstruct the EPs. Further details of the EPD can be found in Ref.~\cite{Adams:2019fpo}. 

The ZDCs and their associated Shower Maximum Detectors (SMDs) are used for determination of the spectator neutron  plane~\cite{Adler:2001fq,SN0448}. 
The ZDCs are Cherenkov-light sampling calorimeters located at forward and backward angles ($|\eta|>6.3$) and are each composed of three identical modules. The SMDs are sandwiched between the ZDC modules and are composed of two planes with scintillator strips aligned with $x$ or $y$ directions perpendicular to the beam. The SMD information thus can be used to measure the centroid of the hadronic shower produced by the spectator neutrons in the ZDCs. The $x$ and $y$ positions of the shower centroid ($\left<X,Y\right>_{\rm ZDCE,W\mathchar`-SMD}$) calculated on an event-by-event basis provide spectator-plane reconstruction (see Refs.~\cite{Adams:2005ca,Adamczyk:2017ird} for details).

We do not use the data from the Beam-Beam Counters (BBC) and the Barrel Electromagnetic Calorimeter (BEMC) in this analysis other than for data quality assurance purposes. The time-dependence of the $Q$-vectors from the BBCs are studied to identify bad runs. The number of TPC tracks matched to the BEMC ($N_{\rm trk}^{\rm BEMC\mathchar`-matched}$) is also examined as a function of time to identify outlier runs.

\subsection{Blinding of data sets and preparation for analysis}

The recommendation to perform a blind analysis of the isobar data was initially made by the Nuclear and Particle Physics Program Advisory Committee at Brookhaven National Laboratory~\cite{pac2017}. The procedure to blind the isobar data is determined and implemented well before the actual data taking. The raw data are made inaccessible to the analysts to eliminate possible unconscious biases. 

A total of five institutional groups within the collaboration perform blind analyses of the isobar data. The analysts from each group focus on a specific analysis method described in Sec.~\ref{sec:observables}. Substantial overlap of some analyses helps to cross check the results.
The details of the blinding procedure and data structure are decided by an Analysis Blinding Committee (ABC), consisting of STAR members who are not part of the team of analysts.  The ABC works in close collaboration with the data production team to provide the analysts with access only to data in which species-specific information is disguised or removed, until the final un-blinded analysis step. Before the final step ABC also makes sure that the information provided to the analysts to perform quality assurance (QA) of the data do not reveal the species identity.

\subsection{Methods for isobar blind analysis\label{sec:BlindMethod}}

\begin{figure*}[h]
    \centering
    \includegraphics[width=0.9\textwidth]{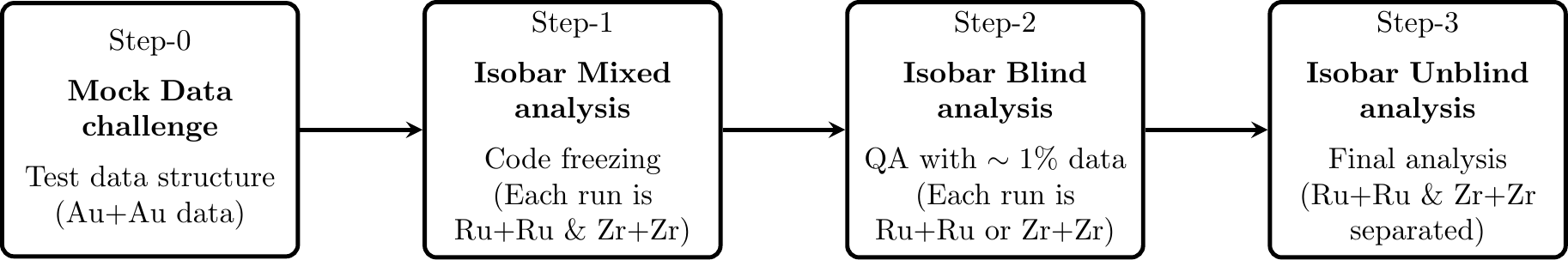}
    \caption{Flowchart to illustrate the steps of the isobar blind analysis~\cite{Tribedy:2020npn}. This is based on the procedure for the isobar blind analysis outlined in Ref.~\cite{STAR:2019bjg}.}
    \label{fig_cartoon_blindanalysis}
\end{figure*}

The detailed procedure for the blind analysis of isobar data is outlined in Ref.~\cite{STAR:2019bjg} and is strictly followed by the analysts. Shown in Fig.~\ref{fig_cartoon_blindanalysis}, the blind analysis procedure includes a mock-data challenge to perform a closure test and three main steps: 1) isobar-mixed analysis, 2) isobar-blind analysis, and 3) isobar-unblind analysis~\cite{Tribedy:2020npn}. 

In the zeroth step preceding the blind analysis, the analysts  participated in a mock-data challenge. The purpose of this step is to familiarize the analysts with the data structures that have been designed for the blind analysis and the techniques to access the data. Feedback is also provided to the ABC to ensure feasibility of the analysis blinding process. Data for Au+Au collisions at $\sqrt{s_{_{\rm NN}}}=27$ GeV (collected in 2018 after the isobar run) are used for this step.

The first step of this analysis is referred to as the ``isobar-mixed analysis". In this step the majority of the analysis work is done.  Analysts are  provided with a data sample where each ``run" contains events that are a mixed sample of the two species. The analysis teams then perform QA and a complete analysis of the data.  The details of the QA procedure are discussed in the next section. The analysis teams test their analysis code and document their analysis procedures. They are then frozen for the next two steps of the analysis, except for situations as strictly defined at the end of this subsection. An important part of data QA is to reject bad runs and pile-up events. This requires retention of the time ordering of the data. In order to avoid unconscious biases, an automated algorithm for bad run rejection is developed and the corresponding codes are also frozen. The QA algorithm is tested using existing Au$+$Au and U$+$U data. 
In this step the documentation related to the criteria for signatures of the CME in each observable, which we discuss in Sec.~\ref{sec:observables}, is also frozen. From the next steps onwards the analysts can only execute frozen codes. As we discuss later, different groups focus on analysis of specific CME-sensitive observables. In order to check the consistency of the numerical output of the analysis codes from five groups, an exercise is performed in this step. The analysts from different groups are required to estimate a few common observables in the same approach, with exactly the same data, using their own individual codes. The results from different groups are ensured to be numerically identical to each other.

 The second step is referred to as the ``isobar-blind analysis". For this the analysts are provided with files, each of which contain data from a single, but blinded, isobar species to perform run-by-run QA. Every file provided to the analyst contains a limited number of events that is determined to be insufficient to allow an identification of the species or the observation of a statistically significant CME signal. A pseudo run-number is used to hide the identity of the species for each file. The mapping between these pseudo run-numbers and the original ones is not revealed to the analysts. The automated algorithms are then used to identify the runs with stable detector performance and to reject bad runs. 

The final step is referred to as ``isobar-unblind" analysis. In this step, all elements of the data, including species information, are revealed to the analysts and the physics results are produced by the analysts using the previously frozen codes. As mentioned before, analysts from five independent groups participate in the blind analysis. In order to further avoid unconscious biases, analysts from a given group are not allowed to execute their own codes to produce the final results. Instead, a STAR collaborator is identified either from a different blind analysis group or among members not participating in the blind analysis, to run that group's frozen code. 
\emphasize{The findings from this step are directly presented in this paper without alteration.} A brief discussion of post-blinding analysis results is given near the end of the paper in Sec.~\ref{sec:post}.

\subsection{Quality assurance of the blind data}
Unlike conventional QA, the analysis teams do not have access to the full statistics of the recorded data. In accordance with the blind analysis policy, any form of manual selection or rejection of a part of the data sample is not permitted. This makes the QA of the data analysis challenging. In order to avoid unconscious biases and yet perform an effective clean up of data we develop an automated algorithm with predefined criteria for QA. These algorithms perform three major tasks: 1) identify the regions of the data sample or runs with stable detector performance by studying the time dependence of various quantities, 2) identify regions of the data sample with problematic detector performance or outlier runs, and 3) remove pile-up events. 

We study run-by-run variation in the mean value of quantities such as the average multiplicity ($\left< N_{\rm trk}^{\rm offline} \right>$) of tracks from the TPC, basic track level quantities like the distance of closest approach ($\left<{\rm DCA}\right>$), and quantities related to azimuthal acceptance such as mean cosine of the azimuthal angle ($\left<\cos(\phi)\right>$). 
The QA procedure is performed over the entire data sample, and separately for the five analysis groups because each group provides a list of such quantities specific to the analysis. For example, Group-3 and Group-4 use the ZDC for EP analysis and therefore need to carefully study the QA variables for quantities related to the ZDC.  The analyses of other groups that do not use the ZDC do not need to perform QA related to that detector.  Table~\ref{tab:QA} lists the common QA variables and criteria, as well as the analysis-specific ones, to reject bad runs. 

Data collection for the isobar run took eight weeks and two days. During this time, the acceptance of the detector changed due to the temporary failure of electronics modules or other causes. Thus, periods of stable and uniform operation were identified and each stable period was treated separately for acceptance and track weighting corrections. 
To identify jumps or boundaries between stable regions we study QA quantities with time or run numbers. We study the first and second order derivatives of quantities with respect to time. 
The zeros of the first order derivative surrounded by two zeros of the second order derivative defines a run mini-region. From each mini-region we extract the local mean and the weighted error. 
We define regions of stable detector conditions by merging these mini-regions if the mean values of the quantities in adjacent mini-regions are: 1) within five times the weighted error or 2) within one percent of the variation of the local mean. A run is marked as an outlier or bad run in each stable region if the value of the QA quantity is five standard deviations from the local mean. Once the first-round of stable regions are identified and bad runs are removed, the whole process is repeated. Iterations are performed until no additional bad run is identified by the algorithm. The stability of this automated algorithm is tested with existing Au+Au and U+U data sets before the code freeze in step-1 (isobar mixed analysis). 

In the second step of isobar analysis the blind data set is provided to the analysts that includes all the runs for both species (species identity is blinded) but each run contains only approximately 1\% of the entire statistics of that run.
Following the methods of the blind analysis, all the files are named by a pseudo-run-number mapped to the original run-number by the production team to ensure the species are blind to the analysts. 
The analysts prepare the necessary histograms of QA variables with pseudo-run-numbers using the blind data set. 
A non-analyst then helps to re-map the run-numbers, executes the frozen run-by-run QA algorithm and prepares the final lists of bad runs and stable periods for each group. 
These numbers are different for different analysis groups because of the difference in the analysis-specific QA variables (see Table~\ref{tab:QA}). It is important to note that the QA is performed on the combined data set of two species and not on individual species. By the end of the QA, the automated algorithm identified less than 4\% of the data to be discarded from the analysis based on predefined criteria. Since the criteria of pattern recognition to discard the problematic part of the data sample is predefined and frozen prior to the blind analysis, unconscious biases are eliminated.

\begin{table}[]
    \centering
    
\begin{center}

\begin{tabular}{|| c | c  c  c  c  c ||}
\hline
\backslashbox{Variable}{Group} & Group-1 & Group-2 & Group-3 & Group-4 & Group-5 \\ [0.5ex]
\hline\hline
$\langle N_{\rm trk}^{\rm offline} \rangle$ & $\blacksquare$ & $\blacksquare$ & $\blacksquare$ & $\blacksquare$ & $\blacksquare$ \\ 
\hline
$\langle N_{\rm hits}^{\rm TOF} \rangle$ & $\blacksquare$ & $\square$ & $\square$ & $\blacksquare$ & $\square$ \\ 
\hline
$\langle N_{\rm trk}^{\rm TOF\mathchar`-matched} \rangle$ & $\square$ & $\blacksquare$ & $\blacksquare$ & $\blacksquare$ & $\blacksquare$ \\ 
\hline
$\langle p_{T} \rangle$ & $\blacksquare$ & $\square$ & $\blacksquare$ & $\blacksquare$ & $\square$ \\ 
\hline
$\langle \eta \rangle$ & $\blacksquare$ & $\square$ & $\blacksquare$ & $\blacksquare$ & $\square$ \\ 
\hline
$\langle {\rm DCA} \rangle$ & $\blacksquare$ & $\square$ & $\blacksquare$ & $\blacksquare$ & $\square$ \\ 
\hline
$\langle V_z \rangle$ & $\square$ & $\square$ & $\blacksquare$ & $\blacksquare$ & $\square$ \\ 
\hline
$\langle \phi \rangle$ & $\square$ & $\square$ & $\square$ & $\blacksquare$ & $\square$ \\ 
\hline
$\langle N_{\rm fits} \rangle$ & $\square$ & $\square$ & $\blacksquare$ & $\blacksquare$ & $\square$ \\ 
\hline
$\langle Q_{1x} \rangle_{\rm TPC}$ & $\square$ & $\blacksquare$ & $\blacksquare$ & $\square$ & $\blacksquare$ \\ 
\hline
$\langle Q_{1y} \rangle_{\rm TPC}$ & $\square$ & $\blacksquare$ & $\blacksquare$ & $\square$ & $\blacksquare$ \\ 
\hline
$\langle Q_{2x} \rangle_{\rm TPC}$ & $\blacksquare$ & $\blacksquare$ & $\blacksquare$ & $\square$ & $\blacksquare$ \\ 
\hline
$\langle Q_{2y} \rangle_{\rm TPC}$ & $\blacksquare$ & $\blacksquare$ & $\blacksquare$ & $\square$ & $\blacksquare$ \\ 
\hline
$\langle Q_{1x} \rangle_{\rm EPD}$ & $\square$ & $\blacksquare$ & $\square$ & $\square$ & $\blacksquare$ \\ 
\hline
$\langle Q_{1y} \rangle_{\rm EPD}$ & $\square$ & $\blacksquare$ & $\square$ & $\square$ & $\blacksquare$ \\ 
\hline
$\langle Q_{2x} \rangle_{\rm EPD}$ & $\square$ & $\blacksquare$ & $\square$ & $\square$ & $\blacksquare$ \\ 
\hline
$\langle Q_{2y} \rangle_{\rm EPD}$ & $\square$ & $\blacksquare$ & $\square$ & $\square$ & $\blacksquare$ \\ 
\hline
$\langle Q_x \rangle_{\rm BBCE}$ & $\square$ & $\square$ & $\blacksquare$ & $\square$ & $\square$ \\ 
\hline
$\langle Q_y \rangle_{\rm BBCE}$ & $\square$ & $\square$ & $\blacksquare$ & $\square$ & $\square$ \\ 
\hline
$\langle Q_x \rangle_{\rm BBCW}$ & $\square$ & $\square$ & $\blacksquare$ & $\square$ & $\square$ \\ 
\hline
$\langle Q_y \rangle_{\rm BBCW}$ & $\square$ & $\square$ & $\blacksquare$ & $\square$ & $\square$ \\ 
\hline
$\langle X \rangle_{\rm ZDCE\mathchar`-SMD}$ & $\square$ & $\square$ & $\blacksquare$ & $\square$ & $\square$ \\ 
\hline
$\langle Y \rangle_{\rm ZDCE\mathchar`-SMD}$ & $\square$ & $\square$ & $\blacksquare$ & $\square$ & $\square$ \\ 
\hline
$\langle X \rangle_{\rm ZDCW\mathchar`-SMD}$ & $\square$ & $\square$ & $\blacksquare$ & $\square$ & $\square$ \\ 
\hline
$\langle Y \rangle_{\rm ZDCW\mathchar`-SMD}$ & $\square$ & $\square$ & $\blacksquare$ & $\square$ & $\square$ \\ 
\hline
$\langle Q_{1x} \rangle_{\rm ZDC}$ & $\square$ & $\square$ & $\square$ & $\blacksquare$ & $\square$ \\ 
\hline
$\langle Q_{1y} \rangle_{\rm ZDC}$ & $\square$ & $\square$ & $\square$ & $\blacksquare$ & $\square$ \\ 
\hline
$\langle Q_{2x} \rangle_{\rm ZDC}$ & $\square$ & $\square$ & $\square$ & $\blacksquare$ & $\square$ \\ 
\hline
$\langle Q_{2y} \rangle_{\rm ZDC}$ & $\square$ & $\square$ & $\square$ & $\blacksquare$ & $\square$ \\ 
\hline
$\langle N_{\rm trk}^{\rm BEMC\mathchar`-matched} \rangle$ & $\square$ & $\square$ & $\blacksquare$ & $\square$ & $\square$ \\ 
\hline
\hline 
\end{tabular}

\end{center}

    \caption{Common and analysis-specific QA variables and criteria used to reject bad runs. $\blacksquare$ Used, $\square$ Unused. See the texts in Sec.~\ref{sec:detector} for the definition of different variables. The $Q_{nx}$ and $Q_{ny}$ refer to the components of the flow $Q$-vectors that we discuss in the later sections.}
    \label{tab:QA}
\end{table}

Another automated algorithm is implemented prior to the blind analysis to remove pile-up events. Based on studies of previous data sets it is observed that pile-up events lead to satellites in the correlation between the number of tracks from the TPC ($N_{\rm trk}^{\rm offline}$) and the number of TPC tracks matched with TOF ($N_{\rm trk}^{\rm TOF}$). For a given window of $N_{\rm trk}^{\rm TOF}$, the distribution of $N_{\rm trk}^{\rm offline}$ appears to be described by a double negative binomial distribution with two sets of widths and means. The wider distribution corresponds to the pile-up events. For each value of $N_{\rm trk}^{\rm TOF}$ one can reduce the pile-up contribution by applying upper and lower cuts of $3(\sigma+\text{Skewness})$ and $4\sigma$ respectively, around the mean value of the narrow distribution. Such a procedure is implemented in the frozen algorithm and used for pile-up removal in our analysis.

In the final step of the analysis when the isobar data are unblinded we check the distributions of energy deposition in the ZDCs. We find that the Zr+Zr collisions have a significantly larger energy deposition than that of the Ru+Ru collisions, consistent with the larger neutron number in the former. We also check the net-charge distributions from the TPC and find that the Ru+Ru collisions have a larger mean than Zr+Zr collisions. These checks confirm that the two species are correctly separated in the  unblind sample of the data provided to the analysts. 

\subsection{Methodology of uncertainty estimation\label{sec:syst}}

Systematic uncertainties are assessed by varying each of the analysis cuts within a range that is considered as the reasonable maximum range. This way one estimates the quantity $|\Delta|$ which is the absolute difference between the magnitudes of an observable with the default cut and with a particular cut variation. 
The statistical fluctuation on this difference is given by $\sigma_{\Delta}=\sqrt{\sigma_1^2-\sigma_2^2}$, where $\sigma_1$ and $\sigma_2$ are the statistical uncertainties of the two measurements~\cite{Barlow:2002yb}. If $\sigma_{\Delta}$ is larger than  $|\Delta|$, i.e.~the change in the result is consistent with statistical fluctuations, then no systematic uncertainty is considered for this cut variation. Otherwise, the systematic uncertainty is assigned to be $\sigma_{\rm syst}=\sqrt{\Delta^2-\sigma_{\Delta}^2}/\sqrt{12}$. For compound observables, such as the $\Delta\gamma/v_2$, systematic uncertainties are assessed as above, treating the compound observable as a single quantity. This way the (anti-)correlations in the systematic uncertainties in the component variables are automatically taken into account.

All analyses reported in this paper have a common set of cuts and variations for the purpose of systematic uncertainty determination. As noted above, the events used in all analyses are required to have a primary vertex within $-35<V_{z,\textsc{tpc}}<25$~cm. To estimate the systematic uncertainty due to the acceptance dependence on $V_{z,\textsc{tpc}}$, results using only events within $-35<V_{z,\textsc{tpc}}<0$~cm are compared with those from the full $V_{z,\textsc{tpc}}$ range. A maximum DCA of 3~cm and a minimum $N_{\rm fits}$ 
of 16 are required for the TPC tracks to be used in the analysis.
Systematic uncertainties are assessed by varying the maximum DCA from 3~cm to 2~cm and the minimum $N_{\rm fits}$ from 16 to 21.
In addition to the common cuts, each analysis has specific cuts described in the corresponding results subsections in Sec.~\ref{sec:results}. At the end, the systematic uncertainties of all sources are added in quadrature, the value of which is quoted as one standard deviation.

For statistical uncertainty estimations we use the standard error propagation method. We use both analytical and MC (Bootstrap~\cite{Efron:bootstrap}) approaches to examine the  influence of co-variance terms. Such cases may be relevant for primary CME-sensitive quantities like the ratio of $\Delta\gamma/v_2$. We find that the statistical uncertainties in the ratio observable   $(\Delta\gamma/v_2)$ are completely dominated by uncertainties of the numerator (by more than a factor of 50). Furthermore, the covariance between the numerator and denominator is also negligible, simplifying the statistical error calculations.

\section{Centrality determination\label{sec:centrality}}

The centrality determination is made at the beginning of the final step of the isobar analysis, using the unblinded data. This is performed by a team of collaborators who do not take part in the blind analysis of the data, and before any of the observables are measured.

Centrality is defined based on the charged track multiplicity ($N_{\rm trk}^{\rm offline}$) from the TPC within the pseudorapidity acceptance $|\eta|<0.5$.
Each track is required to have a DCA to the primary vertex of less than 3~cm and must be formed from at least 10 ionization points in the TPC gas volume.
The $N_{\rm trk}^{\rm offline}$ depends on the tracking efficiency of the TPC, which in turn
depends on the occupancy of the TPC and hence on the collider luminosity, which is monitored with the ZDC coincidence rate. The $\left<N_{\rm trk}^{\rm offline}\right>$ is found to have a linear dependence on the ZDC coincidence rate.
The parameterization of this dependence is used to correct $N_{\rm trk}^{\rm offline}$ for luminosity effect. 
To this end, $N_{\rm trk}^{\rm offline}$ is first converted to a real number by sampling the range from a half unit below to a half unit above, and the correction is then applied to the real number.
Over the ZDC coincidence rate range of 9.5 kHz to 11.5 kHz, which describes the dominant part of this data set, the luminosity correction to the multiplicity is less than 0.02\% for Ru+Ru collisions and less than 0.29\% for Zr+Zr collisions. 
This luminosity correction is small owing to the very stable beam conditions provided by RHIC during the isobar run. 

The quantity $N_{\rm trk}^{\rm offline}$ is further corrected for the acceptance variation as a function of $V_{z,\textsc{tpc}}$.
To obtain the correction factor, the $N_{\rm trk}^{\rm offline}$ distributions, $P(N_{\rm trk}^{\rm offline})$, are plotted in 2~cm bins of $V_{z,\textsc{tpc}}$ in the range $-35 < V_{z,\textsc{tpc}} < 25$ cm. These multiplicity distributions in heavy-ion collisions have a characteristic sharp decline at large multiplicity values. The location of the half-maximum of this decline is measured by fitting this region with an error function.
The correction factor is determined by making the location of the half-maximum point of the given $V_{z,\textsc{tpc}}$ bin equal to the one at $-1<V_{z,\textsc{tpc}}<1$~cm (the center of the TPC).

\begin{figure*}[htb]
\includegraphics[width=0.48\textwidth]{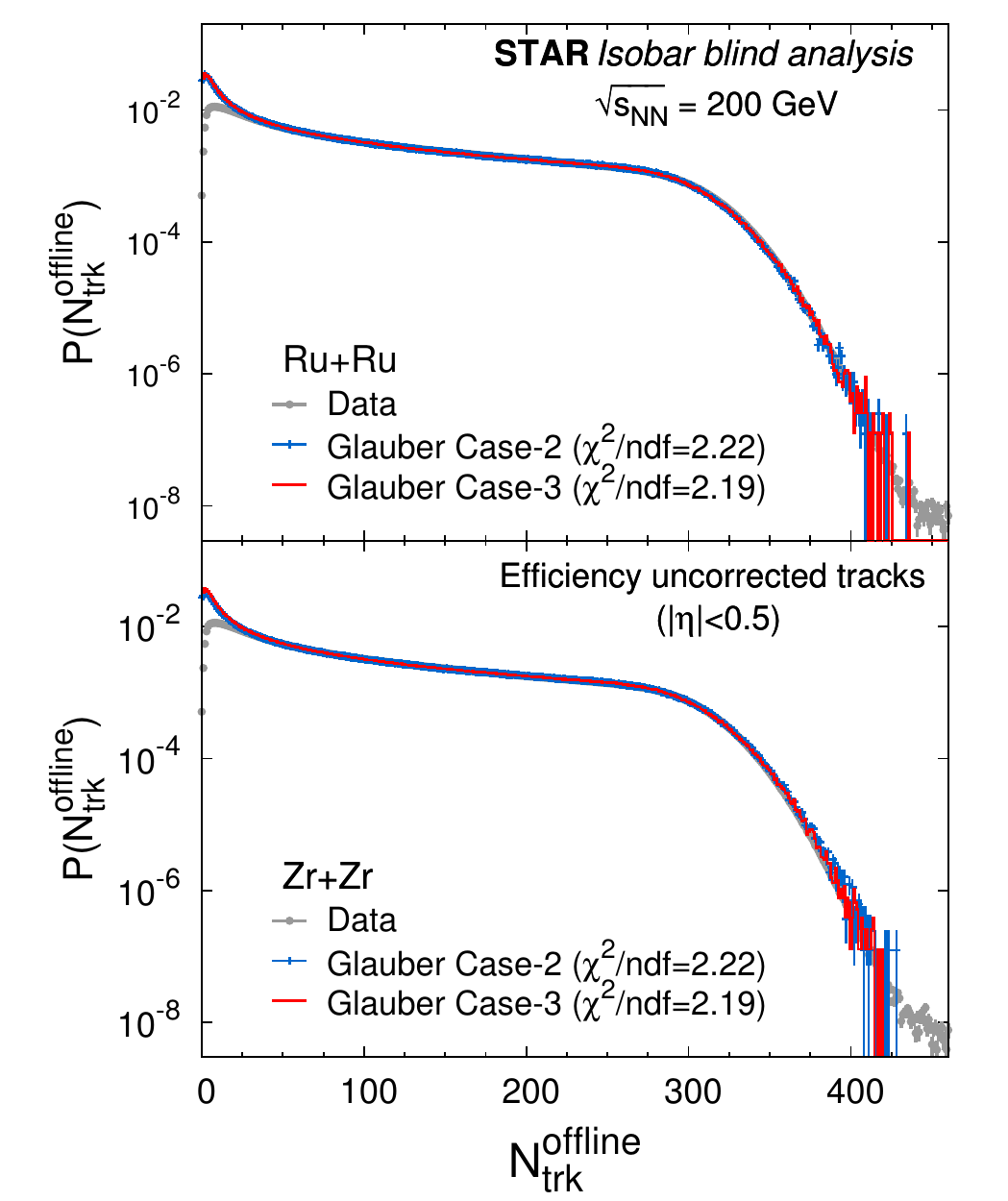}
\includegraphics[width=0.48\textwidth]{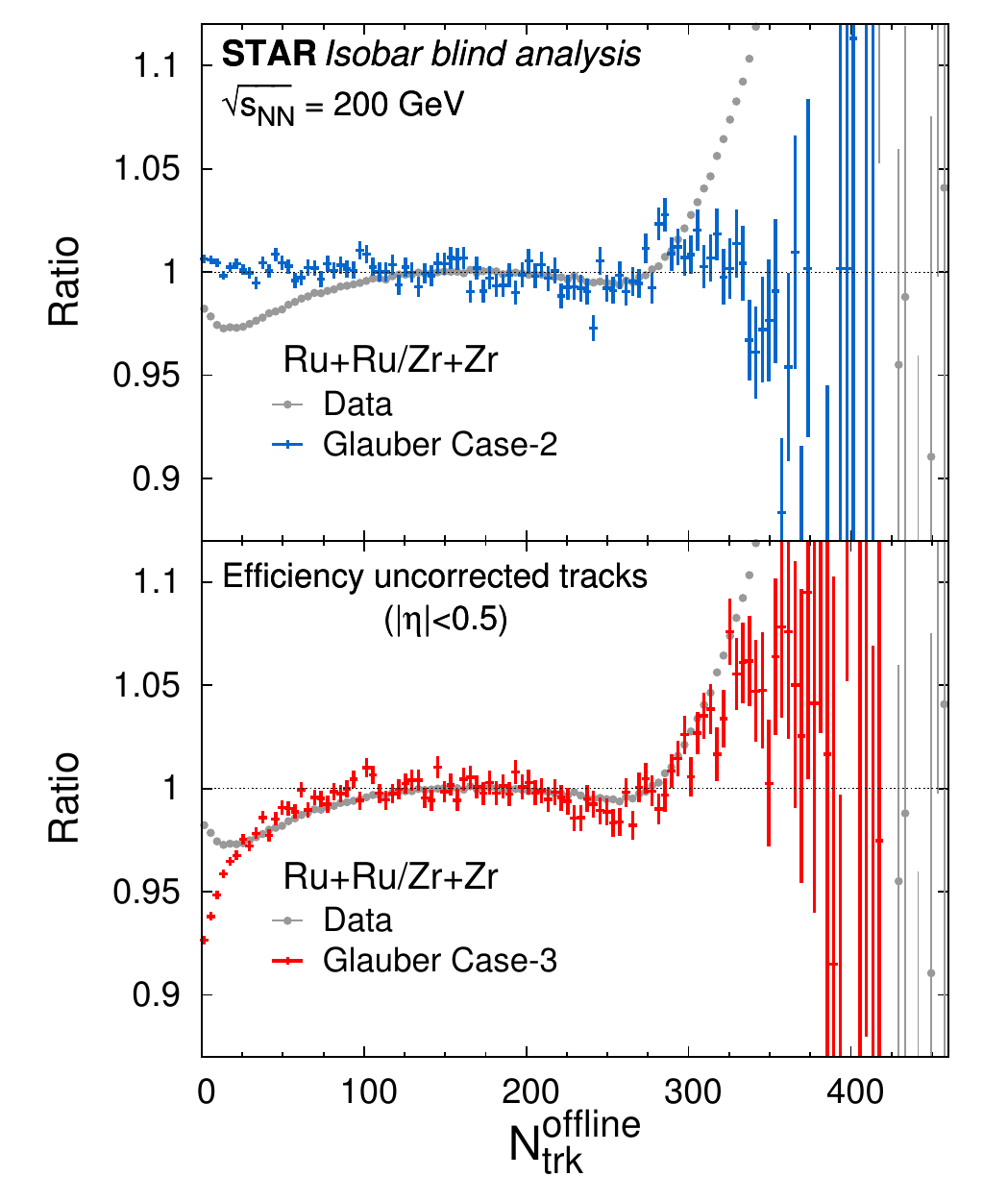}
\caption{
 Distributions of the number of charged particles ($N_{\rm trk}^{\rm offline}$) from the TPC in the pseudorapidity acceptance $|\eta|<0.5$ in Ru+Ru (upper left panel) and Zr+Zr (lower left panel) collisions. The experimental distributions have been corrected for variations in the luminosity and the vertex position $V_{z,\textsc{tpc}}$, and uncorrected for tracking efficiency. 
Fits to the experimental distributions (gray circles) are performed by the two-component Glauber model using two sets of Woods-Saxon parameters in Table~\ref{tab:ws-params} (blue crosses for Case-2 and red crosses for Case-3). The Ru+Ru to Zr+Zr ratio of the experimental data, as well as  those of the Glauber model fit for Case-2 and Case-3 are shown in the upper right and lower right panels, respectively.
The Glauber simulation with the Case-3 nuclear density parameters is used for centrality determination as it provides the best description of the experimental data.
\label{centrality}
}
\end{figure*}

Figure~\ref{centrality} shows the luminosity and $V_{z,\textsc{tpc}}$ corrected distributions $P(N_{\rm trk}^{\rm offline})$ in Ru+Ru and Zr+Zr collisions. 
The centrality classes in this analysis are defined by fitting the $P(N_{\rm trk}^{\rm offline})$ distributions to those obtained from MC Glauber simulations~\cite{Abelev:2008ab,Miller:2007ri}. In Glauber simulations, the probability of a collision at a given impact parameter ($b$) and the corresponding number of participant nucleons ($N_{\rm part}$) and  number of binary nucleon-nucleon collisions ($N_{\rm coll}$) are obtained by MC sampling. The inputs for this calculation are the nuclear thickness function and the inelastic nucleon-nucleon cross section ($\sigma_{_{\rm NN}}^{\rm inel}$) which is taken to be 42~mb for the current case of $\sqrt{s_{_{\rm NN}}}=200$~GeV collisions~\cite{ParticleDataGroup:2020ssz}. 

The nuclear thickness function is the projection of the 3D nuclear density onto the transverse plane (perpendicular to the $z$ axis). It is obtained by sampling nucleons in the incoming nuclei according to the Woods-Saxon (WS) distribution defined in the nucleus rest frame with a spherical coordinate system ($r$ is radial position and $\theta$ is polar angle)~\cite{PhysRev.95.577}:
\begin{equation}
  \rho(r,\theta) = \frac{\rho_0}{1 + \exp\left[ \frac{ r - R \left( 1 + \beta_2 Y^0_2(\theta) \right) } { a } \right] }\,,
\end{equation}
where $R$ is the radius parameter, $a$ is the diffuseness parameter of the nuclear surface, $\beta_2$ is the quadruple deformity parameter, $Y^0_2(\theta)=\frac{1}{4}\sqrt{\frac{5}{\pi}}(3\cos^{2}\theta - 1)$, and $\rho_0$ is the normalization factor. 
Nuclear density distributions of $^{96}_{44}$Ru and $^{96}_{40}$Zr are not accurately known~\cite{Deng:2016knn,Li:2018oec,Hammelmann:2019vwd}. In this work, three sets of WS parameters~\cite{Deng:2016knn,Xu:2021vpn} are investigated. These sets of parameters are listed in Table~\ref{tab:ws-params}. 
The first two sets (Case-1 and Case-2) have the same $R$ and $a$ parameters and different deformations. The parameters are constrained by $e$+$A$ scattering experiments~\cite{Raman:2001nnq, Pritychenko:2013gwa} and calculations based on a finite-range droplet macroscopic model and the folded-Yukawa single-particle microscopic model~\cite{Moller:1993ed}. The charge radius of $^{96}_{44}$Ru, because of its additional protons, is larger than that of $^{96}_{40}$Zr. The neutron and proton density parameters are taken to be the same for both $R$ and $a$, 
so Ru is larger than Zr.
The third set (Case-3) is from recent calculations based on energy density functional theory (DFT), assuming the nuclei are spherical~\cite{Xu:2017zcn,Xu:2021vpn}. The proton and neutron distributions are both calculated, and the overall size of Ru is found to be smaller than Zr because of a significantly thicker neutron skin in the latter. The nucleon distributions are found to be well parameterized by the halo-type WS distributions (i.e.~the neutron $a$ parameter is significantly larger than that for the proton)~\cite{Xu:2021vpn}.

\begin{table*}[hbt]
  \centering
  \caption{The Woods-Saxon parameters used in the Glauber simulations for the centrality determination.}
  \label{tab:ws-params}
  \begin{tabular}{c|ccc|ccc|ccc}
  & \multicolumn{3}{c|}{Case-1~\cite{Deng:2016knn}} 
  & \multicolumn{3}{c|}{Case-2~\cite{Deng:2016knn}} 
  & \multicolumn{3}{c}{Case-3~\cite{Xu:2021vpn}} \\
  Nucleus 
  & $R$~(fm) & $a$~(fm) & $\beta_2$ 
  & $R$~(fm) & $a$~(fm) & $\beta_2$ 
  & $R$~(fm) & $a$~(fm) & $\beta_2$ \\ \hline
  $^{96}_{44}$Ru 
  & 5.085 & 0.46 & 0.158 
  & 5.085 & 0.46 & 0.053 
  & 5.067 & 0.500 & 0 \\
  $^{96}_{40}$Zr 
  & 5.02 & 0.46 & 0.08 
  & 5.02 & 0.46 & 0.217 
  & 4.965 & 0.556 & 0 \\
  \end{tabular}
\end{table*}

In this analysis we use the simple two-component model for multiparticle production~\cite{Kharzeev:2000ph}. Several alternative approaches of multiparticle production have been developed over the years, such as Quark-Glauber~\cite{Eremin:2003qn},  IP-Glasma~\cite{Schenke:2012wb}, \textsc{trento}~\cite{Moreland:2014oya} and Shadowed Glauber~\cite{Chatterjee:2015aja}, that improve the two-component model. These approaches can be investigated in future STAR analyses -- for the current work we stick to the two-component nucleon based MC Glauber  model for simplicity. 
The multiplicity density at a given $b$, with the corresponding $N_{\rm part}$ and $N_{\rm coll}$ from the Glauber calculation for each set of the WS parameters, is parameterized by the two-component model~\cite{Kharzeev:2000ph} as:
\begin{equation}
N_{\rm trk}^{\rm Glauber} = n_{pp}\left[(1-x) N_{\rm part}/2 + xN_{\rm coll} \right]\,,
\label{eq:twocomp}
\end{equation}
where $n_{pp}$ is the average pseudorapidity multiplicity density in zero-bias nucleon-nucleon (NN) collisions, 
and $x$ is the relative contribution to multiplicity from hard processes. The multiplicity given by Eq.~(\ref{eq:twocomp}) is the average multiplicity.
Multiplicity fluctuations are taken into account in the following way. $N_{\rm trk}^{\rm Glauber}$ is considered to be accumulated by $(1-x)N_{\rm part}/2+ xN_{\rm coll}$ (that is rounded to the closest integer) NN collisions. 
In each NN collision, the multiplicity $n$ is obtained by convolution of the negative binomial distribution (NBD) 
\begin{equation}
  P_{\rm NBD}(n_{pp},k;n) = \frac{\Gamma(n+k)}{\Gamma(n+1)\Gamma(k)}\cdot \frac{(n_{pp}/k)^n}{(1+n_{pp}/k)^{n+k}}\,,
\end{equation}
where $\Gamma$ is the gamma function and the fluctuation parameter $k$  controls the sharpness of the large multiplicity tail of the $N_{\rm trk}^{\rm Glauber}$ distribution. 

The Glauber multiplicity distribution obtained in this way is then convolved with a binomial distribution to account 
for the tracking inefficiency and acceptance of the TPC. The net effect depends on the TPC hit occupancy and is modeled as a linear function in the multiplicity~\cite{Abelev:2008ab}.
The final $N_{\rm trk}^{\rm Glauber}$ distribution is then fitted to the experimental $N_{\rm trk}^{\rm offline}$ distribution, with $n_{pp}$, $k$, and $x$ 
as fit parameters. 
The fit is performed simultaneously for Ru+Ru and Zr+Zr datasets with the fit parameters forced to be common for both isobars. Since the peripheral collisions are affected by trigger inefficiency, the fit range is restricted to $N_{\rm trk}^{\rm offline}>50$. 

A simultaneous fit of the $N_{\rm trk}^{\rm offline}$ distributions for the two isobars is performed for each set of the WS parameters for $^{96}_{44}$Ru and $^{96}_{40}$Zr listed in Table~\ref{tab:ws-params}. The first set of parameters (Case-1) is rejected from further analysis because it yields the largest $\chi^2/\text{ndf}$ among the three scenarios. 
The fit results for Case-2 and Case-3 are shown in Fig.~\ref{centrality} (left panels), with similar $\chi^2/\text{ndf}$ values. 
The $P(N_{\rm trk}^{\rm offline})$ distributions shown in Fig.~\ref{centrality} for data are normalized by the number of events. The same is also applied for the Glauber distributions. However, the Glauber distributions are further scaled by an additional factor equal to the ratio of the integrals from $N_{\rm trk}^{\rm offline}=50$ to $500$ taken between the data and Glauber distributions.

In order to further inform the choice of the WS parameters, the ratio of the experimentally measured $N_{\rm trk}^{\rm offline}$ distribution for Ru+Ru to the one for Zr+Zr is compared with the same ratio obtained for the MC Glauber calculations.
These ratios are shown in Fig.~\ref{centrality} (right panels). 
The multiplicity ratio obtained for Case-3 is in a better agreement with the experimental distribution at $N_{\rm trk}^{\rm offline}>$~50, while the ratio for Case-2 deviates from the experimental ratio, particularly in central collisions. Note that the Case-3 fit ratio does not fully describe the data on the large multiplicity tail and there is room for future improvement. The larger multiplicity in central Ru+Ru than in central Zr+Zr collisions is due to the smaller $\sqrt{\left<r^2\right>}$, the root-mean-square (RMS) size (and thus a higher energy density) of the $^{96}_{44}$Ru nucleus compared to the $^{96}_{40}$Zr nucleus, as predicted by DFT~\cite{Xu:2017zcn,Li:2018oec,Li:2019kkh}. If the radius parameter $R$ is set to be smaller for Ru in the WS density parameterization of Case-2 (and Case-1), then the  high multiplicity tails observed in data would also be described~\cite{Li:2018oec}. However, it would still fail to describe the subtle shape in the intermediate multiplicity range observed in data~\cite{Li:2018oec,Xu:2021vpn}. It must be also noted that the non-zero $\beta_2$ parameter for $^{96}_{40}$Zr as used by Case-2 is not compatible with  transition measurements and calculations~\cite{Kremer:2016nnx,Togashi:2016yzs}. Based on the above considerations, the Case-3 WS density parameterization is chosen for our centrality calculations.
The fit corresponds to values of MC Glauber parameters $n_{pp}=2.386$, $k=3.889$, and $x=0.123$.

\begin{table*}[hbt]
  \centering
  \caption{Centrality definition by $N_{\rm trk}^{\rm offline}$ ranges (efficiency-uncorrected multiplicity in the TPC within $|\eta|<0.5$) in Ru+Ru and Zr+Zr collisions at $\sqrt{s_{_{\rm NN}}}=$200~GeV. The first column is the centrality range labels we use throughout the paper. The two centrality columns are the actual centrality ranges which are slightly different because of integer edge cuts used for the centrality determination. The mean $\langle N_{\rm trk}^{\rm offline}\rangle$ values, the mean number of participants ($\langle N_{\rm part}\rangle$), and the mean number of binary collisions ($\langle N_{\rm coll}\rangle$) are also listed. The statistical uncertainties on $\langle N_{\rm trk}^{\rm offline}\rangle$ are all significantly smaller than 0.01. The uncertainties on $\langle N_{\rm part}\rangle$ and $\langle N_{\rm coll}\rangle$ are systematic.}
  \label{tab:centrality}
  \begin{tabular}{c|ccccc|ccccc}
    \hline
   Centrality & \multicolumn{5}{c|}{Ru+Ru} & \multicolumn{5}{c}{Zr+Zr} \\
    label (\%) & 
    Centrality(\%) & $N_{\rm trk}^{\rm offline}$ & $\langle N_{\rm trk}^{\rm offline} \rangle$ & $\langle N_{\rm part} \rangle$ & $\langle N_{\rm coll} \rangle$ &
    Centrality(\%) & $N_{\rm trk}^{\rm offline}$ & $\langle N_{\rm trk}^{\rm offline} \rangle$ & $\langle N_{\rm part} \rangle$ & $\langle N_{\rm coll} \rangle$ \\
    \hline
    0--5   &      0--5.01 & 258.--500. & 289.32 & 166.8$\pm$0.1  &   389$\pm$10   &      0--5.00 & 256.--500. & 287.36 & 165.9$\pm$0.1  &  386$\pm$10   \\
    5--10  &   5.01--9.94 & 216.--258. & 236.30 & 147.5$\pm$1.0  &   323$\pm$5    &   5.00--9.99 & 213.--256. & 233.79 & 146.5$\pm$1.0  &  317$\pm$5    \\
    10--20 &  9.94--19.96 & 151.--216. & 181.76 & 116.5$\pm$0.8  &   232$\pm$3    &  9.99--20.08 & 147.--213. & 178.19 & 115.0$\pm$0.8  &  225$\pm$3    \\
    20--30 & 19.96--30.08 & 103.--151. & 125.84 &  83.3$\pm$0.5  &   146$\pm$2    & 20.08--29.95 & 100.--147. & 122.35 &  81.8$\pm$0.4  &  139$\pm$2    \\
    30--40 & 30.08--39.89 &  69.--103. &  85.22 &  58.8$\pm$0.3  &  89.4$\pm$0.9  & 29.95--40.16 &  65.--100. &  81.62 &  56.7$\pm$0.3  & 83.3$\pm$0.8  \\
    40--50 & 39.89--49.86 &   44.--69. &  55.91 &  40.0$\pm$0.1  &  53.0$\pm$0.5  & 40.16--50.07 &   41.--65. &  52.41 &  38.0$\pm$0.1  & 48.0$\pm$0.4  \\
    50--60 & 49.86--60.29 &   26.--44. &  34.58 &  25.8$\pm$0.1  &  29.4$\pm$0.2  & 50.07--59.72 &   25.--41. &  32.66 &  24.6$\pm$0.1  & 26.9$\pm$0.2  \\
    60--70 & 60.29--70.04 &   15.--26. &  20.34 & 15.83$\pm$0.03 &  15.6$\pm$0.1  & 59.72--70.00 &   14.--25. &  19.34 & 15.10$\pm$0.03 & 14.3$\pm$0.1  \\
    70--80 & 70.04--79.93 &    8.--15. &  11.47 &  9.34$\pm$0.02 &  8.03$\pm$0.04 & 70.00--80.88 &    7.--14. &  10.48 &  8.58$\pm$0.02 & 7.12$\pm$0.04 \\ \hline
    20--50 & 19.96--49.86 &  44.--151. &  89.50 &  60.9$\pm$0.3  &  96.7$\pm$1.0  & 20.08--50.07 &  41.--147. &  85.68 &  58.9$\pm$0.3  & 90.3$\pm$0.9  \\
    \hline
  \end{tabular}
\end{table*}

\begin{figure}[htb]
    \centering
    \includegraphics[width=0.5\textwidth]{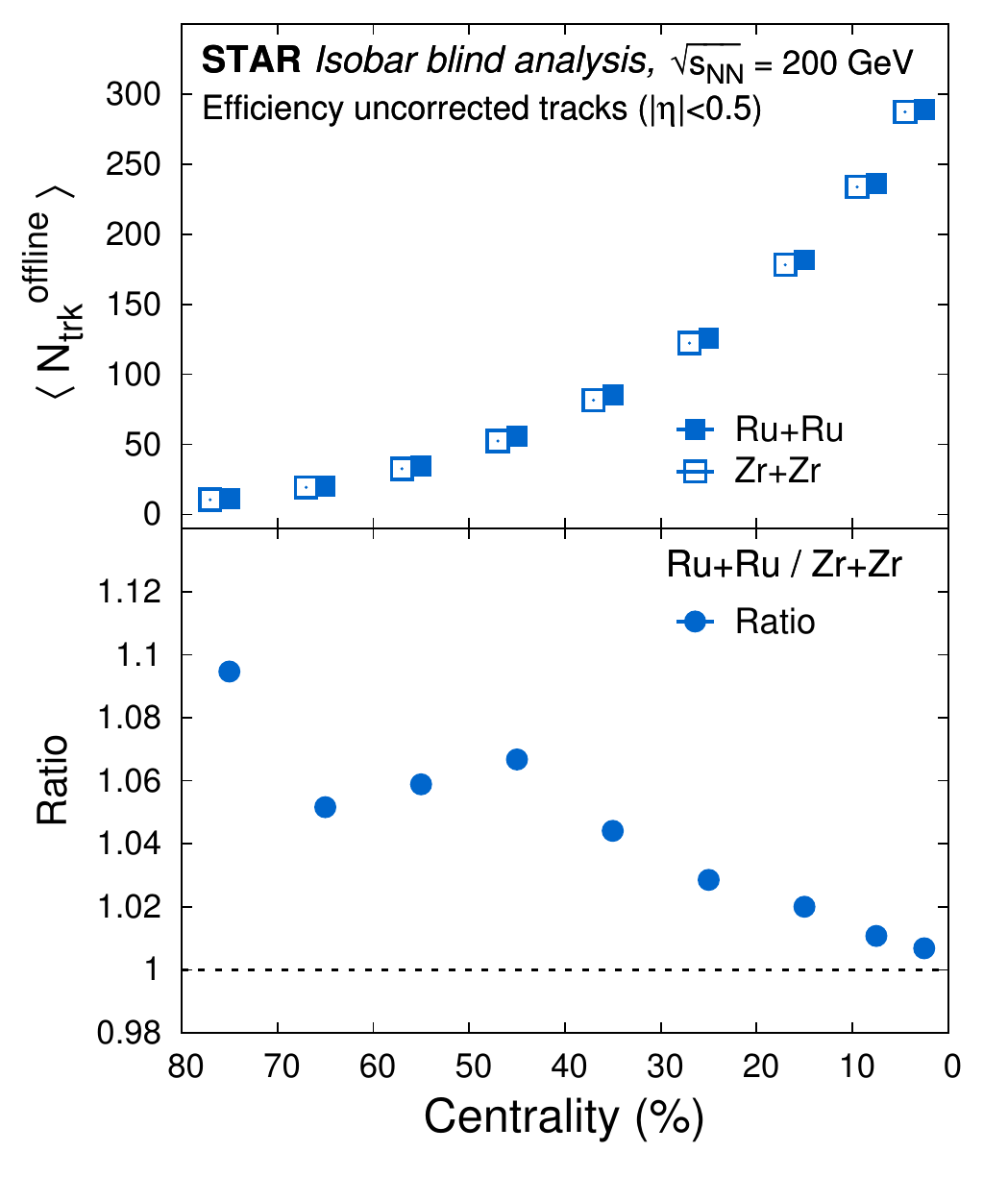}
    \caption{(Upper) The efficiency-uncorrected mean multiplicity $\left<N_{\rm trk}^{\rm offline}\right>$ from the TPC within $|\eta|<0.5$ as a function of centrality in Ru+Ru and Zr+Zr collisions. The centrality bins are shifted horizontally for clarity. (Lower) The ratio of the mean multiplicity in Ru+Ru collisions to that in Zr+Zr collisions in matching centrality. The points include statistical uncertainties that are within the marker size.}
    \label{fig_mean_refmult}
\end{figure}

The centrality of an event is defined by the percentile of the total cross section.
The integer edge cuts are made so that the integrals of the $N_{\rm trk}^{\rm offline}$ distributions would be closest to the 5\% or 10\% mark. For the 0--20\% centrality interval the experimental data are used for integration, while the MC Glauber distributions are used for the remaining range. The reason for this choice is because it is certain that the online trigger is fully efficient for collisions more central than 20\%.

Table~\ref{tab:centrality} lists the centrality definition and the corresponding $\left<N_{\rm trk}^{\rm offline}\right>$, $\left<N_{\rm part}\right>$ and $\left<N_{\rm coll}\right>$ for Ru+Ru and Zr+Zr collisions at $\sqrt{s_{_{\rm NN}}}=$~200~GeV obtained in this work. Throughout this paper, we label the centralities as in the first column of Table~\ref{tab:centrality}. Because of the integer edge cuts in the centrality determination, the actual centrality ranges are slightly different, which are also listed in Table~\ref{tab:centrality} for Ru+Ru and Zr+Zr collisions, respectively. We estimate systematic uncertainties on $\left<N_{\rm part}\right>$ and $\left<N_{\rm coll}\right>$ by varying the input parameters ($R,a$) in the MC Glauber simulation and by varying $n_{pp}$ and $x$ in the two-component model. Figure~\ref{fig_mean_refmult} (upper panel) shows the $\left<N_{\rm trk}^{\rm offline}\right>$ as a function of centrality in the two isobar collision systems. The Ru+Ru/Zr+Zr ratio of the mean multiplicities is shown in the lower panel of Fig.~\ref{fig_mean_refmult}. The mean multiplicity is larger in Ru+Ru collisions than in Zr+Zr collisions of matching centrality. Note that the shape of this ratio as a function of centrality can be affected by the inexact matching of centralities by integer edge cuts on $N_{\rm trk}^{\rm offline}$. The shape may also be influenced by other factors that require further studies.

\section{Observables for isobar blind analysis\label{sec:observables}}
The isobar blind analysis  specifically focuses on the following approaches and corresponding observables. The general strategy is to compare results from the two isobar species to search for a statistically significant difference in the observables used. 
The following subsections describe these approaches and corresponding observables which include: 1) measurements of  the second- and higher-order harmonics of the $\gamma$ correlator, 2) differential measurements of $\Delta\gamma$ (with respect to pseudorapidity gap $\Delta\eta$ and invariant mass $m_{\rm inv}$) to identify and quantify backgrounds, 3) exploiting the relative charge separation across spectator and participant planes, and 4) the use of the $R$ observable to measure charge separation. The first three approaches are based on the aforementioned three-point correlator and the last employs a different approach. \emphasize{For each observable/approach, we predefine a set of the CME signatures prior to the blind analysis, for which a magnitude of high significance must be observed for an affirmative observation of the CME.}

\subsection{$\Delta\gamma$ and mixed harmonics with second and third order event planes}

We rewrite the conventional $\gamma$ correlator (Eq.~(\ref{eq:gamma})) with a more specific notation, 
\begin{equation}
\gamma_{112}=\langle\cos(\phi_{\alpha}+\phi_{\beta}-2\Psi_{2})\rangle\,,
\end{equation}
where $\phi_{\alpha}$ and $\phi_{\beta}$ are the azimuthal angles of particles of interest (POIs) and $\Psi_{2}$ is the second-order flow plane. Here, the subscripts ``1",``1" and ``2" in $\gamma_{112}$ refer to the harmonics associated with the $\phi_{\alpha}$, $\phi_{\beta}$ and $\Psi_2$, respectively. In practice, the flow plane is approximated with the EP ($\rm \Psi_{EP}$) reconstructed with measured particles,
and then the measurement is corrected for the finite EP resolution~\cite{Voloshin:2008dg}.  The charge-dependent backgrounds in $\Delta\gamma_{112}=\gamma_{112}^{\rm OS}-\gamma_{112}^{SS}$ can be broadly understood using the example of resonance decays.
If resonances from the event exhibit elliptic flow, their decay daughters could mimic a signal for  charge separation across the flow plane with a magnitude proportional to $v_2$~\cite{Voloshin:2004vk,Wang:2009kd,Schlichting:2010qia}. Therefore, following Eq.~(\ref{eq:bkgd}), one should study the normalized quantity
\begin{eqnarray}
    \frac{\Delta \gamma_{112}}{v_2}\,,
\label{eq_v2saling}
\end{eqnarray}
 to account for the trivial scaling expected from a purely background scenario. 
The flowing-resonance picture can be generalized to a larger portion of the event, or even the full event,
through the mechanisms of transverse momentum conservation (TMC)~\cite{Pratt:2010zn,Bzdak:2012ia} 
and/or local charge conservation (LCC)~\cite{Schlichting:2010qia}.
In the case of the $\gamma$ correlator this contribution can be written as
\begin{eqnarray}
 \gamma_{112} &=& \langle \cos(\phi_\alpha + \phi_\beta -2 \Psi_2) \rangle \nonumber \\
 &=&\langle\cos(\phi_\alpha-\Psi_2)\cos(\phi_\beta-\Psi_2) \rangle 
-  \langle\sin(\phi_\alpha-\Psi_2)\sin(\phi_\beta-\Psi_2) \rangle \nonumber\\
&=& (\langle v_{1,\alpha}v_{1,\beta}\rangle + B_{\rm IN})- (\langle a_{1,\alpha}a_{1,\beta}\rangle + B_{\rm OUT})\,.
\label{eq:gammaGroup1}
\end{eqnarray}
The CME should dominantly contribute to the $\langle a_{1,\alpha}a_{1,\beta} \rangle$ term. 
The in-plane $\langle v_{1,\alpha}v_{1,\beta} \rangle$ component represents the charge separation unrelated to the magnetic field direction, and
$(B_{\rm IN} - B_{\rm OUT}$) denotes the flow-related background.

Ideally, the two-particle correlator, 
\begin{eqnarray}
\delta &=& \langle \cos(\phi_\alpha -\phi_\beta) \rangle \nonumber \\
&=& (\langle v_{1,\alpha}v_{1,\beta}\rangle + B_{\rm IN}) +(\langle a_{1,\alpha}a_{1,\beta}\rangle + B_{\rm OUT})\,,
\label{eq:delta}
\end{eqnarray}
should also manifest $\langle a_{1,\alpha} a_{1,\beta} \rangle$,
but in reality it could be  dominated by short-range two-particle correlation backgrounds (i.e.~$B_{\rm IN} + B_{\rm OUT}$). 
Similar to $\Delta\gamma_{112}$, we focus on the difference between the opposite-sign and same-sign $\delta$ correlators,
\begin{equation}
\Delta\delta=\delta_{\textsc{os}}-\delta_{\textsc{ss}}\,.
\label{eq_deltadelta}
\end{equation}
The background contributions due to the LCC and TMC have a similar characteristic structure that involves the coupling between $v_2$ and $\Delta\delta$~\cite{Schlichting:2010qia,Pratt:2010zn,Bzdak:2010fd,Bzdak:2012ia}. 
This motivates the study of the normalized quantity of $\Delta \gamma$ scaled by $v_2$ and $\Delta \delta$, defined as:  
\begin{equation}
  \kappa_{112} \equiv \frac{\Delta \gamma_{112}}{v_2\, \Delta \delta}\,.
\label{kappa112}
\end{equation}
The observation of the CME requires  $\kappa_{112}$ to be larger than $\kappa^{\rm TMC/LCC}_{112}$. While a reliable estimate of $\kappa^{\rm TMC/LCC}_{112}$ is still elusive, the comparison of $\gamma_{112}$ (and $\kappa_{112}$) between isobar collisions might give a more definite conclusion on the CME signal.

It is intuitive to introduce some variations in the $\gamma$ correlator  to understand the background mechanisms in $\gamma_{112}$~\cite{Sirunyan:2017quh}, such as
\begin{equation}
\gamma_{123} = \langle \cos(\phi_\alpha + 2\phi_\beta -3{\rm \Psi_{3}}) \rangle\,. 
\end{equation}
This correlator is expected to be insensitive to the CME, because the correlation is negligible between the magnetic field and the third harmonic plane, $\Psi_3$. However, background due to flowing resonances along the $\Psi_3$ plane can contribute to this observable. In analogy to Eq.~(\ref{eq:bkgd}) one can write: 
\begin{equation}
    \Delta\gamma_{123,\rm bkgd} = \frac{4N_{\rm 2p}}{N^2}\left<\cos(\phi_{\alpha}-2\phi_{\beta}-3\phi_{\rm 2p})\right>v_{3,{\rm 2p}}\,.
    \label{eq:bkgd123}
\end{equation}
Therefore, similar to Eq.~(\ref{eq_v2saling}) we also study the scaled quantity 
\begin{equation}
 \frac{\Delta \gamma_{123}}{v_3}\,.  
\end{equation}

Although the direct comparison of $\Delta\gamma_{112}/v_2$ and $\Delta\gamma_{123}/v_3$ is 
hard to interpret for a given system~\cite{Choudhury2020, Schenke:2019pmk}, it is useful to contrast signal and background scenarios by comparing each quantity between the two isobar systems. When compared between the two isobars, in contrast to  $\Delta\gamma_{112}/v_2$ which is driven by differences in both signal and background, $\Delta\gamma_{123}/v_3$ will only be driven by the background difference. 
Since Ru+Ru has a larger magnetic field than Zr+Zr, the CME expectation for mixed-harmonic measurements would be: 
\begin{eqnarray}
&&\frac{(\Delta\gamma_{112}/v_2)^{\rm Ru+Ru}}{(\Delta\gamma_{112}/v_2)^{\rm Zr+Zr}}>1 \,,\label{eq:dg_ratio}\\
&&\frac{(\Delta\gamma_{112}/v_2)^{\rm Ru+Ru}}{(\Delta\gamma_{112}/v_2)^{\rm Zr+Zr}}>\frac{(\Delta\gamma_{123}/v_3)^{\rm Ru+Ru}}{(\Delta\gamma_{123}/v_3)^{\rm Zr+Zr}} \,,\label{eq:dgdg_ratio}\\
&&\frac{(\Delta\gamma_{112}/v_2)^{\rm Ru+Ru}}{(\Delta\gamma_{112}/v_2)^{\rm Zr+Zr}}>\frac{(\Delta\delta)^{\rm Ru+Ru}}{(\Delta\delta)^{\rm Zr+Zr}} \,.
\label{eq:kappa_ratio1}
\end{eqnarray}
The last condition (Eq.\ref{eq:kappa_ratio1}) can be re-written as \begin{eqnarray}
&&\frac{\kappa_{112}^{\rm Ru+Ru}}{\kappa_{112}^{\rm Zr+Zr}}>1\,.
\label{eq:kappa_ratio}
\end{eqnarray}
In general, the algebra relating $\kappa$, $\Delta\gamma$, $v_2$, and $\Delta\delta$ relies on the symmetry assumption of   $\langle\sin(\phi_\alpha-\phi_\beta)\sin n(\phi_\beta-\phi_c)\rangle=0$, with ``$c$" labeling the particle used for EP reconstruction~\cite{Sirunyan:2017quh} and $n$ representing the harmonic order. One can circumvent this assumption by introducing a slight variant of $\kappa$ that measures the factorization breaking:
\begin{eqnarray}
k_n = \frac{\Delta \langle \cos(\Delta\phi_{\alpha \beta})\cos(n\Delta\phi_{\beta c}) \rangle}{ v_{n}^2\{2\} \Delta \delta_{\alpha\beta} }\,.
\label{eq_kn}
\end{eqnarray}
Here the first ``$\Delta$" in the numerator denotes the difference between opposite-sign and same-sign measurements of the quantity inside the  average. The quantity $\Delta \phi_{\alpha \beta}=\phi_\alpha-\phi_\beta$ denotes the relative azimuthal angle between charge-carrying particles, whereas the quantity $\Delta \phi_{\beta c}=\phi_\beta-\phi_c$ is the relative difference between one of the charge-carrying particles and the particles used for EP reconstruction. The quantity $\Delta \delta_{\alpha \beta}$ in the denominator has the same definition as Eq.~(\ref{eq_deltadelta}). The quantity $v_n\{2\}$ is the $n$-th order harmonic anisotropy coefficients estimated using two-particle correlations. The CME is expected to cause an excess charge separation perpendicular to the $\Psi_2$ plane, whereas the background-driven charge separations along the $\Psi_2$ and $\Psi_3$ planes are proportional to $v_2$ and $v_3$, respectively. Under these assumptions, one expects the case for the CME to be:
\begin{eqnarray}
 \frac{k_2^{\rm Ru+Ru}}{k_2^{\rm Zr+Zr}} > \frac{k_3^{\rm Ru+Ru}}{k_3^{\rm Zr+Zr}}\,.
 \label{eq:k2k3signature}
\end{eqnarray}
 
For simplicity, the notation $\gamma$ is used in place of $\gamma_{112}$ in the following subsections (Sec. IV B-E).

\subsection{Relative pseudorapidity dependence of $\Delta\gamma$}

The relative pseudorapidity dependence of azimuthal correlations is widely studied to identify sources of long-range components that are dominated by early-time dynamics. They are contrasted to late-time correlations that are restricted by causality to appear as short-range correlations~\cite{Dumitru:2010iy}. The same approach can be extended to charge-dependent correlations which provide the impetus to explore the dependence of $\Delta\gamma$ on the pseudorapidity gap between the charge-carrying particles
$\Delta\eta_{\alpha\beta}=|\eta_\alpha-\eta_\beta|$ in $\langle\cos(\phi_{\alpha} (\eta_\alpha)+\phi_{\beta}(\eta_\beta)-2\Psi_{\rm RP}) \rangle$.
Such measurements have been performed in STAR with Au$+$Au and U$+$U data~\cite{STAR:2009tro,Tribedy:2017hwn}. The possible sources of short-range correlations due to photon conversion to $e^+e^-$, HBT, and Coulomb effects can be identified and described as Gaussian peaks at small $\Delta\eta_{ab}$, the width and magnitude of which strongly depend on centrality and system size~\cite{STAR:2011ryj}. Going to more peripheral centrality bins, it becomes harder to identify such components as they overlap with sources of di-jet fragmentation that dominate both same-sign and opposite-sign correlations. Decomposing different components of $\Delta\gamma$ via study of $\Delta\eta_{ab}$-dependence is challenging, although a clear sign of different sources of correlations is visible in the change of shape of individual same-sign and opposite-sign measurements of the $\gamma$ correlator~\cite{Tribedy:2017hwn}. Nevertheless, these differential measurements of $\Delta\gamma$ in isobar collisions offer the prospects for studying the $\Delta\eta$ dependence of the CME. By comparing the differential measurements in Ru+Ru and Zr+Zr, it may be possible to extract the $\Delta\eta$ distribution of the CME signal, thus providing deeper insight into the origin of the phenomenon. The magnetic field driven CME signal is expected to dominate the long-range component $\Delta\eta_{\alpha\beta}>1$ of the $\Delta\eta$ dependence like other early stage phenomena while the background due to resonance decay are expected to be short-range $\Delta\eta_{\alpha\beta}<1$~\cite{Dumitru:2010iy}. In a CME scenario we expect the long-range component in the case of Ru+Ru collisions to be larger than that of Zr+Zr.

\subsection{Invariant mass dependence of $\Delta\gamma$ \label{sec:minv}} 

Since resonances present a large background source to the CME, the study of invariant mass ($m_{\rm inv}$) dependence of the measured signal is natural and was first introduced in Ref.~\cite{Zhao:2017nfq}. If we perform the analysis using pairs of pions, differential measurement of $\Delta\gamma$ with respect to $m_{\rm inv}$ should show peak-like structures similar to those in the relative pair multiplicity difference,
\begin{equation}
    r=(N_{\textsc{os}}-N_{\textsc{ss}})/N_{\textsc{os}}\,,
    \label{eq:r}
\end{equation}
if backgrounds from neutral resonances dominate the measurement. Here $N_{\textsc{os}}$ and $N_{\textsc{ss}}$ are the numbers of opposite-sign and same-sign pion pairs, respectively. Indeed, similar peak structures are observed and an analysis utilizing the $m_{\rm inv}$ dependence and the ESE technique has been performed to extract the possible fraction of the CME signal in Au+Au collisions~\cite{Adam:2020zsu}. 
A similar analyses can be applied separately to the individual Ru+Ru and Zr+Zr data to extract a CME fraction in each system. Such an analysis will be performed in future work.

In this analysis we focus on contrasting the two isobar systems. We may gain insight into the mass dependence of the CME by combining the measurements in Ru+Ru and Zr+Zr collisions. Assuming in this blind analysis that the physics background is proportional to $v_2$ only (i.e.~everything else is identical between the two isobar systems except  $v_2$), we have
\begin{equation}
    \Delta\gamma^{\rm Ru+Ru}-a'\Delta\gamma^{\rm Zr+Zr}
    =\Delta\gamma_{\textsc{cme}}^{\rm Ru+Ru}-a'\Delta\gamma_{\textsc{cme}}^{\rm Zr+Zr}\,,
    \label{eq:minvDg}
\end{equation}
where 
\begin{equation}
    a'=v_2^{\rm Ru+Ru}/v_2^{\rm Zr+Zr}\,.
    \label{eq:minvaprime}
\end{equation}
The quantity $a'$ can be safely assumed to be independent of $m_{\rm inv}$,  because the two isobar systems are similar.
A CME signature would be a positive measurement of the l.h.s.~of Eq.~(\ref{eq:minvDg}):
\begin{equation}
    \Delta\gamma^{\rm Ru+Ru}-a'\Delta\gamma^{\rm Zr+Zr}>0\,.
    \label{eq:minvsign}
\end{equation}
Because the mass dependence of the CME signal is unlikely to  differ between Ru+Ru and Zr+Zr collisions, such a measurement would give unique insight on the mass dependence of the CME. Note Eq.~(\ref{eq:minvDg}) is valid for other independent variables besides $m_{\rm inv}$, such as the $\Delta\eta$ described in the previous subsection.

\subsection{$\Delta\gamma$ with spectator and participant  planes (approach-I)} \label{sec:Group3SPPP}
This analysis makes use of the fact that the magnetic field driven signal is more correlated to the RP, in contrast to flow-driven backgrounds which are maximal along the particpant plane (PP). 
The idea was first published in Ref.~\cite{Xu:2017qfs} and later discussed in Ref.~\cite{Voloshin:2018qsm}. It requires measurement of $\Delta\gamma$ with respect to the plane of produced particles, a proxy for the PP, as well as with respect to the plane of spectators, a good proxy for RP. In STAR, the two measurements can be done using $\Psi_2$ from the TPC and $\Psi_1$ from the ZDCs, respectively. 

The approach is based on three main assumptions: 1) the measured $\Delta\gamma$ has contributions from signal and background, which can be decomposed as $\Delta\gamma=\Delta\gamma_{\rm bkg}+\Delta\gamma_{\rm sig}$, 2) the background contribution to $\Delta\gamma$ should follow the scaling $\Delta\gamma_{\rm bkg}\{\textsc{zdc}\}/\Delta\gamma_{\rm bkg}\{\textsc{tpc}\}=v_{2}\{\textsc{zdc}\}/v_{2}\{\textsc{tpc}\}$, and 3) the signal contribution to $\Delta\gamma$ should follow the scaling $\Delta\gamma_{\rm sig}\{\textsc{zdc}\}/\Delta\gamma_{\rm sig}\{\textsc{tpc}\}=v_{2}\{\textsc{tpc}\}/v_{2}\{\textsc{zdc}\}$. 
The first one has been known to be a working assumption, widely used for a long time~\cite{Kharzeev:2015znc,Zhao:2019hta}. 
The second one is borne out by the fact that backgrounds come from particle correlations whose sources are $v_2$ modulated (see Eq.~(\ref{eq:bkgd}))~\cite{Voloshin:2004vk,Wang:2009kd,Bzdak:2009fc,Schlichting:2010qia}. The beauty of the method is that, because the TPC and ZDC measurements are performed in identical events, all other factors contributing to $\Delta\gamma$ (such as resonance decay correlations and multiplicity dilution) cancel except $v_2$. Nevertheless, non-flow effects could potentially spoil the scaling which requires quantitative investigations~\cite{Feng:2021pgf}.
The validity of the third assumption is studied and demonstrated in Ref.~\cite{Xu:2017qfs}. 
The reciprocal stems from fluctuations of RP and PP, whose relative azimuthal angle may be quantified by $a=\langle\cos2(\Psi_{\textsc{pp}}-\Psi_{\textsc{rp}})\rangle$~\cite{Alver:2008zza}.

Using all three assumptions, one can extract the fraction of possible CME signal in a fully data-driven way~\cite{Xu:2017qfs},
\begin{eqnarray}
 f_{\textsc{cme}}=\frac{\Delta\gamma_{\textsc{cme}}\{\textsc{tpc}\}}{\Delta\gamma\{\textsc{tpc}\}}=\frac{A/a-1}{1/a^2-1}\,,
 \label{eq:fcme}
\end{eqnarray}
where
\begin{eqnarray}
 A=\Delta\gamma\{\textsc{zdc}\}/\Delta\gamma\{\textsc{tpc}\}\,,
 \label{eq:A}
\end{eqnarray}
and the $a$ parameter can be determined by
\begin{eqnarray}
    a=v_2\{\textsc{zdc}\}/v_2\{\textsc{tpc}\}\,.
    \label{eq:a}
\end{eqnarray}
The $f_{\textsc{cme}}$ given by Eq.~(\ref{eq:fcme}) is the fraction of CME contribution to the $\Delta\gamma\{\textsc{tpc}\}$ with respect to the TPC EP.

Such an analysis has been applied to existing Au+Au data, and a CME signal fraction of the order of 10\% has been extracted with a significance of 1--3$\sigma$~\cite{STAR:2021pwb}. We apply the same analysis to the isobar data as part of the blind analysis. The
 $f_{\textsc{cme}}$ is extracted in each isobar system separately. 
The case for the CME in this analysis would be
\begin{equation}
    f_{\textsc{cme}}^{\rm Ru+Ru}>f_{\textsc{cme}}^{\rm Zr+Zr}>0\,.
\end{equation}

One can get an additional constraint on $f_{\textsc{cme}}^{\rm Ru+Ru}$ and $f_{\textsc{cme}}^{\rm Zr+Zr}$. Assuming in this blind analysis that the physics background is proportional to $v_2$ only, 
\begin{equation}
    (1-f_{\textsc{cme}}^{\rm Ru+Ru})\Delta\gamma^{\rm Ru+Ru}/v_2^{\rm Ru+Ru}=(1-f_{\textsc{cme}}^{\rm Zr+Zr})\Delta\gamma^{\rm Zr+Zr}/v_2^{\rm Zr+Zr}\,,
\end{equation}
we obtain
\begin{equation}
    f_{\textsc{cme}}^{\rm Ru+Ru}=\left(\frac{a'}{A'}\right)f_{\textsc{cme}}^{\rm Zr+Zr}+\left(1-\frac{a'}{A'}\right),
    \label{eq:ff}
\end{equation}
where 
\begin{eqnarray}
 A'=\Delta\gamma^{\rm Ru+Ru}/\Delta\gamma^{\rm Zr+Zr}\,,
 \label{eq:Aprime}
\end{eqnarray}
and 
$a'$ is again given by Eq.~(\ref{eq:minvaprime}).
The quantity $a'/A'$ is the double ratio of
\begin{equation}
    a'/A'=\left(\Delta\gamma^{\rm Zr+Zr}/v_2^{\rm Zr+Zr}\right) / \left(\Delta\gamma^{\rm Ru+Ru}/v_2^{\rm Ru+Ru}\right)\,.
    \label{eq:Aa}
\end{equation}
The individual measurements of   $f_{\textsc{cme}}^{\rm Ru+Ru}$ and $f_{\textsc{cme}}^{\rm Zr+Zr}$ by Eq.~(\ref{eq:fcme}) and the constraint on their relationship by Eq.~(\ref{eq:ff}) give quantitatively an allowed region of the CME signal fractions.

\subsection{$\Delta\gamma$ with participant and spectator planes (approach-II)} \label{sec:Group4_PPSP}

The main objective is to obtain the double ratio $(\dgamma/v_2)_{\Ru}/(\dgamma/v_2)_{\Zr}$.
As discussed in Ref.~\cite{Voloshin:2018qsm} an evaluation of the ratio 
$\dgv$ does not require knowledge of the reaction plane resolution,
which reduces the systematic uncertainty.  It also ``normalizes'' the
$\gamma$ correlator to the elliptic flow value (which is proportional to background) and thus can be used for a
direct comparison of the signals in different isobar collisions, even
if the values of elliptic flow are slightly different in the two systems.  
Thus, the double ratio $(\Delta\gamma/v_2)_{\rm Ru+Ru}/(\Delta\gamma/v_2)_{\rm Zr+Zr}$, and specifically its deviation from unity, can be directly used for a qualitative detection of the CME signal. 
To extract the CME signal in this approach the double ratio is fit with the equation:
\be 
\frac{\dgv_{\Ru}}{\dgv_{\Zr}}=1+f_{\rm \sss CME}^{\sss
  \Zr}[(B_{\sss \Ru}/B_{\sss \Zr})^2-1], 
\label{eq:dblratio}
\ee 
where $f_{\textsc{cme}}^{\Zr}$ is the CME fraction in the $\dgamma$
correlator measured in \Zr\ collisions, and $(B_{\sss \Ru}/B_{\sss \Zr})$
is the ratio of the magnetic field strengths in \Ru\ and \Zr\
collisions. By default this ratio is taken as the ratio of the nuclear
charges, but can be varied to take into account the uncertainties
related to the magnetic field determination.

For a non-zero CME signal it is expected that the double ratio
$(\dgamma/v_2)_{\Ru}/(\dgamma/v_2)_{\Zr}$ would be greater than unity,
as the CME signal in \Ru\ collisions is expected to be about 15\% larger
than in \Zr\ collisions~\cite{Voloshin:2010ut,Deng:2016knn}, and the background difference should be significantly smaller.

For the separate estimates of the CME signal in each of the isobar
collisions, the $\gamma$ correlator and elliptic flow can be also measured using STAR's two ZDC-SMD $\Psi_1$ event planes (spectator planes):
\begin{eqnarray}
(\Delta\gamma/v_2)_{\rm ZDC} 
&=& 
\frac{\Delta
  \langle\cos(\phi_{\alpha}+\phi_{\beta}-\Psi_1^{\rm W}-\Psi_1^{\rm E})\rangle}
{\langle\cos(2\phi-\Psi_1^{\rm W}-\Psi_1^{\rm E})\rangle}, 
\label{eq:dg_zdc}
\end{eqnarray}
where $\Psi_1^{\rm W(E)}$ is the event plane determined with ZDC-SMD in the west (east) side of STAR and the west side corresponds to the backward direction. 
Then this can be used for calculations of the double ratios:
\begin{eqnarray}
  \frac{\displaystyle (\Delta\gamma/v_2)_{\rm spectator}}
{\displaystyle (\Delta\gamma/v_2)_{\rm participant}} = \frac{\displaystyle(\Delta\gamma/v_2)_{\rm ZDC}}
{\displaystyle (\Delta\gamma/v_2)_{\rm TPC}} =
  \frac{ \displaystyle
  \Delta
  \langle\cos(\phi_{\alpha}+\phi_{\beta}-\Psi_1^{\rm W}-\Psi_1^{\rm E})\rangle
/\langle\cos(2\phi-\Psi_1^{\rm W}-\Psi_1^{\rm E})\rangle
}
{ \displaystyle \Delta \langle\cos(\phi_{\alpha}+\phi_{\beta}-2\phi_c)\rangle
/ \langle\cos(2\phi_\alpha-2\phi_c)\rangle  }.
\label{eq:dblratio}
\end{eqnarray}

To extract the signal, one has to make further assumptions~\cite{Voloshin:2018qsm}. Following the
most plausible scenario of the magnetic field oriented on average
perpendicular to the spectator plane, the CME fraction, $f^{\textsc{tpc}}_{\textsc{cme}}$, can be extracted via fitting of the results with the equation:
\begin{equation}
\frac{\dgv_{\rm ZDC}}{\dgv_{\rm TPC}} = 1 + f^{\rm \sss
  TPC}_{\rm \sss CME} \left(
\frac{v_{2}^2\{\rm TPC\} }{v_2^2{\rm \{ZDC\}}} - 1
\right).
\label{eq:Group4_fcmeSP0}
\end{equation}
While the calculation of the double ratio, l.h.s. of Eq.~\eqref{eq:Group4_fcmeSP0}, does not require knowledge of the reaction plane resolutions, the quantitative estimate of $f^{\textsc{tpc}}_{\textsc{cme}}$  from the double ratio requires $v_2$ values corrected for the reaction plane resolution. 
For the correlations relative to the sum of the first harmonic ZDC event planes the corresponding event plane resolution can be extracted directly from the data as $\langle \cos(\Psi_1^{\rm W}-\Psi_1^{\rm E}) \rangle$.

\subsection{$R_{\Psi_{2}}$ variable} 

The $R_{\Psi_2}(\Delta S)$ variable provides an alternate way of measuring charge separation. It is obtained by taking the ratio of two sets of correlation functions \cite{Magdy:2017yje,Magdy:2018lwk} defined as:
\begin{equation}
R_{\Psi_2}(\Delta S) = C_{\Psi_2}(\Delta S)/C_{\Psi_2}^{\perp}(\Delta S).
\label{eq:4}
\end{equation}
Here the correlation functions $C_{\Psi_2}(\Delta S)$ and $C_{\Psi_2}^{\perp}(\Delta S)$ quantify charge separation $\Delta S$, perpendicular and parallel (respectively) to the $\Psi_{2}$ EP. The suffix ``$\perp$" is motivated by the direction of the $\vec{B}$ field. Since the $\vec{B}$ field is nearly perpendicular to the $\Psi_{2}$ EP, $C_{\Psi_2}(\Delta S)$ and $C_{\Psi_2}^{\perp}(\Delta S)$ measure charge separation approximately parallel and perpendicular (respectively) to the $\vec{B}$ field. These correlation functions are further obtained from the ratios of two distributions~\cite{Magdy:2017yje,Magdy:2018lwk};
\begin{equation}
C_{\Psi_{2}}(\Delta S) = \frac{N_{\text{real}}(\Delta S)}{N_{\text{shuffled}}(\Delta S)}. 
\label{eq:5}
\end{equation}
Here, $N_{\text{real}}(\Delta S)$ is the distribution of the quantity $\Delta S$ that measures the event-by-event average of the charge separation: 
\begin{equation}
\Delta S =  \frac{{\sum\limits_1^{n^+} {w^{+}_i\sin (\Delta {\varphi_{2} })} }}{\sum\limits_1^{n^+} {w^{+}_i}} 
   - \frac{{\sum\limits_1^{n^-} {w^{-}_i\sin ( \Delta {\varphi_{2}  })} }}{\sum\limits_1^{n^-} {w^{-}_i}}.
\label{eq:6}
\end{equation}
Here, ${n^-}$ and ${n^+}$ are the numbers of negatively and positively charged particles in an event, $w^{\pm}_i$ are weights that account for non-uniformity of the azimuthal acceptance of the TPC and $\Delta {\varphi_{2}}= \phi - \Psi_{2}$. The distribution $N_{\text{shuffled}}(\Delta S)$ is obtained in a way similar to that for $N_{\text{real}}(\Delta S)$ but after random reassignment (shuffling) of the charges of the reconstructed tracks in each event. 
This randomization makes $N_{\text{shuffled}}(\Delta S)$ insensitive to charge-dependent correlations and ensures identical event property between the numerator and the denominator in Eq.~(\ref{eq:5}).
The correlation function $C_{\Psi_{2}}^{\perp}(\Delta S)$ is constructed in a way similar to Eq.\ref{eq:5} by replacing $\Psi_{2}$ with $\Psi_{2}+\pi/2$. Both $C_{\Psi_{2}}(\Delta S)$ and $C_{\Psi_{2}}^{\perp}(\Delta S)$ have nearly Gaussian shapes around $\Delta S=0$. 

The final variable $R_{\Psi_2}(\Delta S)$, obtained according to Eq.\ref{eq:4} measures the relative charge separation between parallel and perpendicular directions to the $\vec{B}$ field. CME-driven charge separation along the $\vec{B}$ field is expected to lead to concave-shaped $R_{\Psi_2}(\Delta S)$ distributions with stronger CME signals leading to narrower (more concave) distributions~\cite{Magdy:2017yje}. The width of $R_{\Psi_2}(\Delta S)$ reflects the magnitude of charge separation, which is also influenced by particle number fluctuations and resolution of the EP. Both increase the width ($\mathrm{\sigma_{\Delta_{Sh}}}$) of the R-variable. The effect of the particle number fluctuations can be accounted for by scaling $\Delta S$ by the width $\mathrm{\sigma_{\Delta_{Sh}}}$ of the $N_{\text{shuffled}}$ distribution, i.e., $\Delta S^{'} = \Delta S/\mathrm{\sigma_{\Delta_{Sh}}}$. This re-scaled distribution of $\Delta S^{\prime}$ can be further corrected for EP resolution. This is done by using a parametrized function ${\delta_{R} = \text{Res}\times  e^{(1-\text{Res})^2}}$to correct $\Delta S^{'}$ to $\Delta S^{''}=\Delta S^{'} \times \delta_{R}$. Here $\text{Res}$ is the EP resolution. This approach of correction has been verified using simulation studies~\cite{Magdy:2017yje} and data-driven tests.

After the analysis code was frozen, the $R_{\Psi_3}$~\cite{Bozek:2017plp,Magdy:2017yje} observable, constructed
to be insensitive to CME, was found to have a programming error that failed to convert some integers to floats, along with an issue in azimuthal periodicity (for more details see Ref.~\cite{Feng:2018chm}).  Consequently, $R_{\Psi_3}$ results are not included in this paper.

Assuming collisions of Ru+Ru produce stronger magnetic field than that of  Zr+Zr, a signature for CME-driven charge separation would be indicated by the observation
\begin{equation}
    1/\sigma_{R_{\Psi_2}}^{\rm{Ru+Ru}} > 1/\sigma_{R_{\Psi_2}}^{\rm{Zr+Zr}}\,,
    \label{eq:Rsignature}
\end{equation} 
where $\sigma_{R_{\Psi_2}}$ is the Gaussian width of the respective $R(\Delta S'')$ distribution.

\section{Isobar blind analysis results\label{sec:results}}
\begin{figure}[htb]
    \centering
    \includegraphics[width=0.45\textwidth]{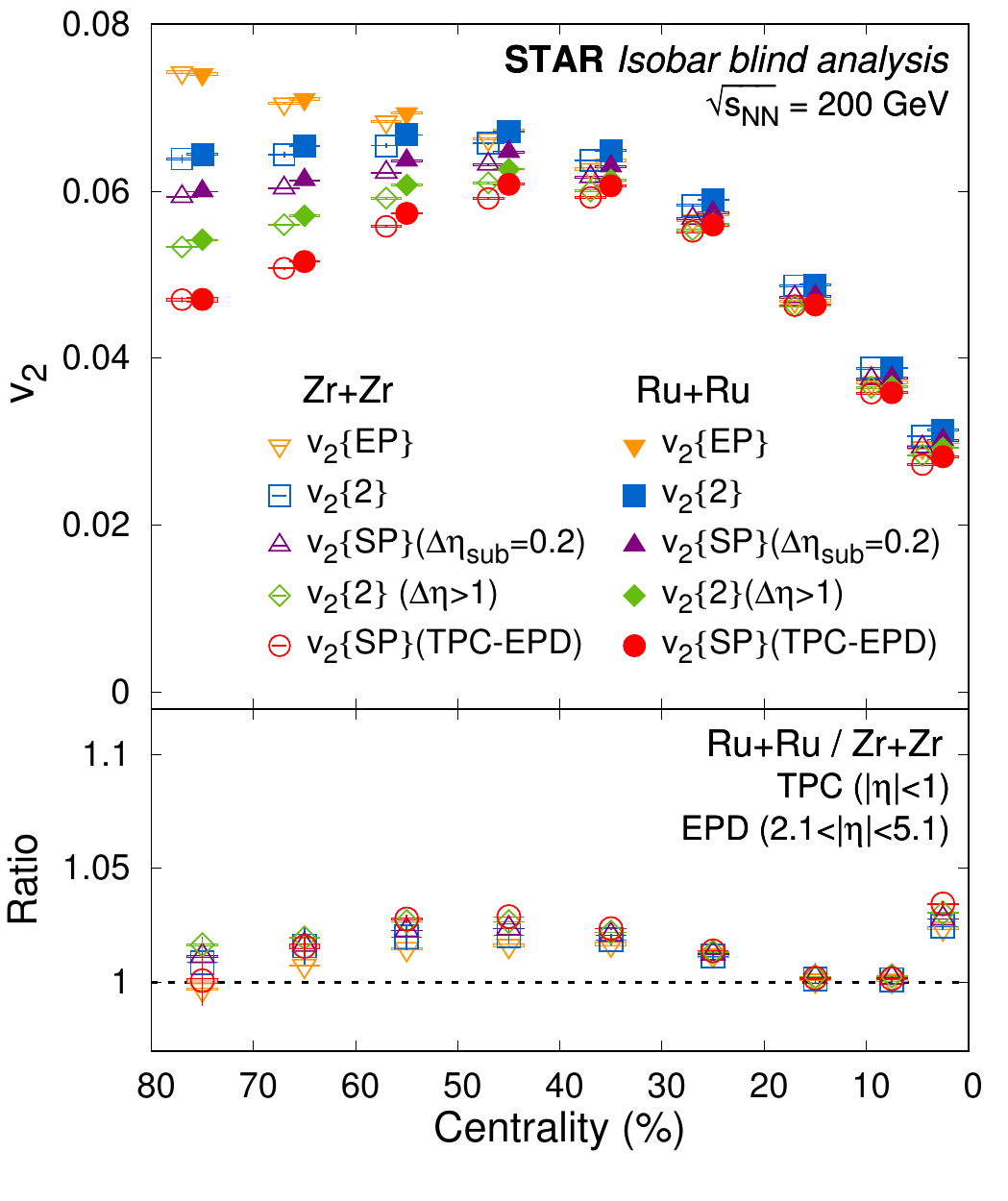}
    \includegraphics[width=0.45\textwidth]{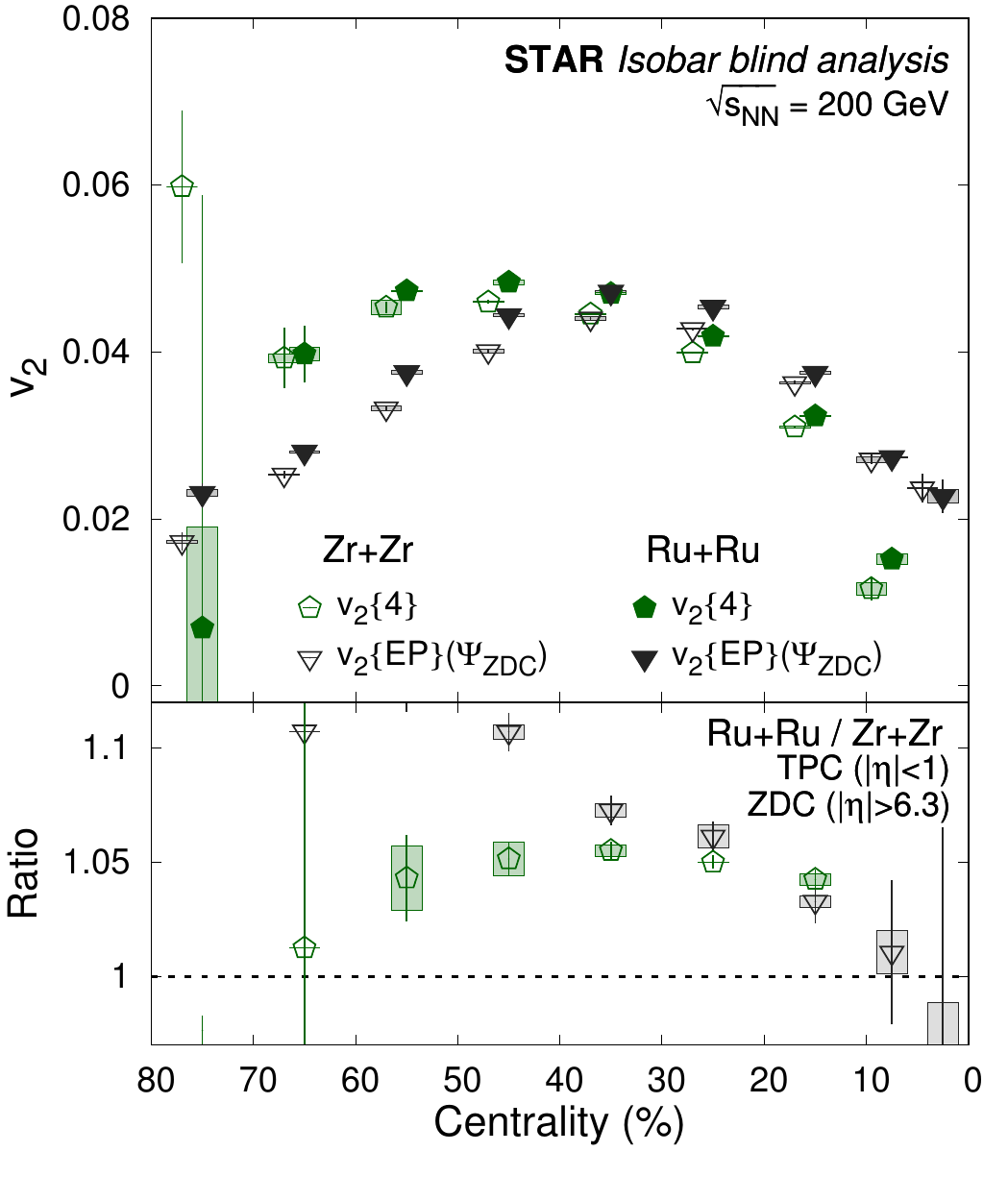}
    \caption{(Left) Elliptic anisotropy $v_2$ measurements using different methods in isobar collisions at $\sqrt{s_{_{\rm NN}}}=200$~GeV as a function of centrality using TPC and EPD detectors. In the upper panels, the solid and open symbols represent measurements for Ru+Ru and Zr+Zr collisions, respectively. The data points are shifted along the $x$ axis for clarity. The lower panels show the $v_2$ ratios in Ru+Ru over Zr+Zr collisions. The statistical uncertainties are represented by lines and systematic uncertainties by boxes. (Right) The same showing measurements for four particle correlations using TPC and EP determined from ZDC. The data points are shifted horizontally for clarity.}
    \label{fig_v2_all}
\end{figure}

\begin{figure}
\centering
\includegraphics[width=0.48\textwidth]{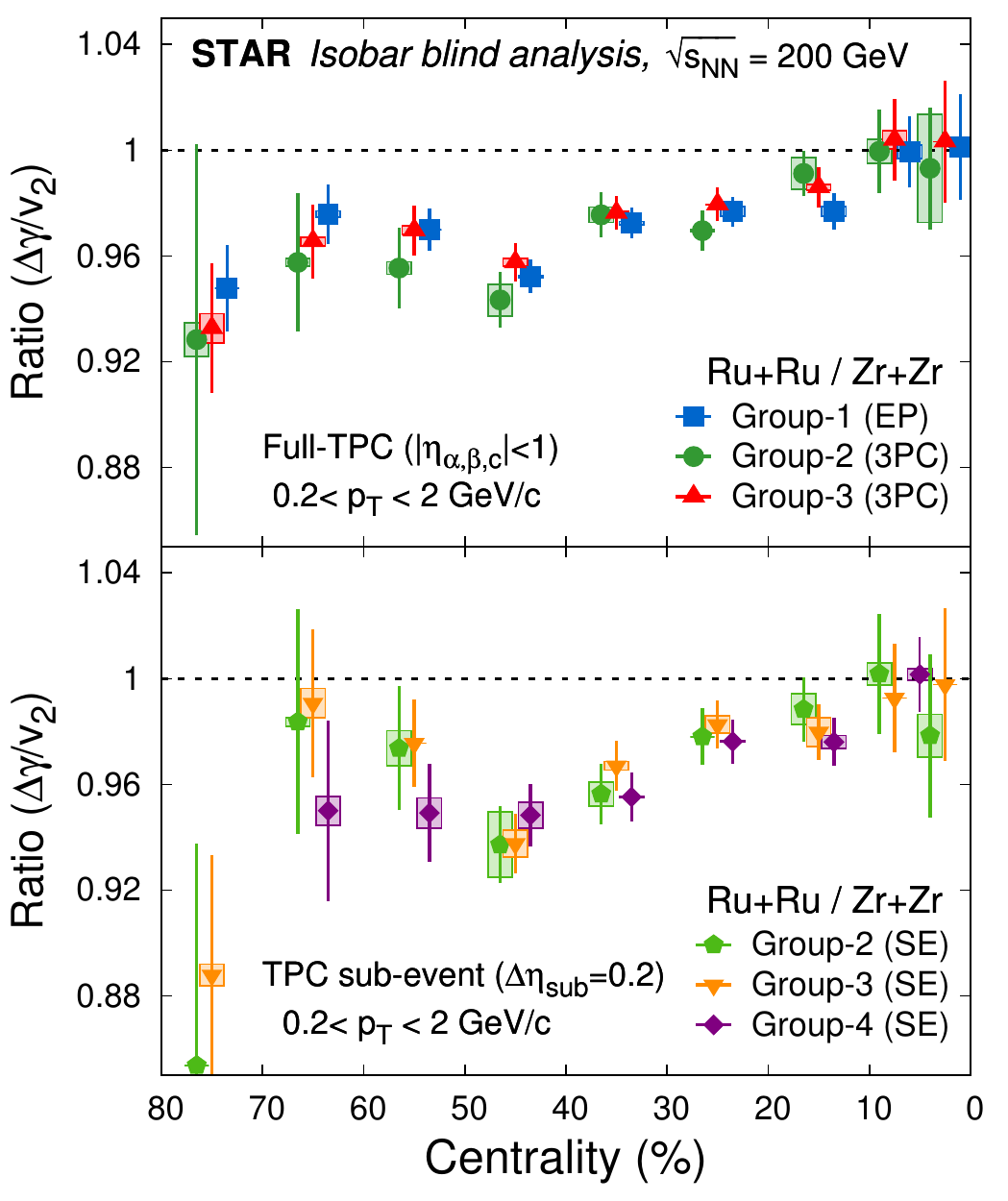}
\caption{Ru+Ru over Zr+Zr ratio of the primary CME-sensitive observable $\Delta \gamma/v_2$ estimated using different methods and by independent analysis groups. The vertical lines represent statistical uncertainties while the  rectangular boxes represent systematic uncertainties. The upper panel shows the results using the entire TPC acceptance estimated using event-plane (EP) and three-particle correlations (3PC) methods without any $\eta$ gaps. The lower panel shows the results using a sub-event (SE) method with gap ($\Delta\eta_{\rm sub}$) of 0.2. 
Note the most central data point from Group-4 is for 0--10\% centrality.
The centrality bins are shifted horizontally for clarity.}
    \label{fig_deltagamma_v2_compilation}
\end{figure}

The major background in the primary CME-sensitive observable, $\Delta\gamma_{112}$, is due to elliptic flow, $v_2$. Therefore, we first examine the $v_2$ measurements from the blind analysis. The upper panels of Fig.~\ref{fig_v2_all} show a compilation of the $v_2$ results obtained by different analysis groups using different methods and reported in the following subsections. For clarity, not all results from all groups are shown in Fig.~\ref{fig_v2_all}. The $v_2$ values from different methods are expected to be different~\cite{Poskanzer:1998yz}, and we found consistency in results among different groups using the same method.
The lower panels of Fig.~\ref{fig_v2_all} show the $v_2$ ratio between Ru+Ru  and Zr+Zr collisions. All the ratios, except noticeably the $v_2\{4\}$ and $v_2\{\Psi_{\textsc{zdc}}\}$ ratios, fall on a common curve. The ratios are above unity by 2--3\% in mid-central collisions and fall off towards peripheral and central collisions, with the exception of the top 5\% centrality bin, where the ratios are also above unity by a few percent. The central-collision results are likely due to a larger quadrupole deformity in $^{96}_{44}$Ru compared with $^{96}_{40}$Zr, which needs future investigation. The above-unity ratio in mid-central collisions may originate from the different nuclear structures between the two isobars as predicted by the DFT calculations~\cite{Xu:2017zcn,Li:2018oec}. These $v_2$ ratios imply different magnitudes of the CME backgrounds in the two isobar systems, and this effect is taken into account in the $\Delta\gamma_{112}/v_2$ and $\kappa_{112}$ observables described in Sec.~\ref{sec:observables}. 
It is often advantageous to study the CME observables with $\Psi_{\textsc{zdc}}$,  which measures the spectator plane, and is more correlated with the magnetic field than the participant plane. However, the $v_2\{\Psi_{\textsc{zdc}}\}$ ratio is significantly larger than unity; this comes primarily from the better alignment of the spectator and participant planes in Ru+Ru than in Zr+Zr collisions, as predicted by the DFT nuclear structure calculations~\cite{Xu:2017zcn}. 
This would imply that the advantage in using the isobar difference or ratio to search for the CME with $\Psi_{\textsc{zdc}}$ is limited~\cite{Xu:2017zcn}.
The qualitative consistencies between $v_{2}\{\Psi_{\textsc{zdc}}\}$ and $v_{2}\{4\}$ and between their respective Ru+Ru/Zr+Zr ratios are observed as expected in a Gaussian model description of flow fluctuations~\cite{Voloshin:2007pc}.

In the following subsections, we discuss the results of the CME-related observables from the five independent analysis groups, each focusing on a few specific observables related to measurements from specific detectors. The detailed implementations differ among the groups with regards to estimation of harmonic flow vectors, re-weighting, the pseudorapidity gap to reduce non-flow, and correction of non-uniform acceptance. While focusing on various aspects, four of the five groups have analyzed the $\Delta\gamma/v_2$ observable. Figure~\ref{fig_deltagamma_v2_compilation} compares the $\Delta\gamma/v_2$ measurements with both the full-event and sub-event methods. The statistical uncertainties are largely correlated among the groups because the same initial data sample  is analyzed; the results are not identical because of the analysis-specific event selection criteria (see Table.\ref{tab:QA}) and the slightly different methods. Using the Barlow approach~\cite{Barlow:2002yb}, we have verified that the results from different groups are consistent within the statistical fluctuations due to those differences. Moreover, the final conclusion on the observability of the CME is consistent among all five analysis groups.

In addition to the centrality dependence results reported in the following subsections, in order to have the best statistics, we also quote the final results for the Ru+Ru over Zr+Zr ratio observables for the centrality range of 20--50\%. The choice of this centrality range is determined by two considerations. One is that the mid-central collisions present the best EP reconstruction resolution as well as the most significant magnetic field strengths (hence the possibly largest CME signal difference between the isobar species). The other consideration is that the online trigger efficiency starts to deteriorate from the 50\% centrality mark towards more-peripheral collisions (see Sec.~\ref{sec:centrality}).
A compilation of results from different groups is presented in the summary subsection~\ref{sec:summary}.

\subsection{$\Delta\gamma$ measurements with TPC event plane (Group-1)}

The flow plane for a specific pseudorapidity range is unknown for each event. In practice, we estimate an $n^{\rm th}$-harmonic flow plane 
with the azimuthal angle ($\Psi_n$) of the flow vector $\overrightarrow{Q_n} = \big(\sum_i^N w_i\cos(n\phi_i), \sum_i^N w_i\sin(n\phi_i)\big)$, where $\phi_i$ represents the azimuthal angle of a detected particle, and $w_i$ is a weight (often set to $p_T$) to optimize the EP resolution. For example, the $v_n$ measurement with respect to the full TPC EP is denoted by
\begin{equation}
v_n\{{\rm TPC~EP}\} = \langle \cos (n\phi-n\Psi_n^{\rm TPC})\rangle\,.
\label{eq_v2ep}
\end{equation}
The corresponding $\gamma_{112}$ correlator is represented by
\begin{eqnarray}
\gamma_{112}\{\rm TPC~ EP\}= \langle \cos(\phi_\alpha+\phi_\beta-2\Psi_2^{\rm TPC})\rangle\,.
\label{eq_gmmaep}
\end{eqnarray}
The two-particle $\delta$ correlator is estimated in the same way as defined in Eq.~(\ref{eq:delta}). To account for the detector non-uniformity, both $\phi$ and $\Psi_2^{\rm TPC}$ have been corrected by the shifting method~\cite{Poskanzer:1998yz}, such that they have uniform distributions.

In this subsection, the POIs (with azimuthal angle represented by $\phi$ in Eq.~(\ref{eq_v2ep}) or $\phi_{\alpha,\beta}$ in Eq.~(\ref{eq_gmmaep})) are  taken from the TPC acceptance of $|\eta|<1$.
By default, the full EP over the same $\eta$ range is used for the $v_2$ and $\Delta\gamma_{112}$ measurements, with no $\eta$ gap between the EP and the POIs or between the two POIs. For each POI or POI pair, the full EP is re-estimated by excluding the POI or POI pair to remove self-correlation. This approach yields the smallest statistical uncertainties, with the largest possible number of POIs and the highest possible EP resolution.
The systematic uncertainties due to the lack of an $\eta$ gap are expected to be canceled to a large extent in the ratio between the two isobar systems,  and this idea has been corroborated by the $v_2$ ratios in Fig.~\ref{fig_v2_all}, and will be further tested in the following discussions of the results using finite $\eta$ gaps. 

\begin{figure*}
\begin{minipage}[c]{0.48\textwidth}
\centering
\includegraphics[width=\textwidth]{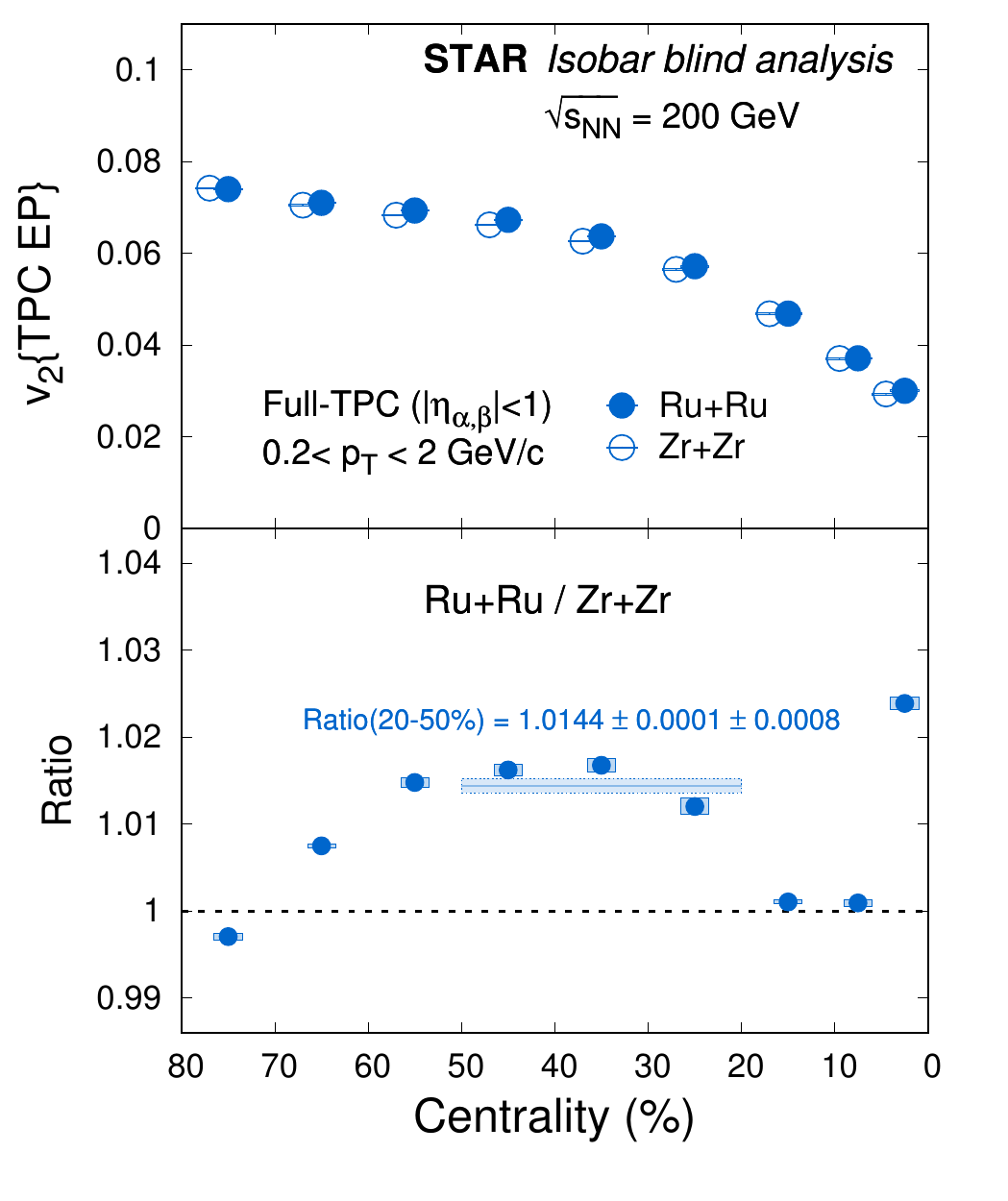}
\caption{$v_2$ measured with the full TPC EP for Ru+Ru and Zr+Zr collisions at $\sqrt{s_{_{\rm NN}}} = 200$ GeV (upper panel) and the ratio of Ru+Ru to Zr+Zr (lower panel). The centrality bins are shifted horizontally for clarity. The border-less horizontal bands denote the statistical uncertainties. The horizontal bands with the dashed border represent the systematic uncertainties.}
\label{fig:Group1_v2}
\end{minipage}
\hspace{0.02\textwidth}
\begin{minipage}[c]{0.48\textwidth}
\centering
\includegraphics[width=\textwidth]{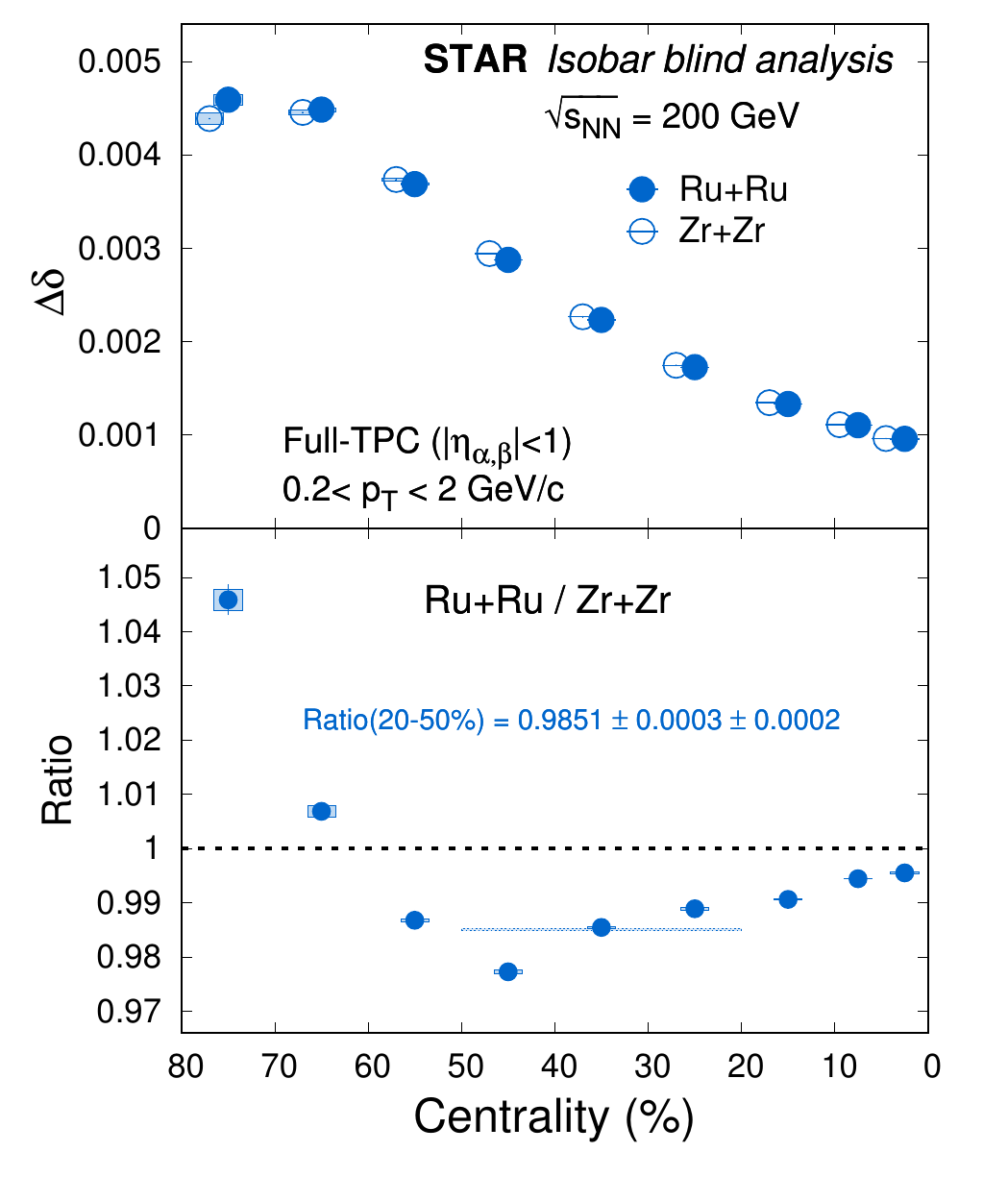}
\caption{ $\Delta\delta$ measured for Ru+Ru and Zr+Zr collisions at $\sqrt{s_{_{\rm NN}}} = 200$ GeV (upper panel) and the ratio of Ru+Ru to Zr+Zr (lower panel). The centrality bins are shifted horizontally for clarity. The border-less horizontal bands denote the statistical uncertainties. The horizontal bands with the dashed border represent the systematic uncertainties.}
\label{fig:Group1_Ddelta}
\end{minipage}
\end{figure*}

\begin{figure*}
\begin{minipage}[c]{0.48\textwidth}
\centering
\includegraphics[width=\textwidth]{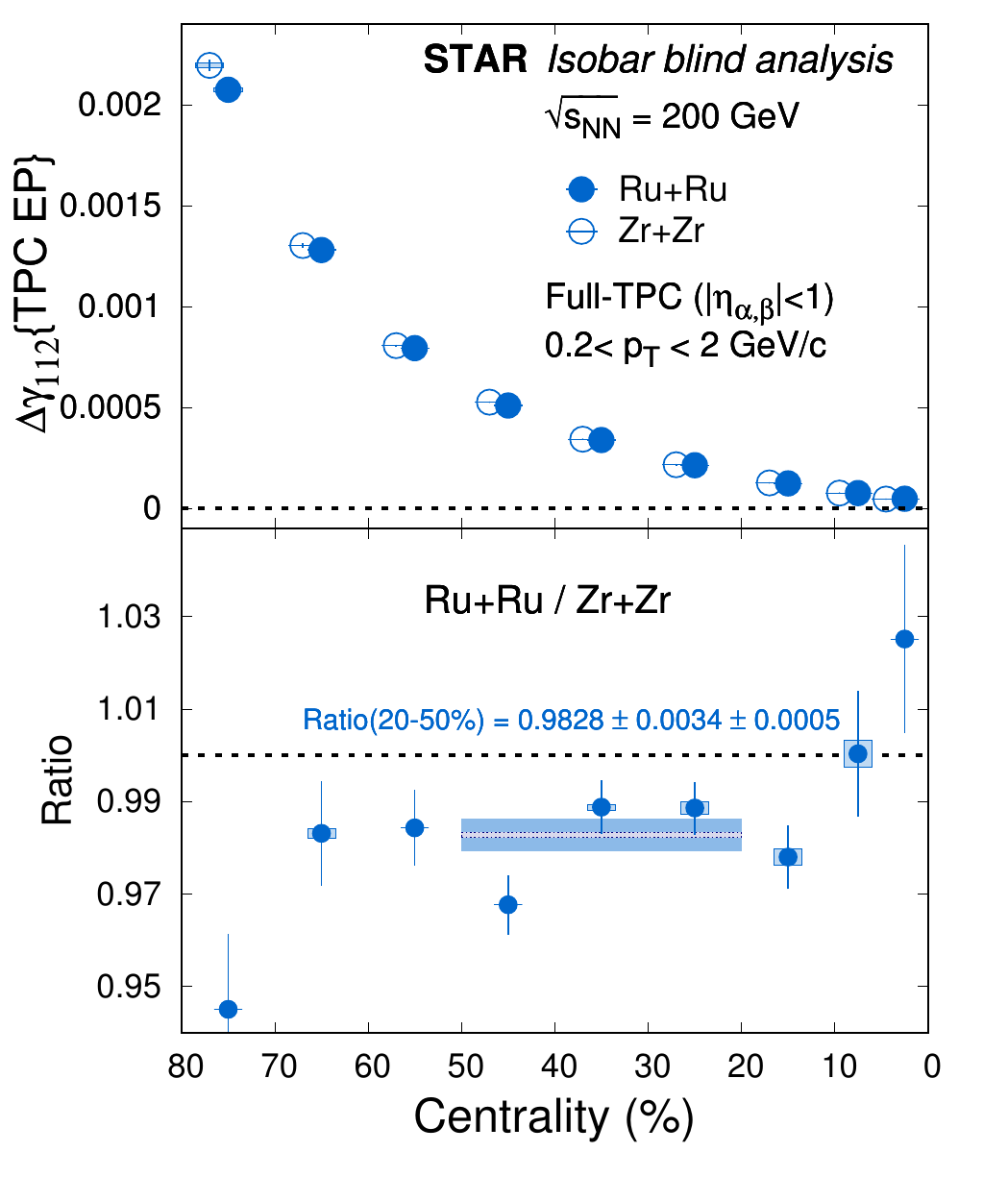}
\caption{ $\Delta\gamma_{112}$ measured with the full TPC EP for Ru+Ru and Zr+Zr collisions at $\sqrt{s_{_{\rm NN}}} = 200$ GeV (upper panel) and the ratio of Ru+Ru to Zr+Zr (lower panel). The centrality bins are shifted horizontally for clarity. The border-less horizontal bands denote the statistical uncertainties. The horizontal bands with the dashed border represent the systematic uncertainties.  
    }
    \label{fig:Group1_Dgamma}
\end{minipage}
\hspace{0.02\textwidth}
\begin{minipage}[c]{0.48\textwidth}
    \centering
\includegraphics[width=\textwidth]{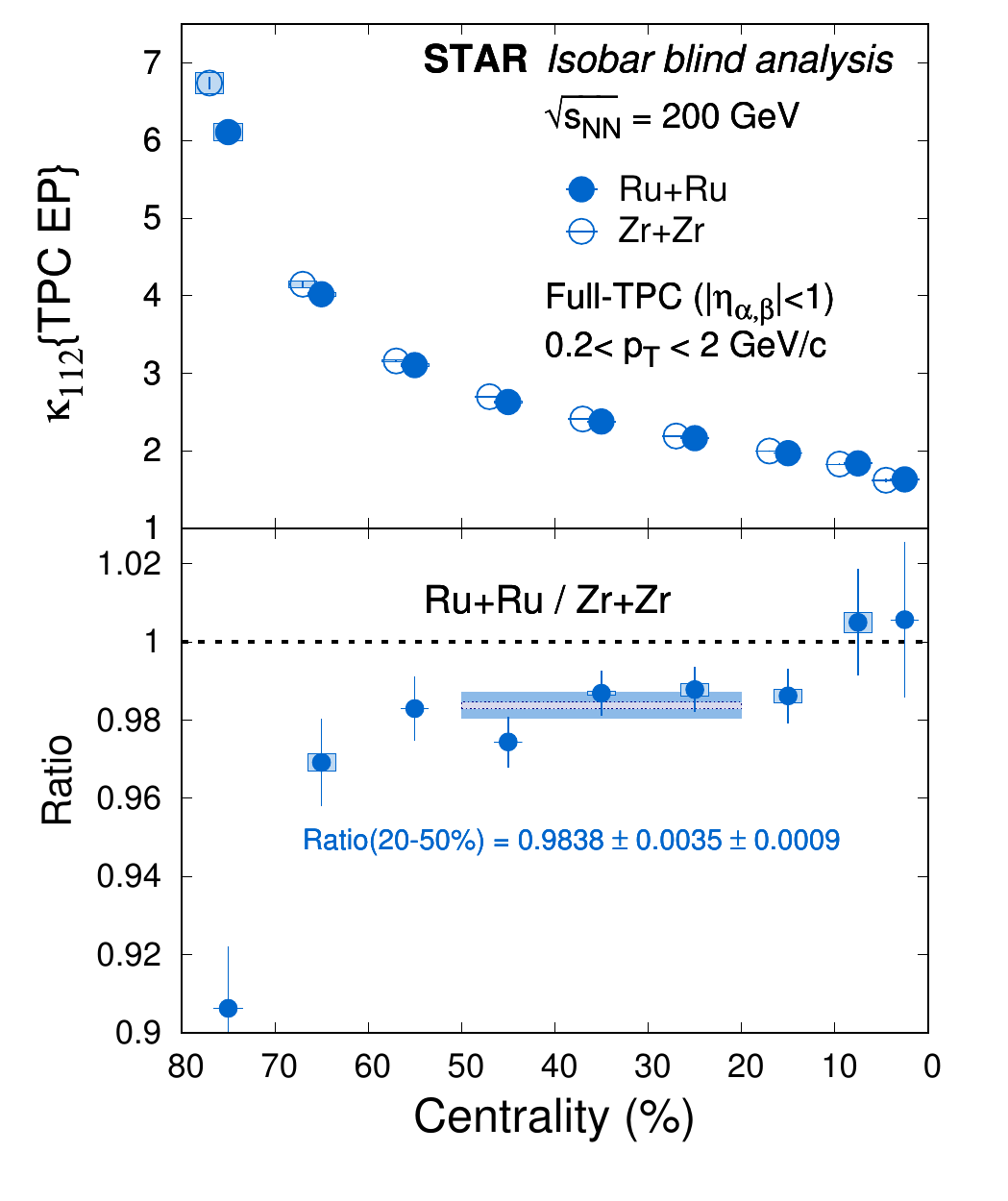}
\caption{ $\kappa_{112}$ measured with the full TPC EP for Ru+Ru and Zr+Zr collisions at $\sqrt{s_{_{\rm NN}}} = 200$ GeV (upper panel) and the ratio of Ru+Ru to Zr+Zr (lower panel). The centrality bins are shifted horizontally for clarity. The border-less horizontal bands denote the statistical uncertainties. The horizontal bands with the dashed border represent the systematic uncertainties.}
\label{fig:Group1_kappa}
\end{minipage}
\end{figure*}

Figure~\ref{fig:Group1_v2}
shows $v_2\{\rm TPC~ EP\}$ as a function of centrality for Ru+Ru and Zr+Zr collisions at $\sqrt{s_{_{\rm NN}}} = 200$ GeV in the upper panel, and the ratio of Ru+Ru to Zr+Zr in the lower panel. The $v_2$ ratio averaged over the 20--50\% centrality range is $1.0144\pm0.0001 (\rm stat.)\pm0.0008 (\rm syst.)$. Given the statistical and systematic uncertainties, this value is significantly above unity, and we consider two potential origins: (a) the two nuclei could have different nuclear density parameters, and (b) non-flow contributions could be different in the two systems. Scenario (b) can be examined using the measurements with various $\eta$ gaps: the mean value of the $v_2$ ratio becomes 1.0146, 1.0149 and 1.0161 for the two-particle cumulant method ($v_2\{2\}$ defined in Eq.~(\ref{eq_v2pc})) with no $\eta$ gap, $\Delta\eta_{\alpha\beta}>0.05$ and $\Delta\eta_{\alpha\beta}>0.2$, respectively. Here $\Delta\eta_{\alpha\beta}$ is the $\eta$ gap between particles $\alpha$ and $\beta$. 
Since the $v_2$ ratio is consistently above unity, we exclude the non-flow explanation. Therefore, the isobar data indicate that the $^{96}_{44}$Ru and $^{96}_{40}$Zr nuclei have different nuclear density distributions, yielding a larger eccentricity in Ru+Ru than in Zr+Zr collisions at a given centrality~\cite{Xu:2017zcn}. This results in the $v_2$ ratio in the lower panel of Fig.~\ref{fig:Group1_v2} being larger than unity.

Figure~\ref{fig:Group1_Ddelta}
shows $\Delta\delta$ vs centrality for Ru+Ru and Zr+Zr collisions at $\sqrt{s_{_{\rm NN}}} = 200$ GeV in the upper panel, and the ratio of Ru+Ru to Zr+Zr in the lower panel. There is no $\eta$ gap between the two POIs. The $\Delta\delta$ ratio averaged over the 20--50\% centrality range is $0.9851\pm0.0003 (\rm stat.)\pm0.0002 (\rm syst.)$, below unity with high measured significance.
The central value of the $\Delta\delta$ ratio changes to 0.9846 and 0.9833 with $\Delta\eta_{\alpha\beta}>0.05$ and $\Delta\eta_{\alpha\beta}>0.2$, respectively. Thus the short-range correlations have a very small impact on the $\Delta\delta$ ratio.

Figure~\ref{fig:Group1_Dgamma}
shows $\Delta\gamma_{112}$ as a function of centrality measured with the full TPC EP for Ru+Ru and Zr+Zr collisions at $\sqrt{s_{_{\rm NN}}} = 200$ GeV in the upper panel, and the ratio of Ru+Ru to Zr+Zr in the lower panel. By default, no $\eta$ gap is applied between the two POIs or between the EP and the POIs. The $\Delta\gamma_{112}$ ratio averaged over the 20--50\% centrality range is $0.9828\pm0.0034(\rm stat.)\pm0.0005(\rm syst.)$.
When a finite $\eta$ gap is applied between the two POIs, the central value of the $\Delta\gamma_{112}$ ratio becomes
0.9822 and 0.9825 with $\Delta\eta_{\alpha\beta}>0.05$ and $\Delta\eta_{\alpha\beta}>0.2$, respectively. Therefore, the $\Delta\gamma_{112}$ ratio is insensitive to the short-range correlations.

Figure~\ref{fig:Group1_kappa}
shows $\kappa_{112}$ vs centrality measured with the full TPC EP for Ru+Ru and Zr+Zr collisions at $\sqrt{s_{_{\rm NN}}} = 200$ GeV in the upper panel, and the ratio of Ru+Ru to Zr+Zr in the lower panel.
The default $\kappa_{112}$ ratio averaged over the 20--50\% centrality range is $0.9838\pm0.0035(\rm stat.)\pm0.0009(\rm syst.)$, which changes to 0.9827 and 0.9831 with $\Delta\eta_{\alpha\beta}>0.05$ and $\Delta\eta_{\alpha\beta}>0.2$, respectively. We conclude that the CME signature predefined in Eq.~(\ref{eq:kappa_ratio}) is not observed in this blind analysis of the isobar data. It is noteworthy that we have reached a precision better than 0.4\% on these measurements of the ratio between Ru+Ru and Zr+Zr collisions.

After unblinding of the isobar species, we observe  the multiplicity difference between the two isobar systems at a given centrality, as shown in Table~\ref{tab:centrality}. Although the effects of the multiplicity mismatch are largely canceled in the ratio of $\Delta\gamma_{112}$ over $v_2\Delta\delta$, there could still be residual contributions driving the $\kappa_{112}$
ratio below unity, which needs further investigation.
Additional discussions on the multiplicity mismatch can be found in Sec.~\ref{sec:post}.

\subsection{Mixed harmonic measurements (Group-2)}

While the analysis from the previous group focuses on the EP method, in this subsection: 1) we focus on measurements of harmonic coefficients and charge sensitive correlations using two-particle, three-particle correlations and the scalar-product method, and 2) we further extend the correlation measurements by requiring one of the particles from the forward EPD.

We measure harmonic flow coefficients $v_{n}\{2\}$ from the full TPC using two-particle correlations, where 
\begin{equation}
    v^2_{n=2,3}\{2\} (|\eta|<1)=\left<\cos(n\phi_1-n\phi_2)\right>\,.
    \label{eq_v2pc}
\end{equation}
In this $v^2_{n}\{2\}$ measurement from the TPC, we put a cut of $\Delta\eta_{1,2}>0.05$ to mitigate effects of two track merging and $e^+e^-$ due to photon conversion. For $v^2_{n}\{2\}$ measurements, we remove the short-range component due to HBT, Coulomb effects using a double Gaussian fit as described in Ref.~\cite{STAR:2016vqt}. We also estimate harmonic coefficients without such Gaussian subtraction but using a cut of $\Delta \eta_{1,2}>1$ in Eq.~(\ref{eq_v2pc}). In this paper we denote such measurements as $v_{n}\{2\}(\Delta\eta>1)$. In addition we also estimate $v_n$ using sub-event methods $v_n^2\{\rm  SP\}=$ $\langle Q_{n,a}Q^{*}_{n,b}\rangle$, where the $Q$-vectors $Q_{n,a}$ and $Q_{n,b}$ are taken from two  halves of TPC around $\eta=0$ separated by a pseudorapidity gap of  $\Delta\eta_{\rm sub}=0.2$. We denote such measurements as $v_{n}\{\rm SP\}(\Delta\eta_{\rm sub}=0.2)$. 

We present measurements of data from the new EPD detector ($2.1<|\eta| < 5.1$). 
We estimate the elliptic and triangular anisotropy of particles at mid-rapidity with respect to the forward PPs in the EPD by
\begin{equation}
v_{n=2,3}\{ {\rm SP} \}({\rm TPC \!-\! EPD})
\equiv\left<\cos\left(n\phi - n\Psi_n^{\rm EPD}\right)\right>
=\frac{\langle Q_{n,{\rm TPC}}Q^{^*}_{n,{\rm EPDE}} + Q_{n,{\rm TPC}}Q^{^*}_{n,{\rm EPDW}}  \rangle}{2 \sqrt{ \langle  Q_{n,{\rm EPDE}}Q^{^*}_{n,{\rm EPDW}} } \rangle} \,,
    \label{eq_vnepd}
\end{equation}
using the scalar-product (SP) method where $Q$ and $Q^{*}$ denote the $Q$-vectors and their complex conjugates~\cite{STAR:2002hbo}. 
\begin{figure*}[t]
    \centering
    \includegraphics[width=0.45\textwidth]{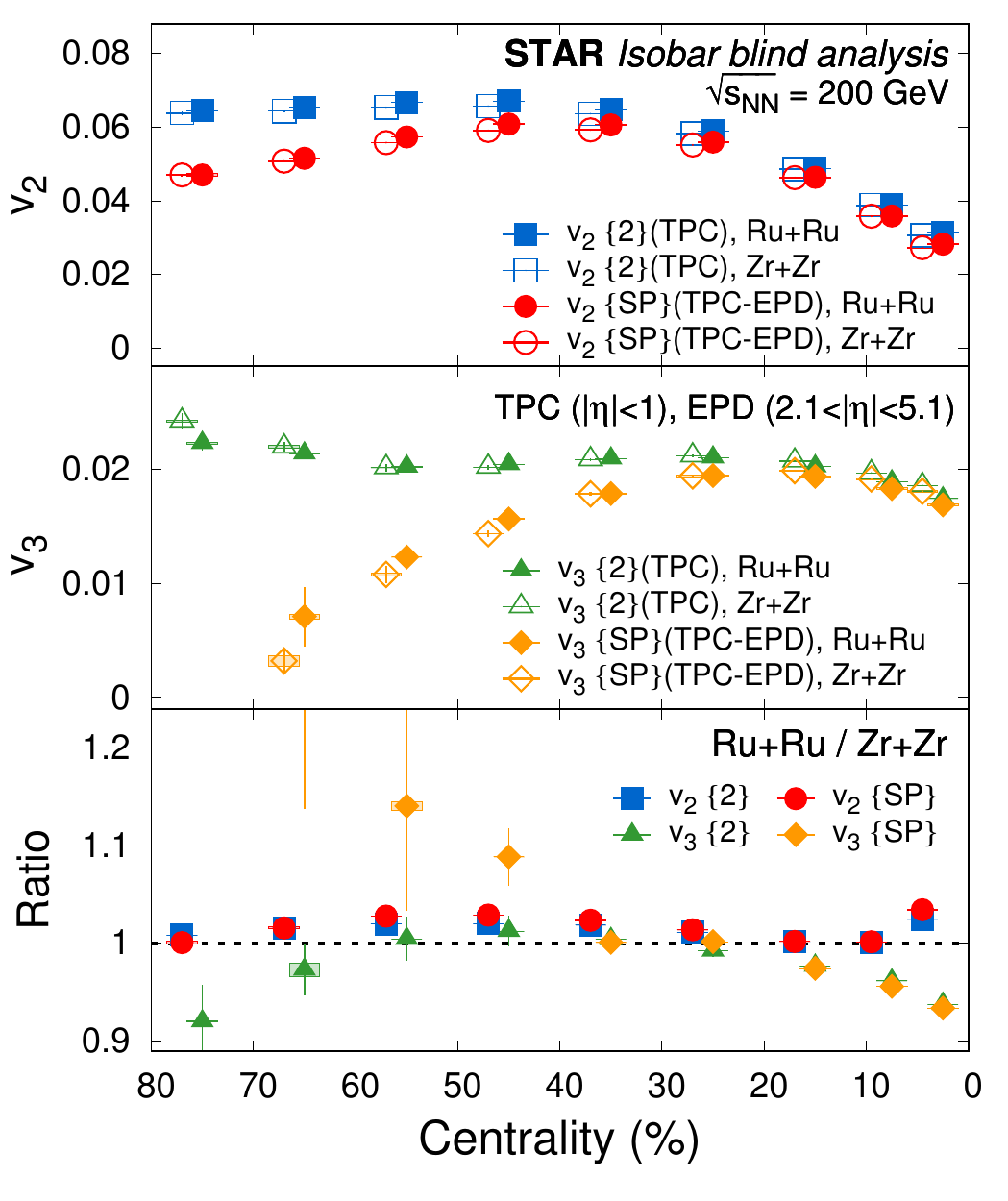}
    \includegraphics[width=0.45\textwidth]{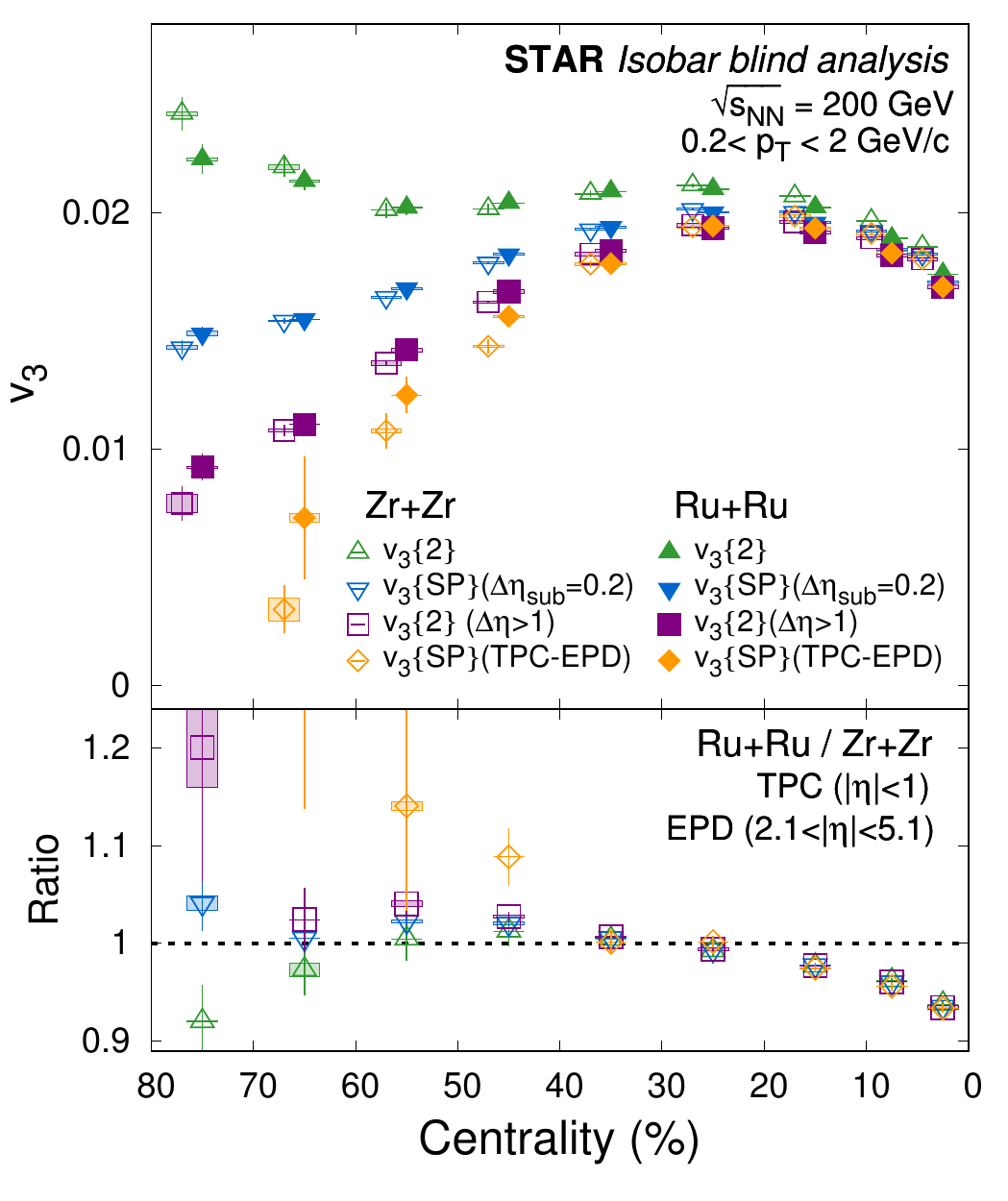}
    \caption{(Left) Elliptic and triangular anisotropies measured for the two systems using the combination of TPC and EPDs. The boxes represent systematic uncertainties and the lines represent statistical errors. (Right) Compilation of $v_3$ using different methods and cuts on pseudorapidity separation. Results are shown for individual systems in open symbols for Zr+Zr and solid symbols for Ru+Ru. Results are also shown for the ratio of Ru+Ru to Zr+Zr in open symbols. The centrality bins are shifted horizontally for clarity.}
    \label{Group2_1}
\end{figure*}

The upper and middle panels of Fig.~\ref{Group2_1} show the centrality dependence of $v_2$ and $v_3$, respectively, with the  two aforementioned approaches. The measurements of these flow harmonics using only TPC and TPC-EPD are noticeably different, especially  in peripheral events for $v_2$, and   in mid-central and peripheral collisions for $v_3$. A possible explanation for such an observation could be the pseudorapidity dependence of non-flow, de-correlation and flow fluctuations~\cite{Bozek:2010vz,Voloshin:2007pc}. Owing to low multiplicity and poor resolution of the third-order EP, EPDs do not allow for the $v_3$ measurements  beyond 60-70\% centrality. A compilation of $v_3$ results is shown in Fig.~\ref{Group2_1} (right) to demonstrate the effect of pseudorapdity separation between POI and EP (or between two POIs). 

The lower panel presents ratios of the flow harmonics for Ru+Ru over Zr+Zr collisions, with a few interesting features. First, the $v_2$ ratio in the most central events ($0-5\%$) is larger than unity with high significance. As mentioned before,  effects due to nuclear deformation can lead to the difference in the shape even in fully-overlapping collisions, which needs to be confirmed by future studies. Above-unity $v_2$ ratios are also observed in mid-central collisions. This is consistent with the expectation of the eccentricity ratios from nuclear density distributions calculated by DFT~\cite{Xu:2017zcn,Li:2018oec}. Second,  the $v_3$ ratio is significantly below unity in central events, which is counter intuitive, as $v_3$ is supposed to be driven by fluctuations in central collisions. Third,  the $v_3$ ratio significantly deviates from unity in peripheral events, and this deviation has a dependence on pseudorapidity separation between POI and EP. Thus,  we need a better understanding of the possible differences in the nuclear structure and the deformity of the isobars, when comparing the two systems at the same centrality. Further exploration along this direction is beyond the scope of this paper which is primarily focused on the CME blind analysis. These $v_n$ measurements do have implications on the background contribution to CME that is relevant in the scaled charge separation variables. 
\begin{figure*}[hbt]
\begin{minipage}[t]{0.48\textwidth}
\centering
\includegraphics[width=\textwidth]{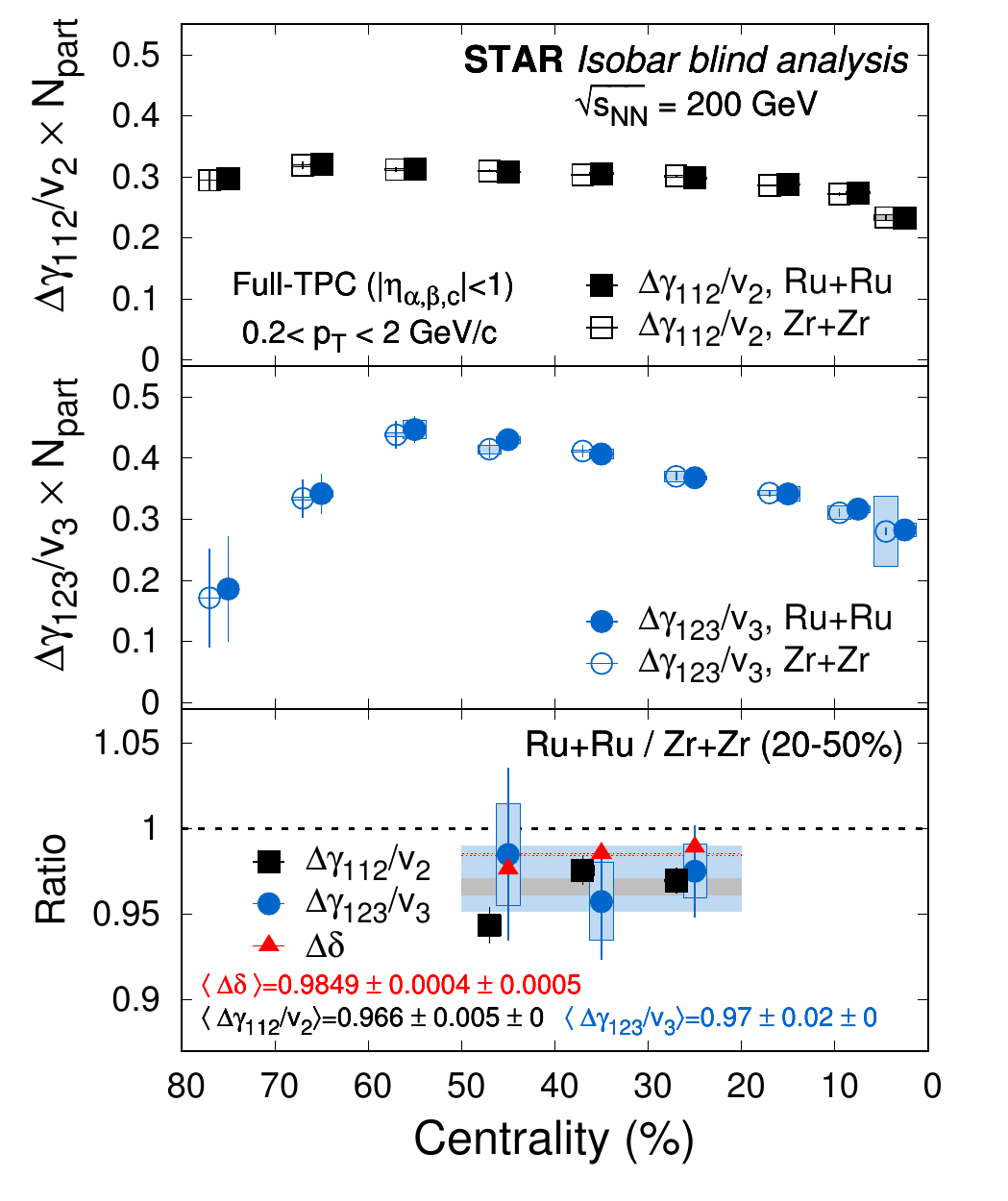}
\caption{Scaled charge separation across the second and third harmonic EPs obtained using all three particles from the TPC acceptance, divided by the anisotropy coefficient. Results are shown for Ru+Ru and Zr+Zr collisions separately on the upper and middle panels over the centrality range of 0-80\%. The centrality bins are shifted horizontally for clarity. The lower panel shows the ratio of various quantities for 20--50\% centrality. The border-less horizontal bands over the 20--50\% centrality range with different colors represent the statistical uncertainties in the combined centrality for different observables. The horizontal bands with the dashed border represent the systematic uncertainties. The $N_{\rm part}$ scaling is applied in the upper two panels to improve the visibility. The $N_{\rm part}$ scaling is not included in the lower panel for the ratios.}
    \label{fig:Group2_2}
\end{minipage}
\hspace{0.02\textwidth}
\begin{minipage}[t]{0.48\textwidth}
   \centering
    \includegraphics[width=\textwidth]{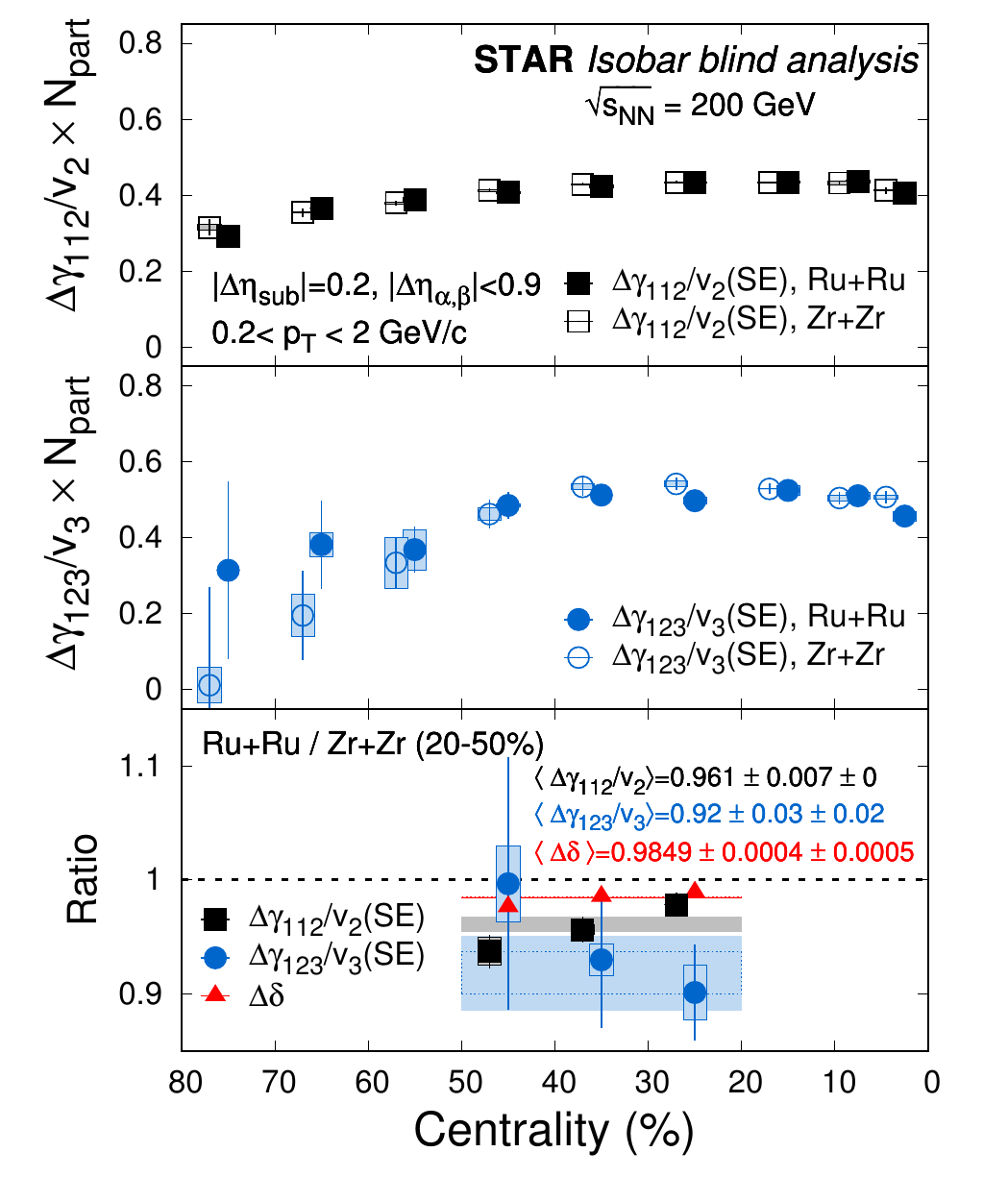}
    \caption{Scaled charge separation across second and third harmonic EPs scaled by the anisotropy coefficient obtained using all three particles from the TPC acceptance but using a sub-event (SE) from $-1<\eta<-0.1$ and $0.1<\eta<1$. Results are shown for Ru+Ru and Zr+Zr collisions separately in the upper and middle panels over the centrality range of 0-80\%. The centrality bins are shifted horizontally for clarity. The lower panel shows the ratio of different quantities for 20--50\% centrality. The border-less horizontal bands over 20--50\% centrality range with different colors represent the statistical uncertainties on the combined centrality for different observables. The horizontal bands with the dashed border represent the systematic uncertainties.  The $N_{\rm part}$ scaling is applied in the upper and middle panels to improve the visibility. The $N_{\rm part}$ scaling is not included in the lower panel for the estimation of ratios.}
    \label{fig:Group2_3}
\end{minipage}
\end{figure*}
 
We perform the measurement of charge separation using the full TPC acceptance ($|\eta|<1$) in the following way
\begin{equation}
\gamma^{\alpha, \beta}_{112} (\eta_{\alpha} , \eta_{\beta}) (|\eta|\!<\!1) 
\equiv \left<  \cos \left( \phi_{\alpha}(\eta_{\alpha}) + \phi_{\beta}(\eta_{\beta}) - 2\Psi_{2}^{|\eta|<1} \right)  \right>  
=  \frac{\left< \cos \left(\phi_{\alpha}(\eta_{\alpha}) + \phi_{\beta}(\eta_{\beta}) - 2\phi_{c}\right)\right> }{v_{2,c}\{2\}} \,.
 \label{eq_112_fulltpc}
\end{equation}
The indices ``$\alpha, \beta, c$" denote three distinctly different particles. The subscripts ``$\alpha, \beta$" denote particle pairs with same (SS) or opposite (OS) sign of electric charges. 
We use the charge-inclusive reference particle `$c$' as a proxy for the elliptic flow plane $\Psi_{2}$ at midrapidity, and the quantity $v_{2,c}\{2\}$ refers to the two-particle elliptic flow coefficient of the reference particle `$c$' that we estimate using two-particle correlations as defined in Eq.~(\ref{eq_v2pc}). 

Similarly with respect to the third harmonic plane, we measure
\begin{equation}
\gamma^{\alpha, \beta}_{123} (\eta_{\alpha} , \eta_{\beta}) (|\eta|\!<\!1) 
\equiv \left<  \cos\left( \phi_{\alpha}(\eta_{\alpha}) + 2\phi_{\beta}(\eta_{\beta}) - 3\Psi_{3}^{|\eta|<1}\right)  \right>  
=  \frac{\left< \cos\left(\phi_{\alpha}(\eta_{\alpha}) + 2\phi_{\beta}(\eta_{\beta}) - 3\phi_{c}\right)\right> }{v_{3,c}\{2\}} \,.
 \label{eq_123_fulltpc}
\end{equation}

Finally we calculate the quantities of interest:  
\begin{equation}
\Delta\gamma_{1mn} = \gamma_{1mn}^{\rm OS} - \gamma_{1mn}^{\rm SS},\, \text{and}, \, \Delta\gamma_{1mn}/v_n \times N_{\rm part}\,.
\end{equation}
The normalization of $\Delta\gamma_{1mn} (m,n=1,2 \text{ or } 2,3)$ by $v_n (n=2,3)$ takes into account the flow-driven background due to resonance decays and local charge conservation~\cite{Voloshin:2004vk,Schlichting:2010qia}. The $N_{\rm part}$ scaling  compensates for the trivial dilution of correlations expected from superposition of independent sources,  and improves the visibility of the data points. 

The upper and middle panels of Fig.~\ref{fig:Group2_2} show the CME-sensitive $\Delta\gamma_{112}/v_2$ and the CME-insensitive $\Delta\gamma_{123}/v_3$ (both multiplied by $N_{\rm part}$), respectively, for individual species. The lower panel presents the ratios of the quantities for the 20--50\% centrality bin in Ru+Ru over Zr+Zr collisions. Note that the ratios do not involve $N_{\rm part}$, whose values are different for the two isobar systems at the same centrality (see Sec.~\ref{sec:centrality}). The ratio of the quantity $\left<\Delta\gamma_{112}/v_2\right>$ is 0.966$\pm$0.005, while the ratio for  $\left<\Delta\gamma_{123}/v_3\right>$ is 0.971 $\pm$ 0.019. The errors quoted here are statistical only. Systematic variation in our measurements for these quantities are not statistically significant, and  are estimated to be zero according to the Barlow approach as described in Section-\ref{sec:syst}. The $\Delta \delta$ ratio in this analysis is $0.9849 \pm 0.0004 {\rm (stat.)} \pm 0.0005 {\rm (syst.)}$, which is consistent with the value ($0.9846 \pm 0.0003\pm 0.0002$) obtained by Group-1 within the statistical and systematic uncertainties.  
In summary, our observation related to charge separation in the 20--50\% centrality is consistent with the following statements: 
\begin{eqnarray}
&&\frac{(\Delta\gamma_{112}/v_2)^{\rm Ru+Ru}}{(\Delta\gamma_{112}/v_2)^{\rm Zr+Zr}}<1\,, \\
&&\frac{(\Delta\gamma_{112}/v_2)^{\rm Ru+Ru}}{(\Delta\gamma_{112}/v_2)^{\rm Zr+Zr}}\approx \frac{(\Delta\gamma_{123}/v_3)^{\rm Ru+Ru}}{(\Delta\gamma_{123}/v_3)^{\rm Zr+Zr}}\,, \\
&&\frac{(\Delta\gamma_{112}/v_2)^{\rm Ru+Ru}}{(\Delta\gamma_{112}/v_2)^{\rm Zr+Zr}}<\frac{(\Delta\delta)^{\rm Ru+Ru}}{(\Delta\delta)^{\rm Zr+Zr}}\,.  
\end{eqnarray}
Therefore, our measurements are not consistent with any of the predefined CME signatures  as set out in Eqs.~(\ref{eq:dg_ratio}), (\ref{eq:dgdg_ratio}), and (\ref{eq:kappa_ratio1}).

\begin{figure*}[hbt]
\begin{minipage}[t]{0.48\textwidth}
\centering
\includegraphics[width=\textwidth]{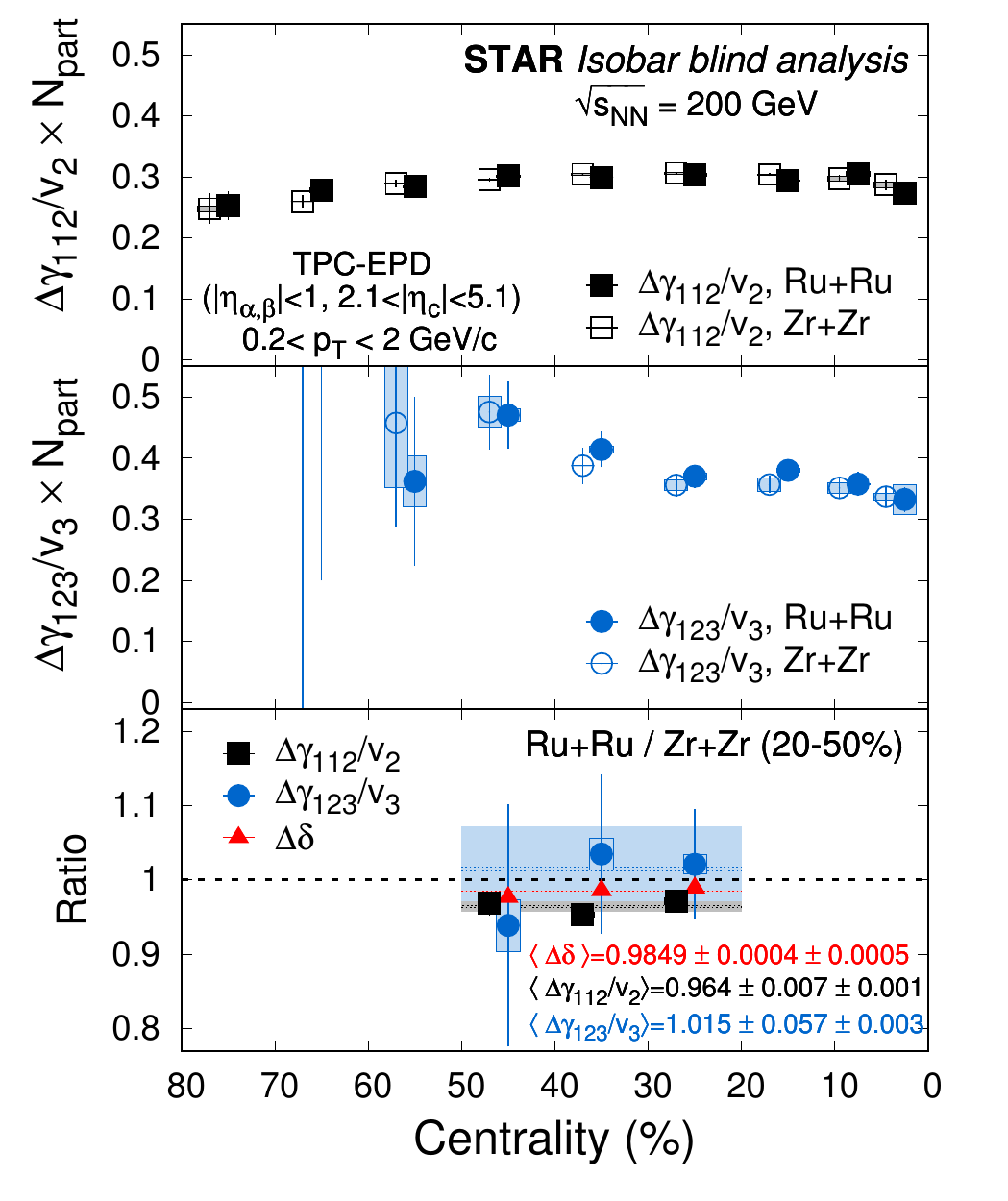}
\caption{Charge separation across the second- and third-order EPs scaled by the anisotropy coefficient obtained using particles from the TPC acceptance and hits from the EPDs. Results are shown for Ru+Ru and Zr+Zr collisions separately on the upper and middle panels over the centrality range of 0-80\%. The centrality bins are shifted horizontally for clarity. The lower panel shows the ratio of different quantities for 20--50\% centrality. The border-less horizontal bands over 20--50\% centrality range with different colors represent the statistical uncertainties on the combined centrality for different observables. The horizontal bands with the dashed border represent the systematic uncertainties. The $N_{\rm part}$ scaling is applied in the upper and middle panels to improve the visibility. The $N_{\rm part}$ scaling is not included in lower panel for the estimation of ratios.}
    \label{fig:Group2_4}
\end{minipage}
\hspace{0.02\textwidth}
\begin{minipage}[t]{0.48\textwidth}
\centering
\includegraphics[width=\textwidth]{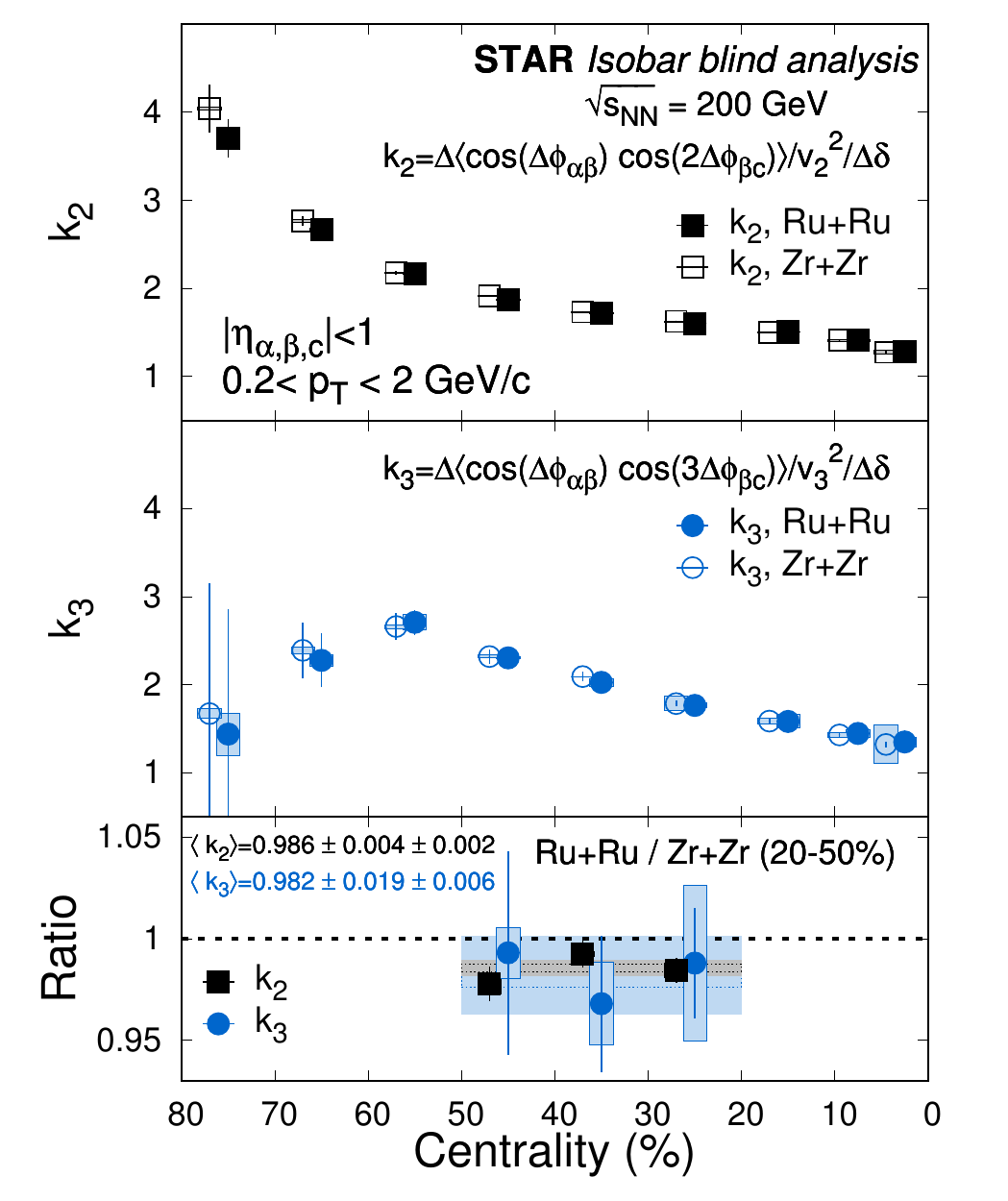}
\caption{Factorization breaking coefficient for the second and third order harmonics measured using particles from the TPC acceptance. Results are shown for Ru+Ru and Zr+Zr collisions separately on the upper and middle panels over the centrality range of 0-80\%. The centrality bins are shifted horizontally for clarity. The lower panel shows the ratio of various quantities for 20--50\% centrality. The border-less horizontal bands over 20--50\% centrality range with different colors represent the statistical uncertainties on the combined centrality for different observables. The horizontal bands with the dashed border represent the systematic uncertainties. }
\label{fig:Group2_5}
\end{minipage}
\end{figure*}

We repeat our analysis with  sub-events (SE) within the STAR TPC, as described in Eq.~(\ref{eq_112_fulltpc}), and
divide the TPC acceptance into two halves, with $0.1\!<\!\eta\!<\!1$ and $-1\!<\!\eta\!<\!-0.1$.
When the charge-carrying particles ``$\alpha$" and ``$\beta$"  are taken from one half,  particle ``$c$" is taken from the other. 
Going from full TPC to sub-event reduces the maximum value of the relative pseudorapidity $\Delta\eta_{\alpha,\beta}$ from 2 to 0.9. 
The sub-events are also used for the $v_n$ estimation. In our blind analysis code we did not include the estimation of $\Delta \delta$ in the acceptance of $0.1\!<\! |\eta|\!<\!1$ which would have been most appropriate. The $\Delta \delta$ ratio used  here are using the full TPC acceptance which is an approximation. The results from this approach are presented in Fig.~\ref{fig:Group2_3}, and show  some noticeable differences from those in Fig.~\ref{fig:Group2_2}, illustrating the influence of using sub-events. However, the final ratios in the lower panel are consistent with those obtained with the full TPC, and hence we still do not observe any of the predefined CME signatures (Eqs.~(\ref{eq:dg_ratio}), (\ref{eq:dgdg_ratio}), and (\ref{eq:kappa_ratio1})) in this measurement. 

We extend our measurement of charge separation with respect to the 2nd harmonic plane from the STAR EPD using
\begin{eqnarray}
\gamma^{\alpha, \beta}_{112}
(\eta_{\alpha}, \eta_{\beta})(\mbox{TPC-EPD}) &\equiv&\langle \mathrm{ \cos( \phi_{\alpha}(\eta_{\alpha}) + \phi_{\beta}(\eta_{\beta}) - 2\Psi_{2}^{\rm EPD})} \rangle\\ 
&=&\frac{ \langle {Q_{\rm 1,TPC}^{\alpha}Q_{\rm 1,TPC}^{\beta}Q^{^*}_{\rm 2,EPDE} + Q_{\rm 1,TPC}^{\alpha}Q_{\rm 1,TPC}^{\beta}Q^{^*}_{\rm 2,EPDW} } \rangle }{2 \sqrt{ \langle { Q_{\rm 2,EPDE}Q^{^*}_{\rm 2,EPDW}} } \rangle }\,. \nonumber  
\end{eqnarray}
Here, we use algebra based on $Q$-vectors~\cite{Bilandzic:2010jr}, defined as $Q_{n}$ = $\sum^{M}_{1} w_{i}e^{in\phi}/\sum^{M}_{1} w_{i}$, where $M$ is the number of particles, and the weight factor $w_{i}$ accounts for the imperfection in the detector acceptance in bins of $\eta$--$\phi$, $p_{T}$ (track-curvature), $V_{z,\textsc{tpc}}$, and centrality. After weight correction, the $Q$-vectors are also ``recentered" as $Q-\left<Q\right>$ for residual acceptance correction. 
Similarly we measure charge separation with respect to a 3rd harmonic plane from the EPD using
\begin{eqnarray}
\gamma^{\alpha, \beta}_{123}
(\eta_{\alpha}, \eta_{\beta})(\mbox{TPC-EPD}) &\equiv&\langle { \cos( \phi_{\alpha}(\eta_{\alpha}) + 2 \phi_{\beta}(\eta_{\beta}) - 3\Psi_{3}^{\rm EPD})} \rangle \\ 
&=&\frac{ \langle {Q_{\rm 1,TPC}^{\alpha}Q_{\rm 2,TPC}^{\beta}Q^{^*}_{\rm 3,EPDE} + Q_{\rm 1,TPC}^{\alpha}Q_{\rm 2,TPC}^{\beta}Q^{^*}_{\rm 3,EPDW} } \rangle }{2 \sqrt{ \langle { Q_{\rm 3,EPDE}Q^{^*}_{\rm 3,EPDW}} } \rangle }\,. \nonumber
\end{eqnarray}

Measurements of charge sensitive variables using the combination of the TPC and the EPD are shown in Fig.~\ref{fig:Group2_4}. Once again these results are consistent with earlier measurements using the full TPC acceptance. Based on the measurements of ratios of variables in Ru+Ru over Zr+Zr collisions, we do not see any of the predefined CME signatures (Eqs.~(\ref{eq:dg_ratio}), (\ref{eq:dgdg_ratio}), and (\ref{eq:kappa_ratio1})). 

The quantities $k_2$ and $k_3$ that measure factorization breaking defined in Eq.~(\ref{eq_kn}) are shown in Fig.~\ref{fig:Group2_5}. Since this measurement involves several higher-order charge-sensitive correlators,  we restrict our measurements to the full TPC only to achieve the best precision. Averaged over  the 20--50\% centrality, the $k_2$ and $k_3$ ratios are  $0.986$ $\pm0.004$ (stat.) $\pm0.002$ (syst.) and  $0.982$ $\pm0.019$ (stat.) $\pm0.006$ (syst.), respectively,  consistent with each other within one standard deviation. In other words we find: 
\begin{eqnarray}
 \frac{k_2^{\rm Ru+Ru}}{k_2^{\rm Zr+Zr}} \approx \frac{k_3^{\rm Ru+Ru}}{k_3^{\rm Zr+Zr}}\,. 
\end{eqnarray}
The predefined CME signature described in Eq.~(\ref{eq:k2k3signature}) is not observed. 

\subsection{Differential $\Delta\gamma$ measurements in pseudorapidity (Group-2)}
\begin{figure*}[htb]
    \centering
    \includegraphics[width=0.48\textwidth]{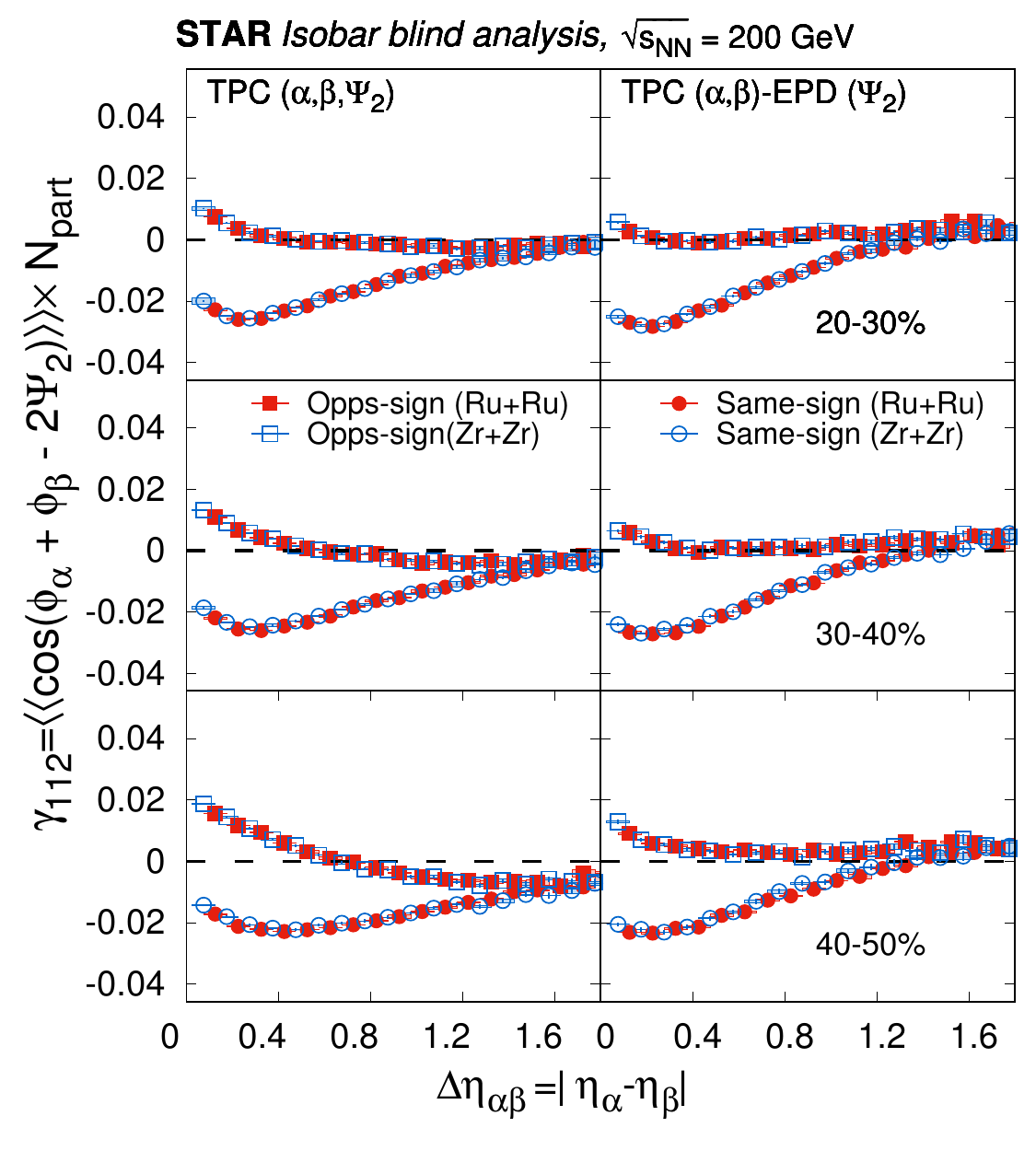}
    \includegraphics[width=0.48\textwidth]{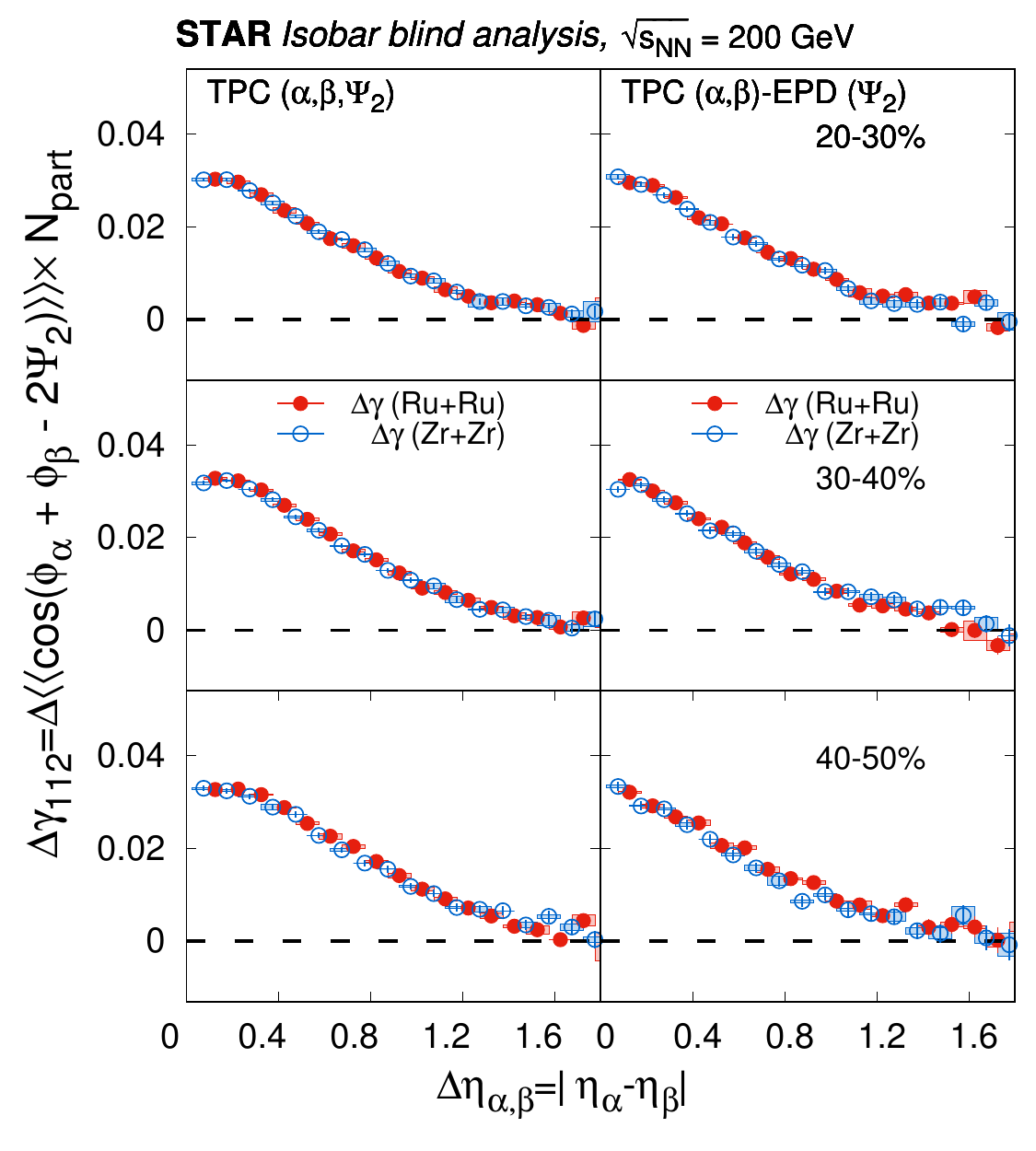}
    \caption{Relative pseudorapidity dependence of the $\gamma$ (left) and $\Delta\gamma$ (right) correlators shown for two species for same-sign (SS) and opposite-sign (OS) correlations. For clarity, the alternate bins in $\Delta\eta$ are shown. The statistical and systematic errors are shown by vertical lines and square boxes, respectively. Within the uncertainty we do not see any species dependence in these measurements.}  
    \label{fig:dedeta_os_ss}
\end{figure*}

Relative pseudorapidity dependence between the charge-carrying particles ($\Delta\eta_{\alpha,\beta}$) of same-sign and opposite-sign $\gamma_{112}$ correlators is shown in Fig.~\ref{fig:dedeta_os_ss} (left) for 20--50\%  Ru+Ru and Zr+Zr collisions. We show two panels in which the third particle or the EP is either obtained from TPC or the EPD. The $\Delta\eta_{\alpha,\beta}$ dependence of $\gamma$ correlator in the individual isobar species have very similar shapes compared to what is reported in the previous STAR measurement in Au+Au collisions~\cite{STAR:2009tro}. Some difference in the shape is observed between measurements using TPC and EPD EPs. The same is also seen for the $\Delta\eta_{\alpha,\beta}$ dependence of $\Delta\gamma_{112}$ shown in Fig.\ref{fig:dedeta_os_ss} (right). Although interesting dependence is observed for the individual distributions we do not see any species dependence within the uncertainties of the current measurements. The expectation for CME was that the long-range part of the $\Delta\eta$ distribution $\Delta\eta_{\alpha,\beta}>1$ will be higher for Ru+Ru collisions. No such observation can be made from the results shown in  Fig.~\ref{fig:dedeta_os_ss}.

\subsection{Differential $\Delta\gamma$ measurements in  invariant mass (Group-3)\label{sec:minvresult}}

In order to isolate the resonance background contributions, we report  measurements of the $\Delta\gamma$ variable, differential in pair invariant mass $m_{\rm inv}$.
This analysis uses the three-particle correlation method to calculate the $\gamma$ correlators~\cite{Abelev:2009ad,Abelev:2009ac},
\begin{equation}
    \gamma=\left<\cos(\phi_\alpha+\phi_\beta-2\phi_c)\right>/v_{2,c}\,,
    \label{eq:abc}
\end{equation}
where the average $\left<\cdots\right>$ runs over all triplets and over all events.
To select good events we require, in addition to those criteria described in Sec.~\ref{sec:detector}, the VPD primary vertex position to be within $| V_{z,\textsc{tpc}} - V_{z,\textsc{vpd}} | < 3$~cm from the one reconstructed by the TPC. The POIs ($\alpha$ and $\beta$) are pions within $0.2<p_T<1.8$~GeV/$c$. They are identified by their specific energy loss in the TPC and their flight time obtained from the TOF detector.  
The $c$ particles are charged hadrons within $0.2<p_T<2.0$~GeV/$c$.
The POIs and particle $c$  are all taken from $|\eta|<1$ (self-correlations are avoided)~\cite{Abelev:2009ad,Abelev:2009ac}.
An $\eta$ gap of 0.05 is applied between the POIs.
No $\eta$ gap is applied between particle $c$ and either of the POIs. 
The $v_{2,c}$ of particle $c$ is calculated from two-particle correlations by the $v_{2}$\{2\} of Eq.~(\ref{eq_v2pc}) where an $\eta$ gap of 1.0 is applied between the two particles.

The systematic uncertainties are assessed according to Sec.~\ref{sec:syst}. In addition, the $\eta$ gap between the POIs (i.e.~between $\alpha$ and $\beta$) is varied from 0.05 (default) to 0 (i.e.~no gap) and 0.2. 
The $\eta$ gap used to determine the $v_{2,c}$ is varied from 1 to 0.5 and 1.4. All systematic uncertainties are added in quadrature.

\begin{figure*}[!hbt]
\begin{minipage}[t]{0.48\textwidth}
    \centering
    \includegraphics[width=\textwidth]{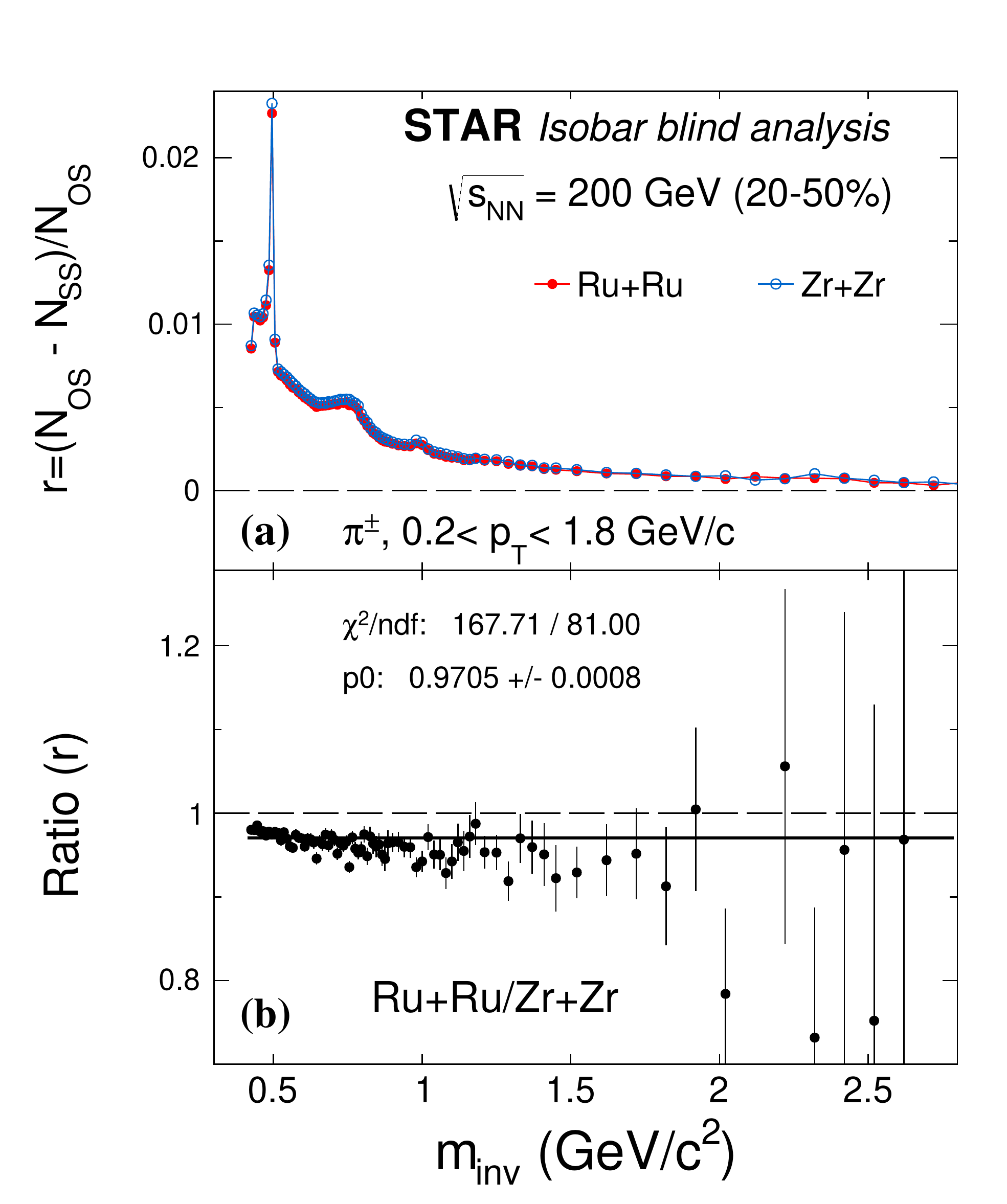}
    \caption{(a) Distributions in the relative pair multiplicity difference, $r=(N_{\textsc{os}}-N_{\textsc{ss}})/N_{\textsc{os}}$, as a function of invariant mass of $\pi^+\pi^-$ pairs in 20--50\% Ru+Ru and Zr+Zr collisions and (b) their ratio. Errors shown are statistical. The solid line in the lower panel is a constant fit to the ratio.}
    \label{fig:r_minv}
\end{minipage}
\hspace{0.02\textwidth}
\begin{minipage}[t]{0.48\textwidth}
   \centering
    \includegraphics[width=\textwidth]{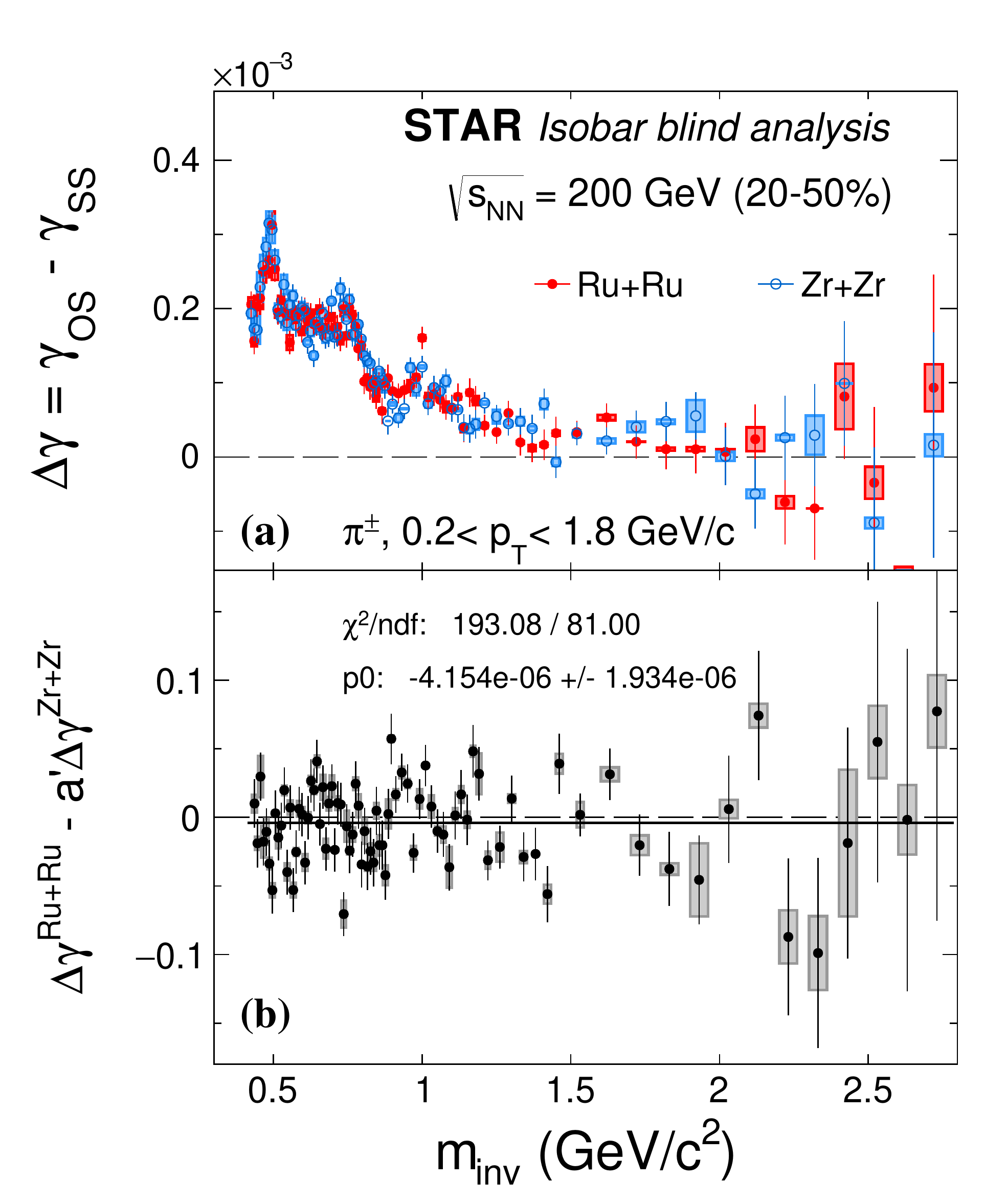}
    \caption{The $\Delta\gamma$ in 20--50\% Ru+Ru and Zr+Zr collisions (a) and their difference defined by Eq.~(\ref{eq:minvDg}) (b) as functions of the $\pi^+\pi^-$ invariant mass $m_{\rm inv}$. The difference in the lower panel would measure the possible CME if the background in $\Delta\gamma$ scales with $v_2$ only ($a'=v_2^{\rm Ru+Ru}/v_2^{\rm Zr+Zr}$ as defined by Eq.~(\ref{eq:minvaprime})). Error bars are statistical and shaded boxes are systematic uncertainties. The solid line in the lower panel is a constant fit to the data.}
    \label{fig:Dg_minv}
\end{minipage}
\end{figure*}

Figure~\ref{fig:r_minv} shows the distributions in the relative pair multiplicity difference of Eq.~(\ref{eq:r}) in Ru+Ru and Zr+Zr collisions in the 20--50\% centrality range in the upper panel and their  ratio in the lower panel. The  ratio has a weak dependence on $m_{\rm inv}$, with an average value in the 20--50\% centrality range of
$r^{\rm Ru+Ru}/r^{\rm Zr+Zr}=0.9705 \pm 0.0008 \mbox{ (stat.)}$. It deviates from unity because the isobar systems do not have the same multiplicity when their centrality defined by cross section percentile is matched (see Sec.~\ref{sec:centrality}). Note that the $r$ is measured with pion pairs; it does not necessarily equal that of charged hadrons. Also note that the $r$ ratio does not necessarily equal the inverse multiplicity ratio because the difference $N_{\textsc{os}}-N_{\textsc{ss}}$ may not strictly scale with multiplicity.

The upper panel of Fig.~\ref{fig:Dg_minv} shows the $\Delta\gamma$ results in Ru+Ru and Zr+Zr collisions in the 20--50\% centrality range as a function of $m_{\rm inv}$. Resonance peaks are observed in $\Delta\gamma$ corresponding to those in $r$ in Fig.~\ref{fig:r_minv}. The lower panel shows the $\Delta\gamma$ difference for the isobars after the $\Delta\gamma$ for Zr+Zr is scaled by the $v_2$ ratio (see Eq.~(\ref{eq:minvDg})). A constant fit to the measured difference in the 20--50\% centrality range yields
$\Delta\gamma^{\rm Ru+Ru}-a'\Delta\gamma^{\rm Zr+Zr}=(-4 \pm 2 \mbox{ (stat.)} \pm 6 \mbox{ (syst.)})\times10^{-6}$. 
The predefined CME signature of a positive value for this difference (Eq.~(\ref{eq:minvsign})) is not observed.

As described in Sec.~\ref{sec:minv}, the predefined CME signature described in Eq.~(\ref{eq:minvsign}) explicitly assumes the $r$ ratio to be unity. Since this assumption is no longer valid for the blind analysis binned in cross-section percentile, as shown in Fig.~\ref{fig:r_minv} lower panel,  
the relevance of the result in Fig.~\ref{fig:Dg_minv}  
to the possible CME needs to be reevaluated.

\subsection{CME fraction utilizing spectator and participant planes:  approach-I (Group-3)}

The CME signal fraction, $f_{\textsc{cme}}$, is extracted from two $\Delta\gamma$ measurements in each of the two isobar systems independently. One measurement is with respect to the second-order harmonic plane reconstructed from mid-rapidity particles measured in the TPC, as a proxy for the PP. The other is with respect to the first-order harmonic plane reconstructed from spectator neutrons measured by the ZDC Shower Maximum Detectors (ZDC-SMDs), as a proxy for the spectator plane. The details of this spectator-participant plane method to extract $f_{\textsc{cme}}$ is described in Sec.~\ref{sec:Group3SPPP}.
To select good events we require, in addition to those criteria described in Sec.~\ref{sec:detector}, the VPD primary vertex position to be within $| V_{z,\textsc{tpc}} - V_{z,\textsc{vpd}} | < 3$~cm from the one reconstructed by the TPC. In this analysis both the full-event and sub-event methods are used as in Ref.~\cite{STAR:2021pwb}. 
The sub-event method is useful to suppress non-flow effects. 

For the full-event analysis, all three particles are charged hadrons taken from $|\eta|<1$. The $\Delta\gamma\{\textsc{tpc}\}$ is calculated by the three-particle cumulant method (Eq.~(\ref{eq:abc})). An $\eta$ gap of 0.05 is applied between the POIs ($\alpha$ and $\beta$); no $\eta$ gap is applied between particle $c$ and either of the POIs. The $v_{2,c}$ used in Eq.~(\ref{eq:abc}) and the $v_2\{\textsc{tpc}\}$ needed by Eq.~(\ref{eq:a}) are equal and are calculated by the two-particle cumulant method of Eq.~(\ref{eq_v2pc}), where no $\eta$ gap is applied between the two particles.

For the sub-event analysis, the $\Delta\gamma\{\textsc{tpc}\}$ and $v_2\{\textsc{tpc}\}$ are calculated by the EP method (Eqs.~(\ref{eq_v2ep}) and~(\ref{eq_gmmaep})). Each TPC event is divided into two sub-events with $-1<\eta<-0.05$ and $0.05<\eta<1$ (thus an $\eta$ gap of 0.1 in between). 
The POIs are charged particles from one sub-event, and the EP is calculated using charged particles from the other sub-event with $0.2<p_T<5$~GeV/$c$.
An $\eta$ gap of 0.05 is applied between the POIs ($\alpha$ and $\beta$).
For EP reconstruction, the azimuthal nonuniformity of the efficiency and acceptance of the TPC is corrected by applying a multi-dimensional $\phi$-dependent weight. The shifting method is performed to further flatten the EP azimuthal distribution.
For $\Delta\gamma\{\textsc{zdc}\}$, the POIs are still from one sub-event to keep the same acceptance as that for $\Delta\gamma\{\textsc{tpc}\}$. 

For both the full-event and sub-event methods, the $\Delta\gamma\{\textsc{zdc}\}$ and $v_2\{\textsc{zdc}\}$ are calculated by the EP method, where the $\gamma$ correlators are given by
\begin{equation}
    \gamma=\left<\cos(\phi_{\alpha}+\phi_{\beta}-2\Psi_{\textsc{zdc}})\right>\,,
\end{equation}
and
\begin{equation}
    v_2\{\textsc{zdc}\}=\left<\cos2(\phi-\Psi_{\textsc{zdc}})\right>\,.
\end{equation}
The ZDC EP angle $\Psi_{\textsc{zdc}}$ is calculated from the $Q$-vector combined from both ZDC-SMDs. Recentering and shifting are applied to flatten the azimuthal distribution of the reconstructed ZDC EP.
The TPC and ZDC EP resolutions are obtained by the conventional method of an iterative procedure~\cite{Poskanzer:1998yz}, using correlations between two $\eta$ sub-events for the former and between the two ZDC-SMDs for the latter.

The systematic uncertainties are assessed according to Sec.~\ref{sec:syst}. In addition, the $\eta$ gap between the POIs is varied from 0.05 (default) to 0 (i.e.~no gap) and 0.2. For the sub-event method, the $\eta$ gap between the sub-events is varied from 0.1 (default) to 0.3. All systematic uncertainties are added in quadrature.
\begin{figure*}[hbt]
    \centering
    \includegraphics[width=0.8\textwidth]{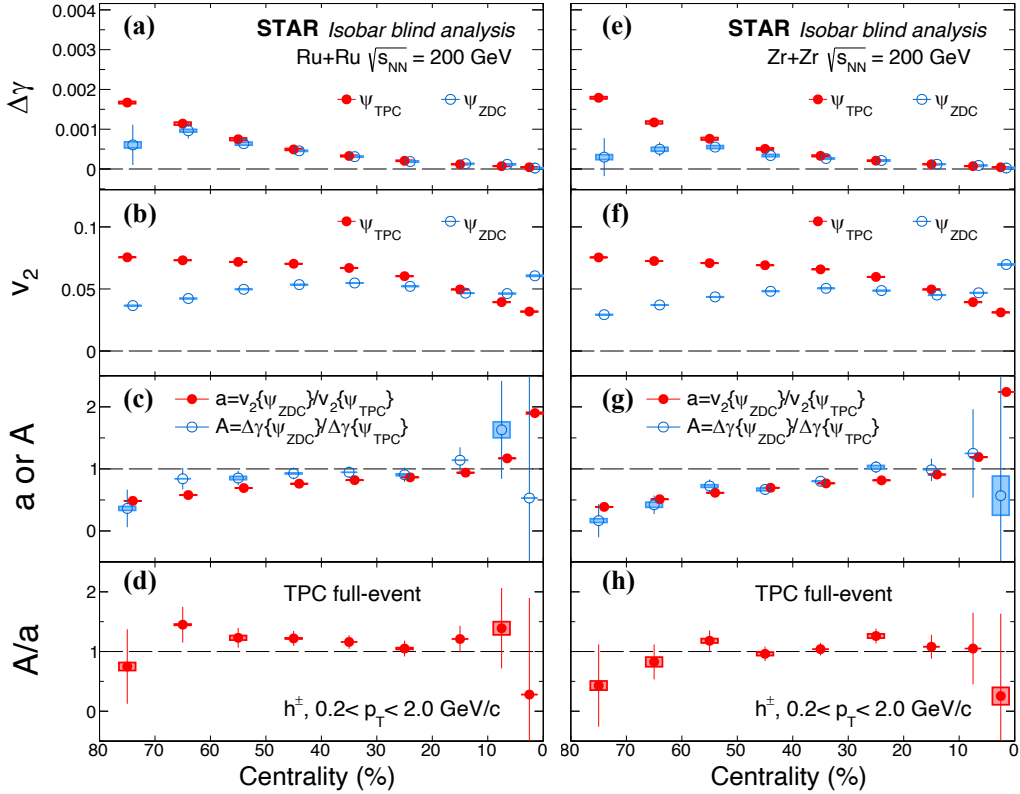}
    \caption{The $\Delta\gamma$ (top row), $v_2$ (second row), $A$ or $a$ (third row), and $A/a$ (bottom row) as functions of centrality in Ru+Ru (left column) and Zr+Zr (right column) collisions. The $\Delta\gamma$ and $v_2$ are calculated with respect to the TPC and ZDC harmonic planes. Results are from the full-event method. Error bars are statistical and shaded boxes are systematic uncertainties. 
    }
    \label{fig:Aa}
\end{figure*}

\begin{figure}[hbt]
    \centering
    \includegraphics[width=0.5\textwidth]{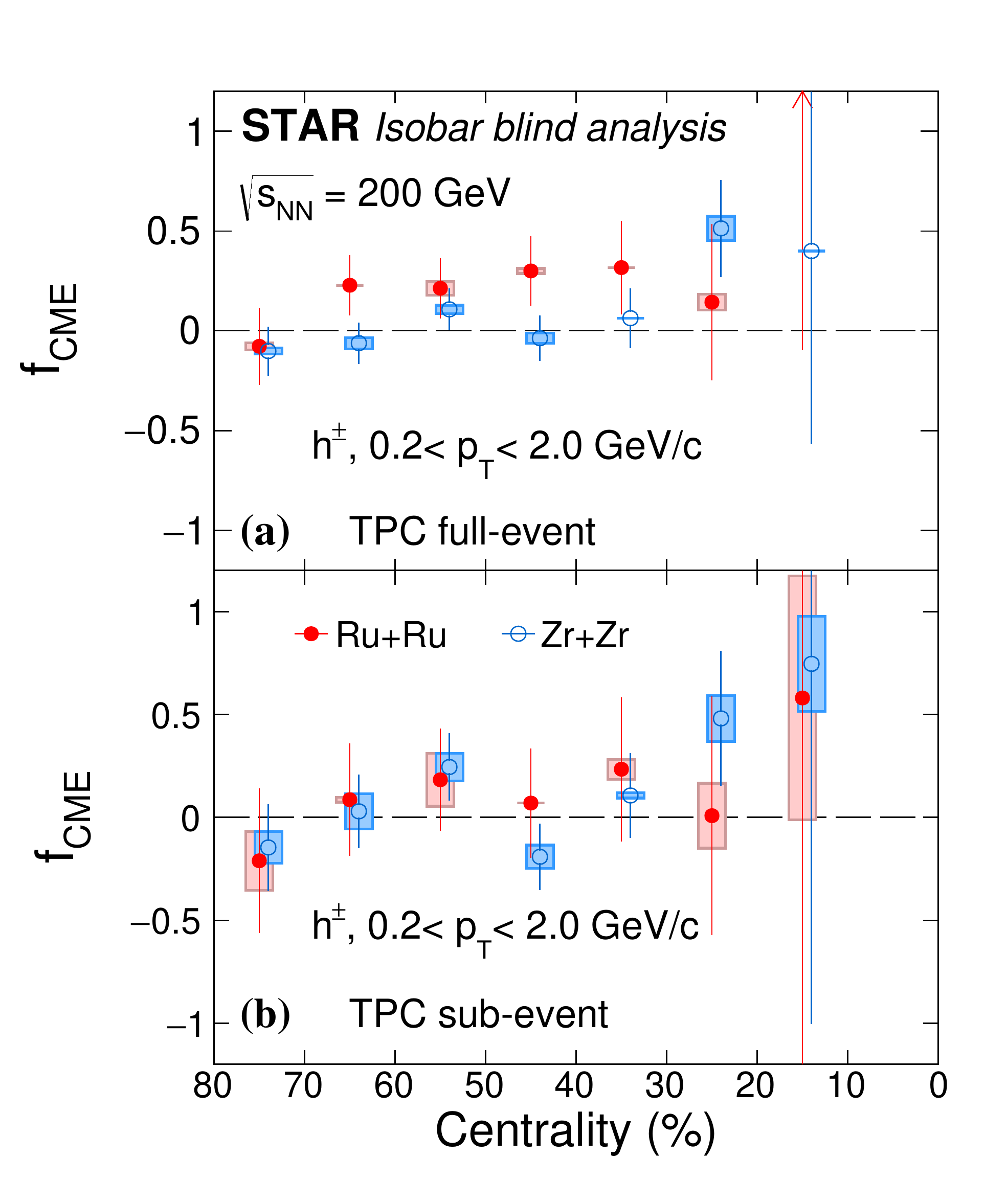}
    \caption{The CME signal fraction in the inclusive $\Delta\gamma$ measurement with respect to the TPC EP, $f_{\textsc{cme}}$, as functions of centrality in Ru+Ru and Zr+Zr collisions at 200~GeV from both the full-event method (a) and the sub-event method (b). Error bars are statistical and shaded boxes are systematic uncertainties.
    }
    \label{fig:fcme}
\end{figure}
\begin{figure*}[hbt]
    \centering
    \includegraphics[width=0.45\textwidth]{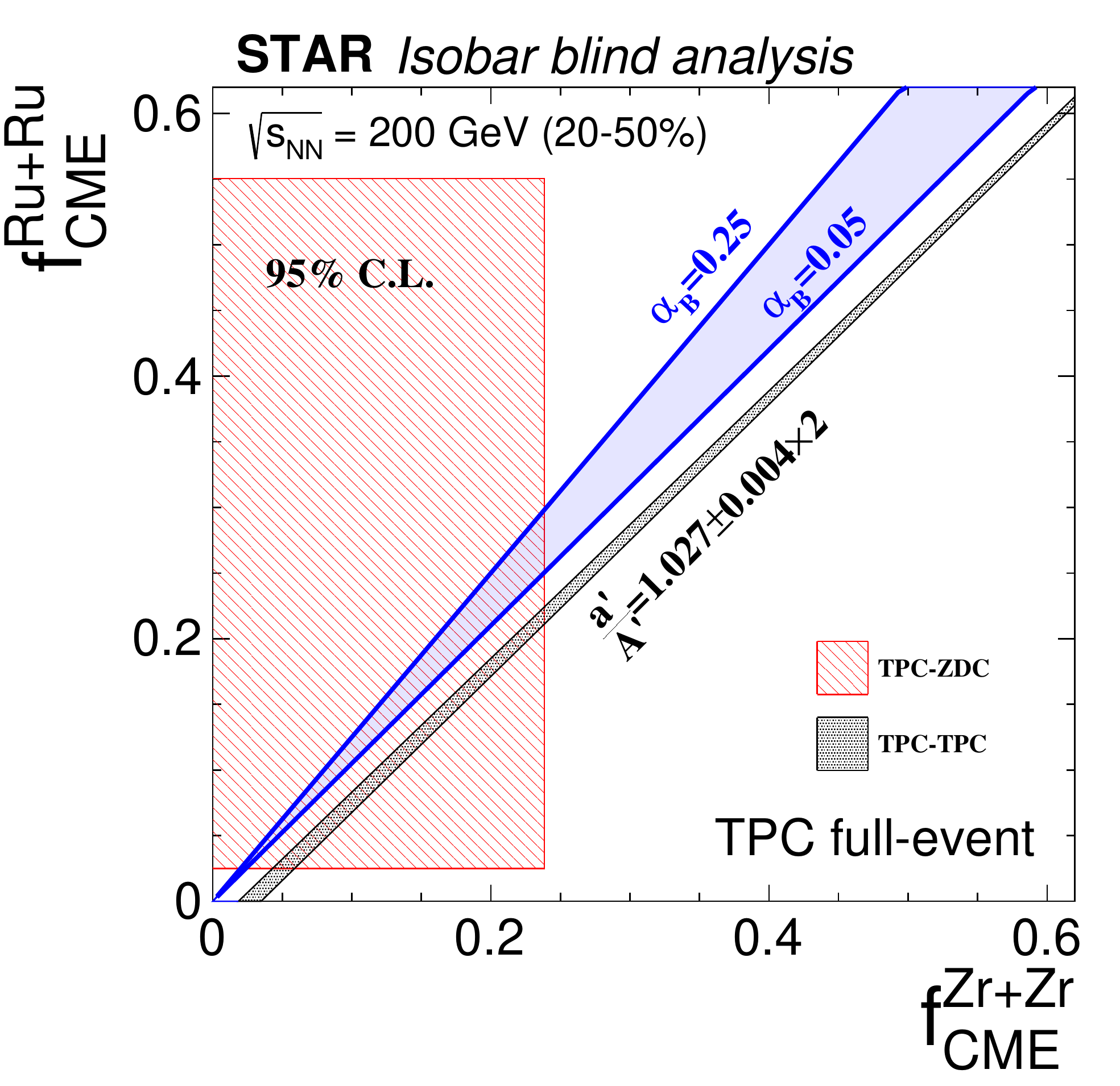}
    \hspace{0.04\textwidth}
    \includegraphics[width=0.45\textwidth]{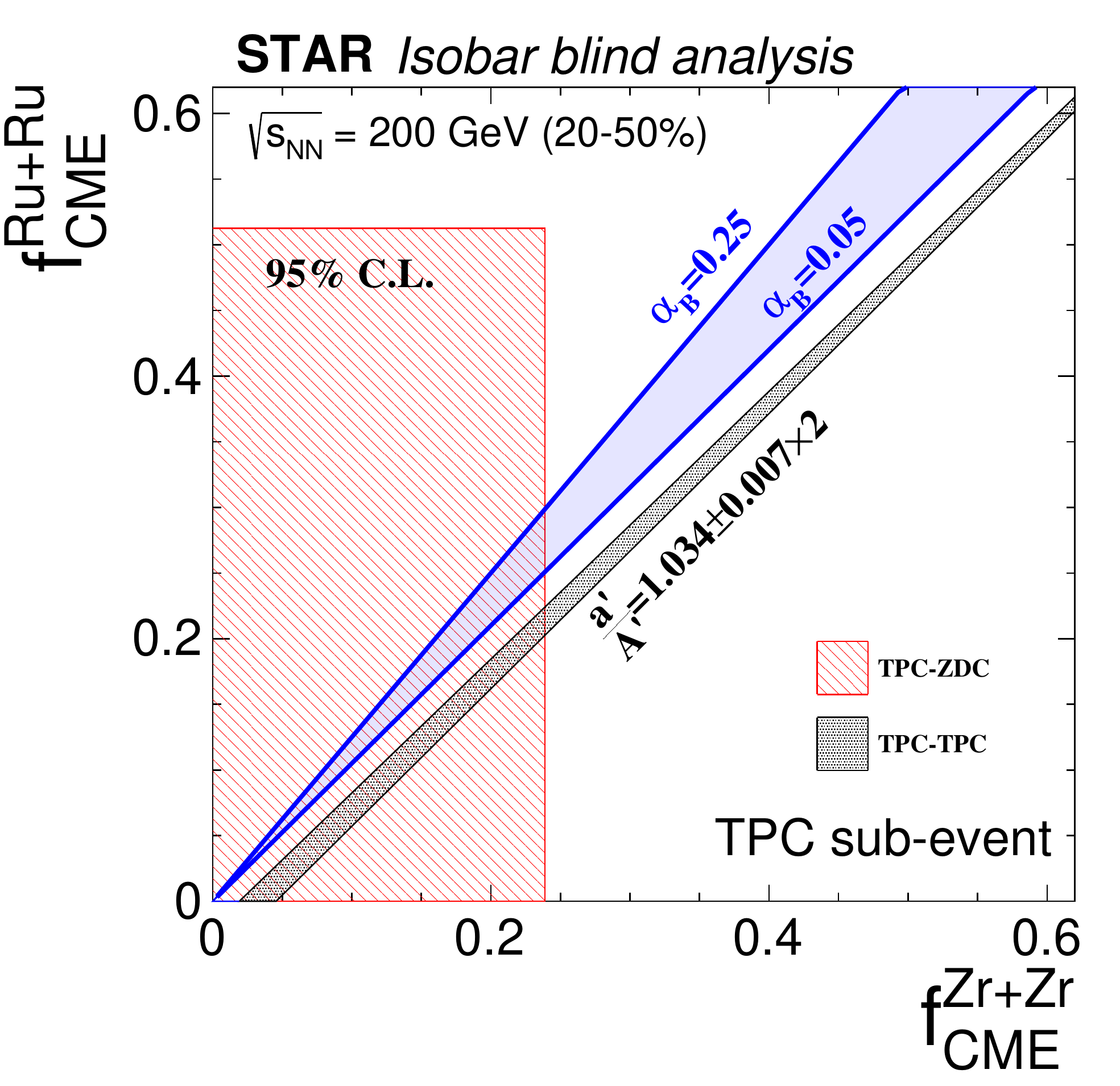}
    \caption{Constraints on $f_{\textsc{cme}}$ in Ru+Ru vs.~$f_{\textsc{cme}}$ in Zr+Zr collisions, both for 20--50\% centrality, with 95\% confidence level ($\pm2\sigma$). The left panel shows the full-event results and the right panel shows the sub-event results. The shaded rectangle on the left is the constraint extracted from each individual isobar collision system using the spectator-participant plane method and Eq.~(\ref{eq:fcme}). The near-diagonal shaded strip would be the constraint extracted from the TPC measurements combining both isobar  systems using Eq.~(\ref{eq:ff}), provided the background in $\Delta\gamma$ scales with $v_2$ only ($a'=v_2^{\rm Ru+Ru}/v_2^{\rm Zr+Zr}$ as defined by Eq.~(\ref{eq:minvaprime})). The two solid lines indicate the expectation from the magnetic field difference: $f_{\textsc{cme}}^{\rm Ru+Ru}/f_{\textsc{cme}}^{\rm Zr+Zr}=(1+\alpha_B)\Delta\gamma^{\rm Zr+Zr}/\Delta\gamma^{\rm Ru+Ru}\approx1+\alpha_B$, where $\alpha_B = 0.15\pm0.05$ is assumed. 
    }
    \label{fig:ff}
\end{figure*}

Figure~\ref{fig:Aa} shows the results from the full-event method as functions of centrality in Ru+Ru (left column) and Zr+Zr (right column) collisions. The $\Delta\gamma$ and $v_2$ with respect to the TPC and ZDC harmonic planes are shown in the first and second rows. 
The third row shows the ratios $A$ (Eq.~(\ref{eq:A})) and $a$ (Eq.~(\ref{eq:a})). The double ratio $A/a$ is shown in the bottom row. The measured $v_2\{\Psi_{\textsc{zdc}}\}$ in central collisions is noticeably large, which is not observed in the results from another group reported in Sec.~\ref{sec:Group4results}. A difference in the analysis method lies in the way to calculate the ZDC harmonic planes. In this analysis, the first-order harmonic $Q$-vectors from the two ZDCs are first combined and then the $\Psi_{\textsc{zdc}}$ is computed. In the analysis in Sec.~\ref{sec:Group4results}, the correlation is performed with the sum of the two first-order harmonic planes separately reconstructed in each ZDC.  Correspondingly, the EP resolutions are calculated in different ways.
The reason for the discrepancy needs further investigation.

We calculate the $f_{\textsc{cme}}$ using Eq.~(\ref{eq:fcme}). This is the fraction of the CME contribution to the $\Delta\gamma\{\textsc{tpc}\}$ with respect to the TPC EP. The results from both the full-event and sub-event methods are shown in Fig.~\ref{fig:fcme} as a function of centrality for the two isobar collision systems. The average $f_{\textsc{cme}}$ values in the 20--50\% centrality range from the full-event method in Ru+Ru and Zr+Zr collisions are 
$f_{\textsc{cme}}^{\rm Ru+Ru} = 0.29 \pm 0.13 \mbox{ (stat.)} \pm 0.01 \mbox{ (syst.)}$ and 
$f_{\textsc{cme}}^{\rm Zr+Zr} = 0.06 \pm 0.08 \mbox{ (stat.)} \pm 0.02 \mbox{ (syst.)}$, 
respectively. The corresponding ratios from the sub-event method are
$f_{\textsc{cme}}^{\rm Ru+Ru} = 0.12 \pm 0.20 \mbox{ (stat.)} \pm 0.00 \mbox{ (syst.)}$ and 
$f_{\textsc{cme}}^{\rm Zr+Zr} = -0.01 \pm 0.12 \mbox{ (stat.)} \pm 0.03 \mbox{ (syst.)}$.
Systematic variations for $f_{\textsc{cme}}^{\rm Ru+Ru}$ are all consistent with statistical fluctuations so a null systematic uncertainty is assigned according to the Barlow prescription~\cite{Barlow:2002yb}.
The large statistical uncertainties are dominated by the $\Delta\gamma$ measurements with respect to the ZDCs which have poor EP resolutions.

Figure~\ref{fig:ff} plots $f_{\textsc{cme}}$ for Ru+Ru collisions on the vertical axis versus $f_{\textsc{cme}}$ for Zr+Zr collisions on the horizontal axis, both at 20--50\% centrality, extracted using both the full-event method (left panel) and sub-event method (right panel). 
An additional constraint is obtained by combining the $\Delta\gamma$ measurements with respect to the TPC EP in both isobar collision systems, as described in Sec.~\ref{sec:Group3SPPP}. This is shown in the near-diagonal shaded strip given by Eq.~(\ref{eq:ff}) using the measured values for the double ratio (Eq.~(\ref{eq:Aa})) of
$a'/A' = 1.027 \pm 0.004 \mbox{ (stat.)} \pm 0.001 \mbox{ (syst.)}$ and 
$1.034 \pm 0.006 \mbox{ (stat.)} \pm 0.003 \mbox{ (syst.)}$,
for the full-event and sub-event methods, respectively.
This would be the correct constraint if the background in $\Delta\gamma$ scales with $v_2$ only (i.e.~the multiplicities are explicitly assumed to be identical between the isobar systems).  
Since this assumption is no longer valid for the blind analysis as function of the cross-section percentile, the near-diagonal strip does not correctly indicate the allowed CME region. Indeed, as shown in Fig.~\ref{fig:ff}, the present near-diagonal strip does not have overlap with the CME region enclosed by the blue solid lines expected from the magnetic field difference.
The relevance of the near-diagonal strip to $f_{\textsc{cme}}$ needs to be revisited in the future by using the properly scaled $a'$.

\subsection{Ratio of $(\dgamma/v_2)$ between two isobar collisions (Group-4)}
\label{sec:Group4results}

One of the main objectives of Group-4 is to obtain the
double ratio $(\dgamma/v_2)_{\ru}/(\dgamma/v_2)_{\zr}$ as a function of centrality.
The quantity $(\dgamma/v_2)$ is calculated as
\begin{eqnarray}
(\Delta\gamma/v_2)_{\rm TPC} &=& \frac{ \Delta
  \langle\cos(\phi_{\alpha} +\phi_{\beta}-2\phi_c)\rangle}{
  \langle\cos(2\phi_\alpha-2\phi_c)\rangle },
\label{eq:dg_tpc}
\end{eqnarray}
where $\Delta$ denotes the difference in the $\gamma$ correlator calculated using opposite and same-charge pairs of particles $\alpha$ and $\beta$.  The
correlator is calculated using the subevents from pseudorapidity windows
$0.1<|\eta|<1.0$ (default) and $0.2<|\eta|<1.0$, with the event plane,
or particle ``$c$'', taken from the opposite pseudorapidity window (e.g., when $-0.1\!>\!\eta_{\alpha,\beta}\!>\!-1.0$ we take $0.1\!<\!\eta_{c}\!<\!1.0$ and vice versa) with
pseudorapidity gaps between the subsevents $\Delta\eta_{\rm sub}=0.2$ (for the default case) and
$\Delta \eta_{\rm sub}=0.3,~0.4$ (for systematic studies). To suppress the non-flow contribution,
$\langle\cos(2\phi_\alpha-2\phi_c)\rangle $ is calculated using the
same-charge particles in the default case and using all charged particles when investigating systematic uncertainties.  All particles are taken from the transverse momentum region
$0.2<p_T<2.0$~GeV/$c$. The results are calculated in 5\% centrality bins and then averaged over a wider centrality range using the inverse of squared statistical uncertainty as a weight when needed.

All quantities in this analysis are obtained with the help of the
recentered $Q$ vectors and presented as ratios, which greatly
reduces the systematic uncertainties. The systematic uncertainty has been estimated from comparison of the results obtained with different $\eta$-gaps between the sub-events, using selection criteria on quality of the TPC tracks, and comparing results from events with the event
vertex from different sides of the TPC center. In addition, in the
estimates of the elliptic flow uncertainties, the results obtained
from correlation of unlike-sign charges are also used. All the
systematic variations are found to be smaller than the statistical
uncertainties.

\begin{figure}[htb]\vspace{-0.2cm}
    \centering
    \includegraphics[width=0.5\textwidth]{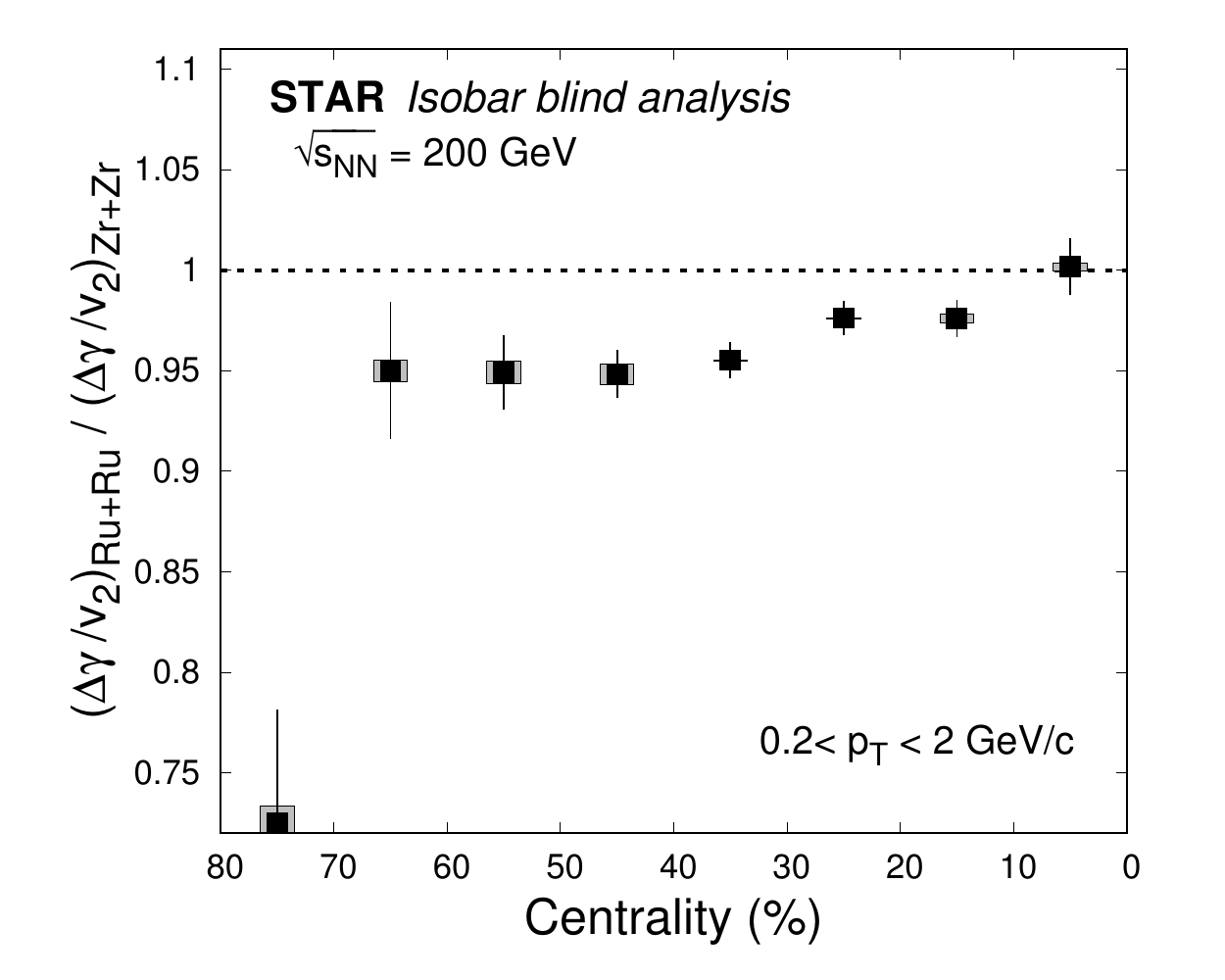}
    \caption{Double ratio $(\dgamma/v_2)_{\ru}/(\dgamma/v_2)_{\zr}  $ as a
      function of centrality for isobar collisions, where shaded boxes represent systematic uncertainties. 
      }\label{fig:Group4_dratioRuZr}
\end{figure}

For a non-zero CME signal the expectation is that the double ratio
$(\dgamma/v_2)_{\ru}/(\dgamma/v_2)_{\zr}$ would be greater than unity,
as the CME signal in \Ru\ collisions is expected to be about 15\% larger
than in \Zr\ collisions. The results of our measurements are presented
in Fig.~\ref{fig:Group4_dratioRuZr}. The plotted ratio is below unity,
which is likely due to a noticeable difference in mean charged
multiplicity in collisions of the two isobar species corresponding to the
same centrality. The multiplicity of charged particles in
\Ru\ collisions is observed to be larger than that in \Zr\ collisions as shown in Fig.~\eqref{fig_mean_refmult}. The drop of the double ratio in most peripheral events is likely due to the sudden change in the multiplicity ratio in the corresponding centrality.

The quantity $(\dgamma/v_2)$ approximately scales with the inverse
of the multiplicity, but no correction for that is anticipated in the
blind analysis. The fraction of the CME signal
contribution to $\dgamma$, if extracted exactly as outlined in the
blind analysis scheme in the 20 to 50\% centrality range would yield a negative value with an uncertainty of about 2\% of the $\dgamma$ magnitude.

\subsection{CME fraction utilizing spectator and participant planes: approach-II (Group 4)}

For the separate estimates of the CME signal in each of the isobar
collisions, the procedure outlined in section~\ref{sec:Group4_PPSP}, Eqs.~\eqref{eq:dg_zdc}--\eqref{eq:Group4_fcmeSP0} was used.

\begin{figure*}[htb]
    \centering
    \includegraphics[width=0.47\textwidth]{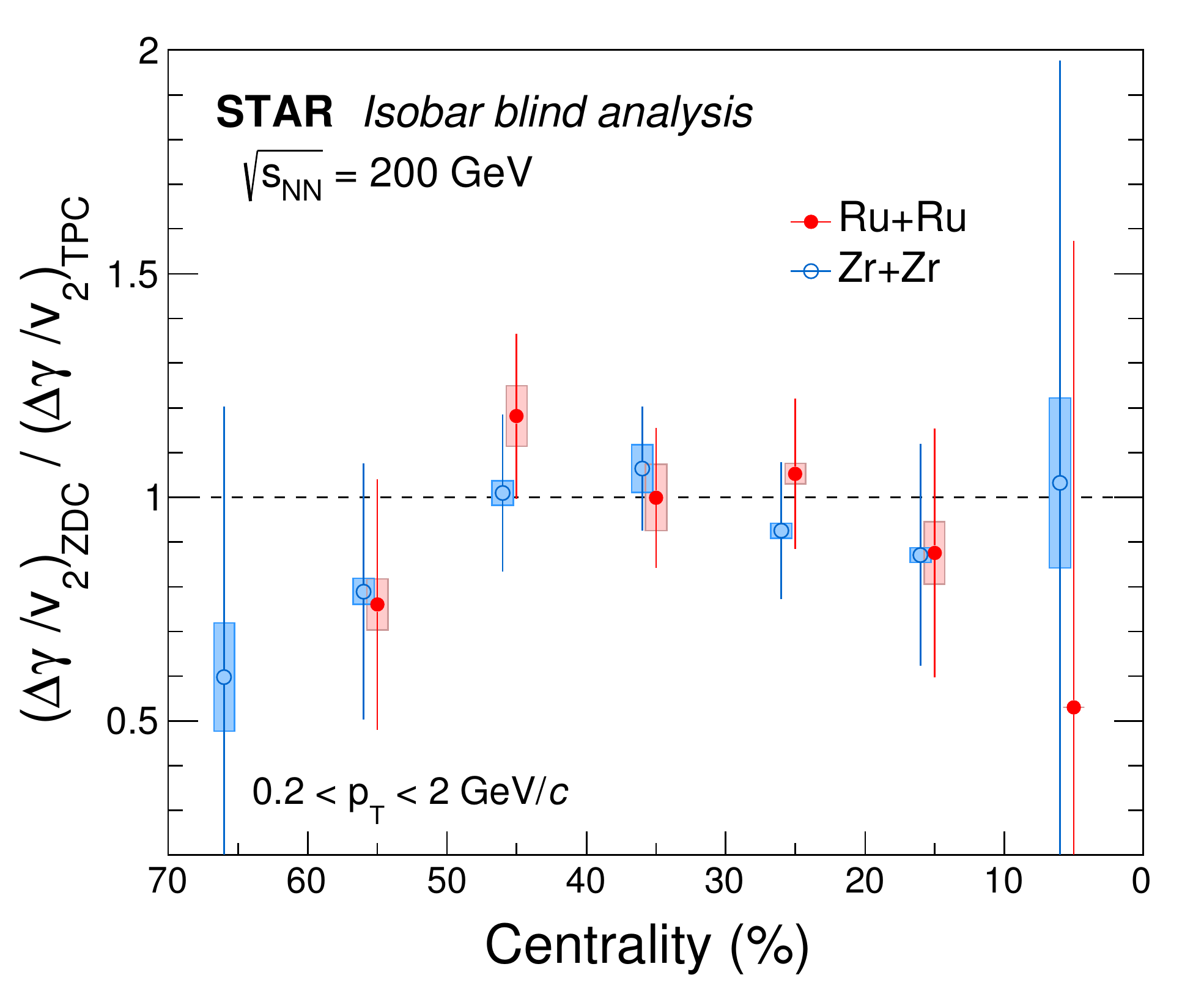}  
    \includegraphics[width=0.47\textwidth]{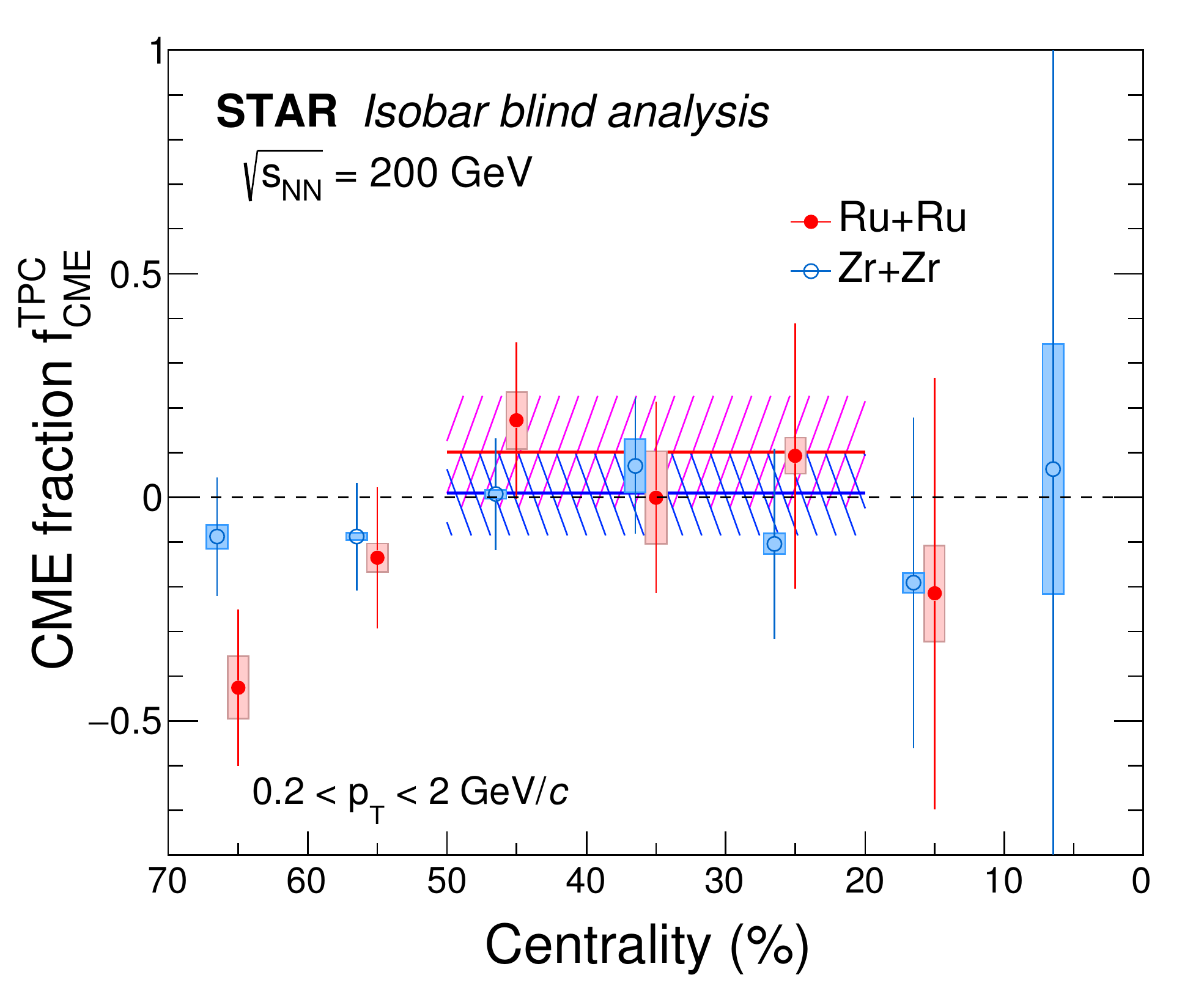}
    \caption{The ratio of $\Delta\gamma/v_2$ (left) and the CME fraction (right) for
      \Ru~and \Zr~collisions from spectator/participant plane analysis. Shaded boxes represent systematic uncertainties and hatched areas represent 1$\sigma$ uncertainties (combined statistical and systematic uncertainties) of the CME fraction for 20--50\% centrality, with the mean values indicated by horizontal solid lines. The data of Zr+Zr collisions are shifted horizontally for clarity.
      }
    \label{fig:Group4_ppsp}
\end{figure*}

\begin{figure}[htb]
    \centering
    \includegraphics[width=0.5\textwidth]{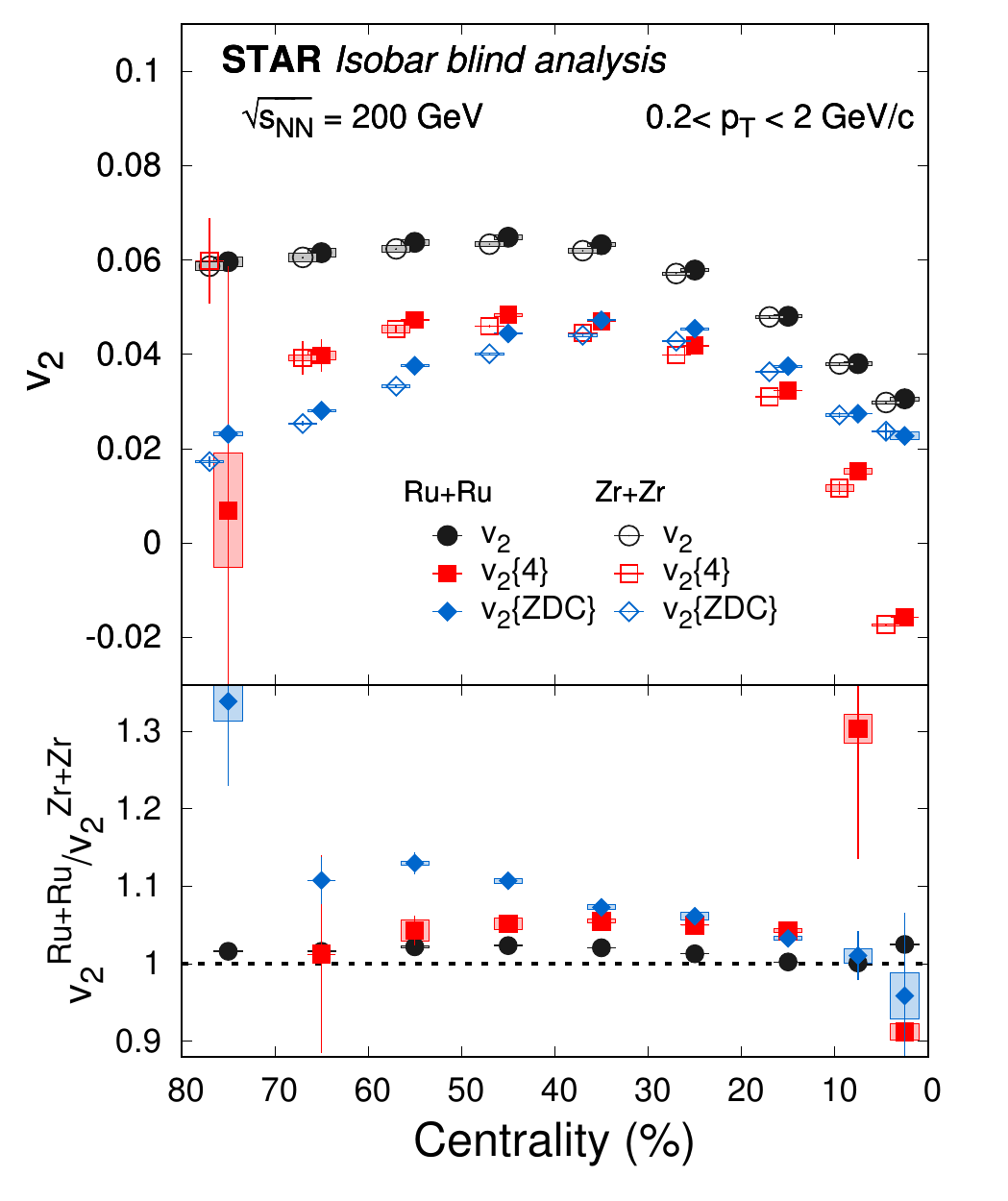}  
    \caption{$v_2\{2\}$, $v_2\{4\}$, and $v_2\{\rm ZDC\}$ for isobar
      collisions as a function of centrality in the top panel.  The
      data of Zr+Zr collisions are slightly shifted along x-axis for better
      visibility.  Ratios of $v_2$ between the two systems are plotted
      in the bottom panel. Open boxes represent systematic uncertainties.}
    \label{fig:Group4_v2}
\end{figure}

The results obtained in this approach are presented in
Fig.~\ref{fig:Group4_ppsp}. We observe that the double ratio,
Fig.~\ref{fig:Group4_ppsp} (left) is very close to unity indicating that the signal is
consistent with zero in both isobar collisions. The fraction of the CME signal 
calculated using Eq.~(\ref{eq:Group4_fcmeSP0}) is presented in
Fig.~\ref{fig:Group4_ppsp} (right), while elliptic flow calculated relative to
the participant (TPC) and spectator (ZDC-SMD) planes is presented in
Fig.~\ref{fig:Group4_v2}. The extracted average CME fraction for 20--50\% centrality is found to be $f_{\rm \sss CME}^{\rm \sss TPC}=0.101\pm0.123~({\rm stat.})\pm0.023~({\rm syst.})$ for Ru+Ru and $f_{\rm \sss CME}^{\rm \sss TPC}=0.009\pm0.088~({\rm stat.})\pm0.033~({\rm syst.})$ for Zr+Zr.
The large statistical uncertainties are dominated by the $\Delta\gamma$ measurements in the ZDCs which have poor EP resolutions. 
The statistical uncertainties on $f_{\textsc{cme}}$ are smaller than those from Group-3 reported in section~\ref{sec:Group3SPPP}, due to a larger difference in $v_2\{\rm ZDC\}$ and $v_2\{\rm TPC\}$ resulting from different approaches of correlating particles at midrapidity with signals from the two ZDCs.

\clearpage
\subsection{$R_{\Psi_{2}}$ measurements (Group-5)\label{sec:Rresult}}

In this part of the analysis, charged particles with transverse momentum $0.2<p_T<2.0$~GeV/$c$ are used to construct $\Psi_{2}$. Each event is subdivided into two sub-events with pseudorapidity $0.1<\eta<1.0$ (West) and $-1.0<\eta<-0.1$ (East) to obtain $\Psi_{2}^{\mathrm{W}}$ (West) and $\Psi_{2}^{\mathrm{E}}$ (East).
Afterward,  $C_{\Psi_{2}}(\Delta S)$, $C_{\Psi_{2}}^{\perp}(\Delta S)$ and $R_{\Psi_{2}}(\Delta S)$  are constructed using charged particles with $0.35<p_T<2.0$~GeV/$c$.
To avoid potential self-correlations, $\Psi_{2}^{\mathrm{E}}$ is used for particles within the $0.1<\eta<1.0$ range and $\Psi_{2}^{\mathrm{W}}$ for particles within the $-1.0<\eta<-0.1$ range. Here the $\Delta S$ distributions associated with the aforementioned quantities are symmetrized around $\Delta S =0$. 
The second $p_T$ selection (beginning at 0.35 GeV/$c$) is chosen to minimize the influence of acceptance effects at low $p_T$ while optimizing the statistics.

The sensitivity of the $R_{\Psi_{2}}(\Delta S^{''})$ distribution to the potential impact from $v_2$-driven background is investigated using event-shape selection via fractional cuts on the magnitude of the second harmonic flow $Q$-vector $q_{2}$ relative to its maximum value $q_{2,\text{max}}$ at fixed multiplicity~\cite{Schukraft:2012ah}. 
This study is motivated by the fact that ${v_2}$ drives background sources of CME and the change in $q_2$ provides a lever-arm to vary $v_2$~\cite{Acharya:2017fau,Zhao:2018ixy}. 
Therefore, the impact of the $v_2$-driven charge separation background can be decreased (increased) by choosing events with smaller (larger) $q_2$ values.

\begin{figure*}[htb]
    \centering
    \includegraphics[width=0.65\textwidth]{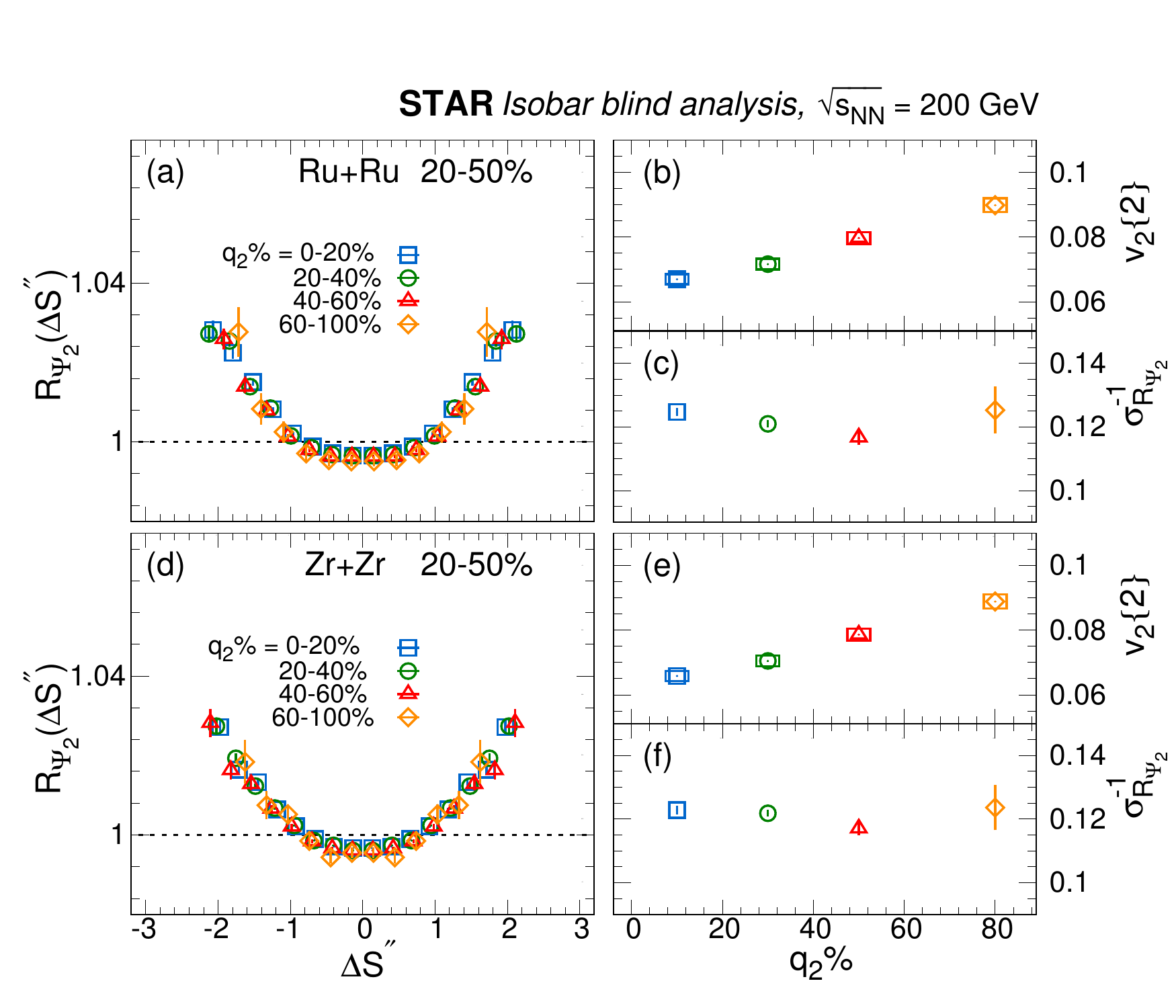}
    \caption{The $q_2$ dependence of the $R_{\Psi_{2}}(\Delta S^{''})$ distributions for Ru+Ru (a) and Zr+Zr (d) for 20--50\% collisions. Panels (b) and (e) show the corresponding $q_2$-dependent $v_2$ values; panels (c) and (f) show the inverse widths (${\sigma^{-1}_{R_{\Psi_2}}}$) for distributions 
in (a) and (d), respectively. The distributions shown in (a) and (d) are symmetrized around $\Delta S^{''} =0$.
} 
    \label{Rcorr:Fig1}
\end{figure*}

Event-shape selection is performed using three sub-events;  $A[\eta < -0.3]$, $B[|\eta| < 0.3]$, and $C[\eta > 0.3]$, following the methods described earlier, and with  $q_2$ selections in sub-event $B$. 
Figure.~\ref{Rcorr:Fig1} shows the $q_2$-selected isobar measurements. The $R_{\Psi_{2}}(\Delta S^{''})$ distributions are given in panels a and d, and the corresponding $v_2$ values, measured using the two sub-event cumulants method~\cite{Jia:2017hbm} and particles with $0.35<p_T<2.0$~GeV/$c$ are shown in panels b and e. The inverse widths (panels c and f) are extracted from the distributions shown in (panels a and d).
Linear fits to the data in panels (b), (c), (e), and (f) indicate that, while $v_2$ shows a $32.0\% \pm 0.01\%$ increase with $q_2$  from $q_2$=0-20\% to 60-100\%, the corresponding inverse width for the $R_{\Psi_2}(\Delta S^{''})$ distributions show an approximate decrease of $7.0\% \pm 4.0\%$.
 Further studies may be needed to understand the physics behind the observed behavior of the widths of $R_{\Psi_2}$ on $q_2$.

\begin{figure*}[htb]
    \centering
    \includegraphics[width=0.9\textwidth]{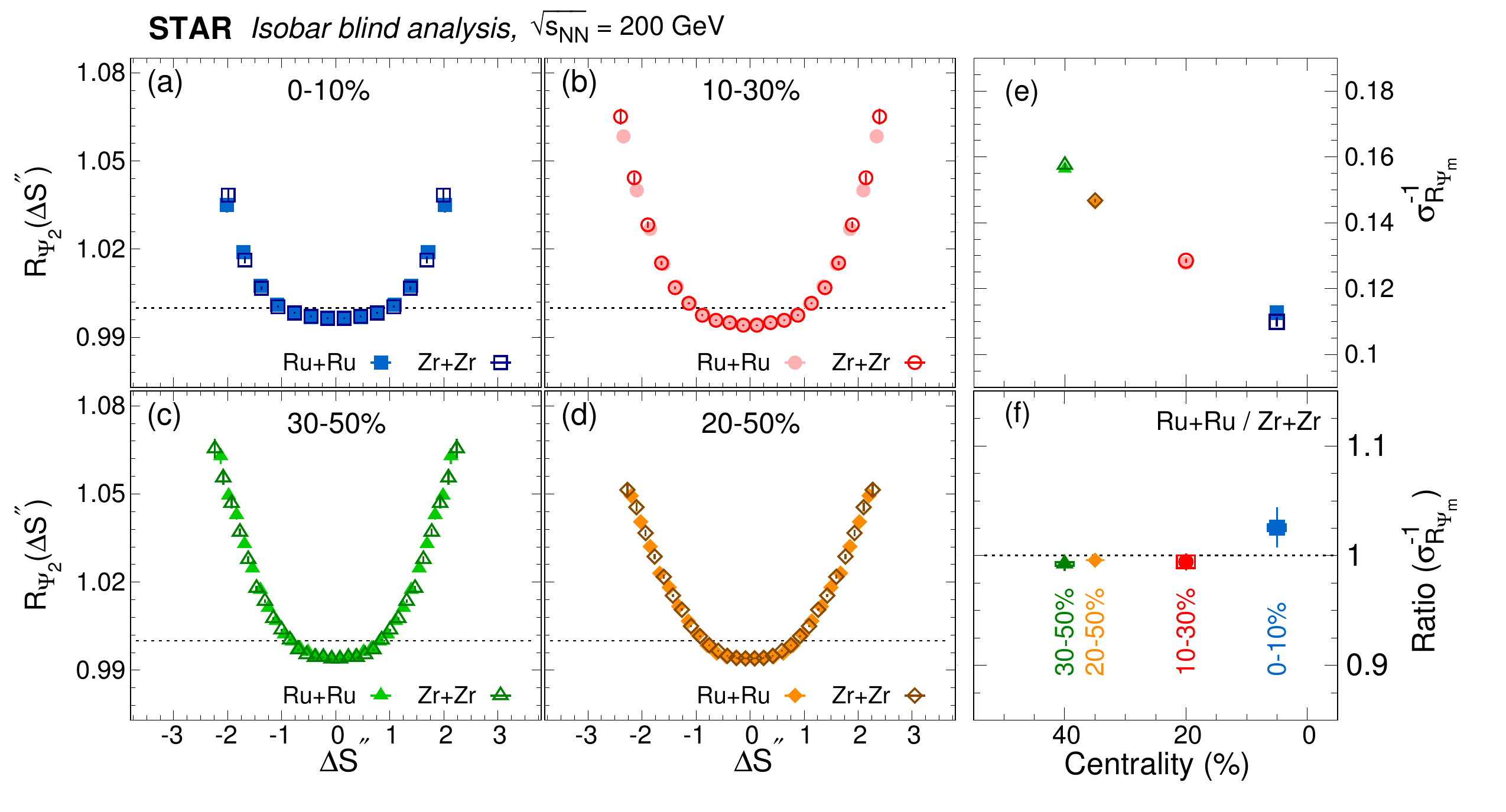}
    \caption{Comparison of the $R_{\Psi_{2}}(\Delta S^{''})$ distributions obtained for charged particles in (a) 0-10\%, (b) 10-30\%, (c) 30-50\% and (d) 20--50\% collisions in Ru+Ru and Zr+Zr collisions at $\sqrt{s_{_{\rm NN}}})=200$ GeV. Panel (e) shows the centrality dependence of the inverse widths ${\sigma^{-1}_{R_{\Psi_2}}}$, extracted from the  $R_{\Psi_{2}}(\Delta S^{''})$ distributions. Panel (f) shows the ratio of the inverse widths of the two isobars. {The distributions shown in (a)-(d) are symmetrized around $\Delta S^{''} =0$}.}.
    \label{Rcorr:Fig2}
\end{figure*}

The  $R_{\Psi_{2}}(\Delta S^{''})$ distributions, extracted for several centrality selections in Ru+Ru and Zr+Zr collisions, 
are shown in Fig.~\ref{Rcorr:Fig2} (a-d). 
They indicate centrality-dependent concave-shaped distributions for $R_{\Psi_2}(\Delta S^{''})$. 
The corresponding inverse widths extracted from these distributions are shown in panel (e). They indicate similar magnitudes for both isobars that increase as collisions become more peripheral.
The difference between the inverse widths for the two isobars is made more transparent in Fig.~\ref{Rcorr:Fig2}(f), where the ratios $\sigma^{-1}_{R_{\Psi_2}}(\rm{Ru+Ru})/\sigma^{-1}_{R_{\Psi_2}}(\rm{Zr+Zr})$ are plotted as a function of collision centrality. Note that the systematic uncertainty is negligible compared to the statistical uncertainties for the 20--50\% selection.

\subsection{Summary and discussions\label{sec:summary}}
\begin{figure*}[htb]
    \centering
    \includegraphics[width=0.95\textwidth]{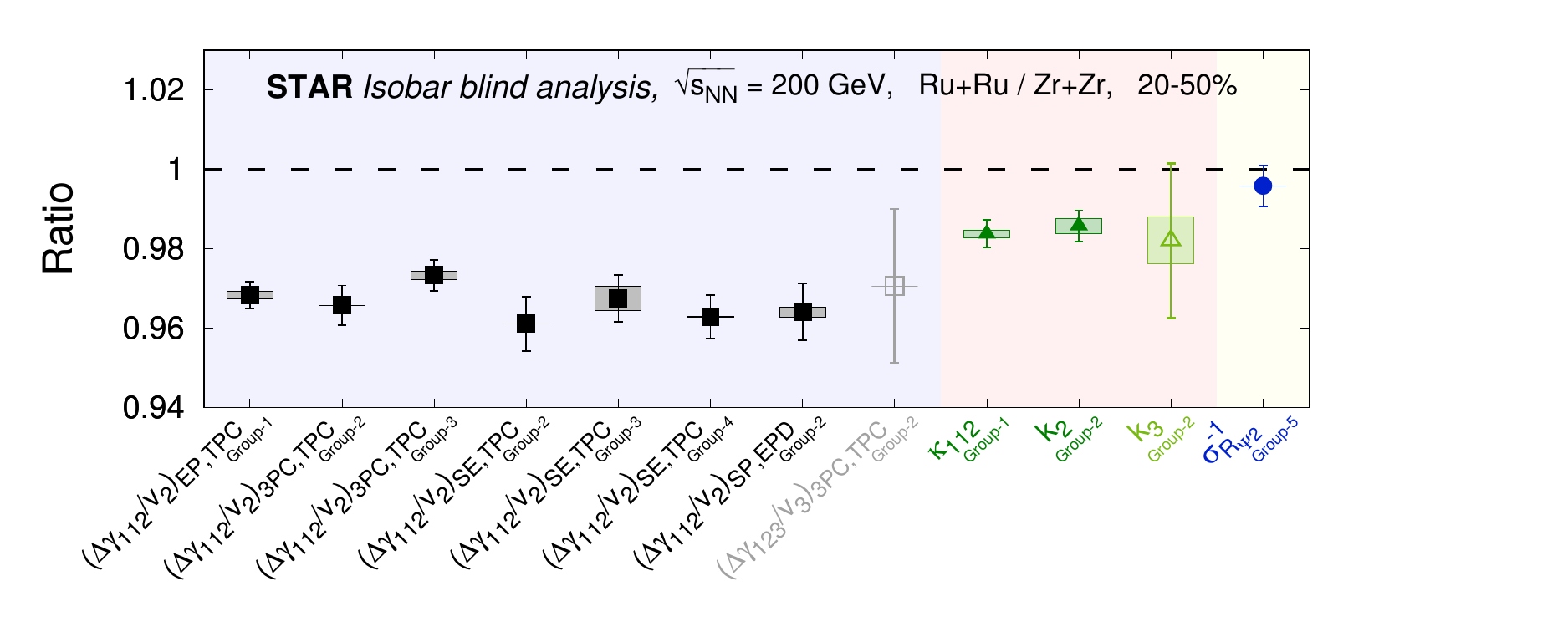}
    \caption{\label{fig_compilation}Compilation of results from the blind analysis. Only results contrasting between the two isobar systems are shown. Results are shown in terms of the ratio of measures in Ru+Ru collisions over Zr+Zr collisions. Solid dark symbols show CME-sensitive measures whereas open light symbols show counterpart measures that are supposed to be insensitive to CME. The vertical lines indicate statistical uncertainties whereas boxes indicate systematic uncertainties. The colors in the background are intended to separate different types of measures. The fact that CME-sensitive observable ratios lie below unity leads to the conclusion that no predefined CME signatures are observed in this blind analysis. 
    } 
    \end{figure*}

The elliptic flow $v_2$ coefficients are found to be larger in Ru+Ru than Zr+Zr collisions, by approximately 2\% in mid-central collisions and by a similar amount in the most central 5\% of collisions.
The shape and magnitude of the $v_2^{\rm Ru+Ru}/v_2^{\rm Zr+Zr}$ ratio as a function of centrality are consistent with the corresponding eccentricity ratio predicted by DFT calculations~\cite{Xu:2017zcn,Li:2018oec}, which can be parameterized by neutron-halo type WS distributions for the $^{96}_{40}$Zr nucleus~\cite{Xu:2021vpn}. Therefore, the current measurements are consistent with the different intrinsic nuclear structures of the two isobars. 
{The $v_2$ difference in central collisions suggests that the $^{96}_{44}$Ru nucleus is more deformed than the $^{96}_{40}$Zr nucleus. However, the ratio of multiplicity distribution is best described by MC-Glauber simulations without intrinsic shapes for both the isobars. Further studies with more sophisticated observables are underway to pin down the nuclear shape difference between $^{96}_{44}$Ru and $^{96}_{40}$Zr.} 
Using the forward detectors EPD and ZDC rather than the TPC to determine the EP leads to a noticeable change in the magnitude of $v_2$ and an even larger change in $v_3$. These changes may primarily be due to 
effects of non-flow, longitudinal de-correlation and flow-fluctuations.
An interesting observation is that the magnitudes of $v_3$ differ with high significance between the two isobars in both peripheral and central collisions, 
which warrants future investigation.

The primary CME-sensitive observable $\Delta\gamma/v_2$ is analyzed by four independent groups. Prior to the blind analysis, the case for observation of a CME signal is predefined to be an excess of $\Delta\gamma/v_2$ in Ru+Ru collisions as compared with Zr+Zr collisions. Results from all groups are inconsistent with this expectation, and therefore no conclusive evidence of the CME is found in this blind analysis. The analysis from one group uses an alternate CME-sensitive measure, namely the $R$ variable. The predefined expectation for the CME for this observable is a larger concavity of the $R$ variable in Ru+Ru collisions compared with Zr+Zr collisions. No such observation is found in the data, and therefore no conclusive evidence of the CME is observed using the $R$ variable in the blind analysis.

Figure~\ref{fig_compilation} presents a compilation of results from the blind analysis for the 20--50\% centrality range. In this figure, the ratio of the value of each observable in Ru+Ru to its value in Zr+Zr collisions is shown; the statistical and systematic uncertainties are shown by lines and boxes, respectively.  Included are results for the CME-sensitive observables $\Delta\gamma/v_2$, $\kappa$, $k$ and $1/\sigma_{R_{\Psi_2}}$ using different detector combinations as well as from  independent analysis groups. 
The ratio values of $\Delta\gamma/v_2$, $\kappa_{112}$, $k_2$, and $1/\sigma_{R_{\Psi_2}}$ 
are all less than or consistent with unity, indicating that the predefined CME signature is not observed in the isobar blind analysis for any of these observables. 
This observation is further corroborated  
by the observation that the CME-insensitive quantities $\Delta\gamma_{123}/v_3$ and $k_3$ have ratios (as shown in the figure) consistent with their second-harmonic CME-sensitive counterparts.

In addition to the integrated quantities shown in Fig.~\ref{fig_compilation}, we have performed differential measurements of $\Delta\gamma$  with $\Delta\eta$ and of $\Delta\gamma$ for pion pairs in invariant mass $m_{\rm inv}$ for both isobar species.  No difference in the shape is observed between the two species in these differential studies. 
The mean value of the variable $r$ that measures the relative excess of opposite-sign relative to same-sign pion pairs at different values of $m_{\rm inv}$ is different for the two isobar species, being smaller in Ru+Ru collisions; this is qualitatively consistent with the charged hadron multiplicity difference in bins of matching centrality between the two isobars. 

The comparison of $\Delta\gamma$ measured with respect to the spectator (measured by the ZDC) and participant (measured by the TPC) planes is used to extract the CME fraction $f_{\textsc{cme}}$ in each individual species. 
Two analysis groups used this method. Group-3 analyzed both the full-event and sub-event correlations, while Group-4 analyzed only the latter. Using the sub-events allows the suppression of non-flow correlations. The sub-event results from the two groups are consistent with each other. The statistical uncertainties on $f_{\textsc{cme}}$ from Group-3 are larger than those from Group-4, due to a smaller difference in $v_2\{\rm ZDC\}$ and $v_2\{\rm TPC\}$ resulting from different approaches of correlating  particles at midrapidity with signals from two ZDCs (see sections~\ref{sec:Group3SPPP} and~\ref{sec:Group4_PPSP}).
All these  results give a CME signal fraction that is consistent with zero with large statistical uncertainties of approximately 10\% (absolute) dominated by the ZDC measurements.

The most recent Au+Au results measured by the spectator and participant plane method from STAR indicate a 
possible CME signal fraction of the order of 10\% with a significance of 1--3$\sigma$~\cite{STAR:2021pwb}. If the CME signal fraction is also 10\% in isobar collisions, then a $3\sigma$ effect would be expected with the current isobar data sample of approximately 2 billion MB events each, according to estimations in Ref.~\cite{Deng:2016knn,starbur17}.
However, it has been pointed out and supported by AVFD simulations that the CME signal fraction may be substantially smaller in isobar collisions compared to Au+Au collisions~\cite{Feng:2021oub}. This would imply a substantially smaller significance in this isobar data sample.

\section{Post blinding\label{sec:post}}
    \begin{figure*}[htb]
    \centering
    \includegraphics[width=0.95\textwidth]{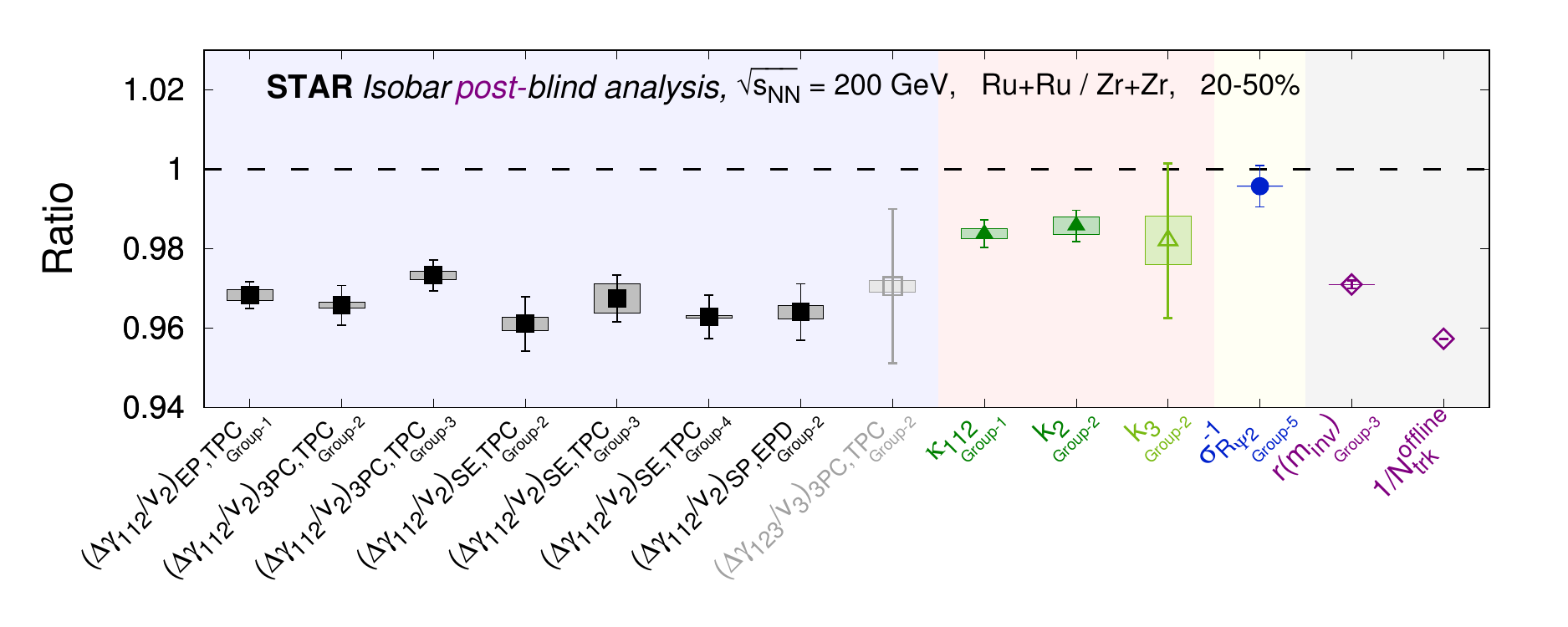}
    \caption{\label{fig_compilation_newalgorithm}Compilation of post-blinding results. This figure is largely the same as Fig.~\ref{fig_compilation} with the following differences: 
    numerical changes in the results from the new run-by-run QA algorithm are treated as an additional systematic uncertainty added in quadrature, and two data points (open markers) have been added on the right to indicate the ratio of inverse multiplicities ($N_{\rm trk}^{\rm offline}$) and the ratio of relative pair multiplicity difference ($r$) as explained in the text.}
\end{figure*}

During the second step of our analysis (the isobar blind analysis) a potential issue was identified related to the predefined criteria of the QA algorithm (as described in Sec.~\ref{sec:BlindMethod}). The condition of being within five times the weighted error or one percent of the variation of the local mean may be too relaxed to identify all the boundaries of stable run periods and outlier runs in some QA variables. When combining the identified run mini-regions, a new algorithm is implemented by 1) removing the ``within one percent of the variation of the local mean" condition, and 2) adding a tolerance of ``within 2-RMS difference", which seems to be more effective for some QA variables such as $N_{\rm fits}$. This new algorithm is again executed in the final step of isobar unblind analysis (Step-3) and all the results using this algorithm are presented in this post-blinding section. No qualitative changes are observed in the final quantities. The numerical changes in the results from this new run-by-run QA algorithm are treated as an additional systematic uncertainty to update Fig.\ref{fig_compilation} and obtain Fig.~\ref{fig_compilation_newalgorithm}. 

Two additional data points are included on Fig.~\ref{fig_compilation_newalgorithm} for the following reasons. Most ratio quantities shown in Fig.~\ref{fig_compilation} or Fig.\ref{fig_compilation_newalgorithm} have magnitudes that are below unity with high significance, whereas in a purely non-CME scenario with controlled backgrounds, the expectation is that these quantities should be consistent with unity. The reason for these ratios being less than unity is, in part, due to the multiplicity difference in the two isobar systems. As documented in Table~\ref{tab:centrality}, the multiplicity distributions are different for the two isobar species to the extent that in bins of matching centrality, the mean multiplicity is around 4\% lower for mid-central Zr+Zr than for mid-central Ru+Ru collisions. The measured magnitudes of most observables, such as $\Delta\gamma$ and $\Delta\delta$, decrease with increasing multiplicity because of the trivial multiplicity dilution for these per-pair quantities. Therefore, the corresponding ratios of these observables between the two isobar systems will become larger, if taken in bins of matching multiplicity. 
Under the approximation that background to $\Delta\gamma$ is caused by flowing clusters with the properties of the clusters staying the same and the number of clusters scaling with multiplicity, the value of $\Delta\gamma$ scales with the inverse of multiplicity~\cite{Abelev:2009ad}, i.e.~$N\Delta\gamma\propto v_2$ with the proportionality presumably equal between the two isobars.  Because of this, it may be considered that the proper baseline for the ratio of $\Delta\gamma/v_2$ between the two isobars is the ratio of the inverse multiplicities of the two systems.  
Analysis with respect to this baseline is not documented in the pre-blinding procedures of this blind analysis, so is not reported  
as part of the blind analysis. We include this inverse multiplicity ratio as the right-most point in Fig.~\ref{fig_compilation_newalgorithm}.

It is interesting to note that ordering among the quantities in their magnitudes is observed in Figs.~\ref{fig_compilation} and~\ref{fig_compilation_newalgorithm}. 
The $\Delta\gamma/v_2$ ratio has a smaller magnitude than the $\kappa$ and $k$ ratios. 
This is consistent with the multiplicity ratio baseline for the former as discussed above and the fact that the trivial multiplicity dependence cancels in the latter so its baseline would be unity. On the other hand, the $R$-variable inverse width $1/\sigma_{R_{\Psi_2}}$ ratio is larger than the $\Delta\gamma/v_2$ ratio. This difference is expected to be driven by: 1) different $p_T$ ranges used for the two quantities, 2) difference in the multiplicity dependence (see, e.g., Ref.~\cite{Choudhury:2021jwd}), and 3) difference in the non-flow contributions. The scaling relations extracted in Ref.~\cite{Choudhury:2021jwd} indicate an approximate relation between $1/\sigma_{R_{\Psi_2}}^2$, multiplicity $N$ and $\Delta\gamma$, which would imply for this analysis  $1/\sigma_{R_{\Psi_2}}^2 \approx N \Delta\gamma$; an estimate based on the measurements from this analysis  indicates this ratio for Ru+Ru over Zr+Zr to be approximately 1.02.

It is not clear that the inverse multiplicity ratio discussed above is the best baseline to use to take into account the multiplicity difference; for example an alternative would be the ratio of excess opposite-sign pairs as quantified by the variable $r$ (see Eq.~(\ref{eq:bkgd}) and Sec.~\ref{sec:minv}).  This 
$r$ ratio (from Sec.~\ref{sec:minvresult} for pion pairs) is also shown in Fig.~\ref{fig_compilation_newalgorithm}. Neither of these baselines would yield the conclusion that a clear CME signal is observed in the analyses presented in Fig.~\ref{fig_compilation_newalgorithm}.

The baselines for the CME-sensitive observables used in this blind analysis, as discussed above, are only general expectations.
The observed multiplicity difference between the isobars requires future CME analyses to better understand the baselines in order to best utilize the  precision demonstrated in this analysis. 

\section{Conclusion}

We report an experimental test of the Chiral Magnetic Effect by a blind analysis of a large statistics data set of isobar $^{96}_{44}$Ru+$^{96}_{44}$Ru and $^{96}_{40}$Zr+$^{96}_{40}$Zr collisions at nucleon-nucleon center-of-mass energy of 200~GeV, taken in 2018 by the STAR Collaboration at RHIC. The backgrounds are reduced using the difference in observables between the two isobar collision systems. The criteria for a positive CME observation are predefined, prior to the blind analysis, as a significant excess of the CME-sensitive observables in Ru+Ru collisions over those in Zr+Zr collisions.
Consistent results are obtained by the five independent groups in this blind analysis.
Significant differences in the multiplicity and flow harmonics are observed between the two systems in a given centrality, indicating that the magnitude of the CME background is different between the two species. 
A precision down to 0.4\% is achieved in the relative magnitudes of pertinent observables between the two isobar systems. 
No CME signature that satisfies the predefined criteria has been observed in isobar collisions in this blind analysis.

\section{Acknowledgement}
We thank the RHIC Operations Group and RCF at BNL, the NERSC Center at LBNL, and the Open Science Grid consortium for providing resources and support.  
We are grateful to Oak Ridge National Laboratory for providing the Ru-96, enriched in a special run, and RIKEN, Japan, for providing critical technology for the Zr-96 beam source.
This work was supported in part by the Office of Nuclear Physics within the U.S. DOE Office of Science, the U.S. National Science Foundation, the Ministry of Education and Science of the Russian Federation, National Natural Science Foundation of China, Chinese Academy of Science, the Ministry of Science and Technology of China and the Chinese Ministry of Education, the Higher Education Sprout Project by Ministry of Education at NCKU, the National Research Foundation of Korea, Czech Science Foundation and Ministry of Education, Youth and Sports of the Czech Republic, Hungarian National Research, Development and Innovation Office, New National Excellency Programme of the Hungarian Ministry of Human Capacities, Department of Atomic Energy and Department of Science and Technology of the Government of India, the National Science Centre of Poland, the Ministry  of Science, Education and Sports of the Republic of Croatia, RosAtom of Russia and German Bundesministerium f\"ur Bildung, Wissenschaft, Forschung and Technologie (BMBF), Helmholtz Association, Ministry of Education Culture, Sports, Science, and Technology (MEXT) and Japan Society for the Promotion of Science (JSPS).

\bibliographystyle{apsrev4-1}
\bibliography{refs}
\end{document}